\documentclass[fleqn,usenatbib]{mnras}

\usepackage{newtxtext,newtxmath}

\usepackage[T1]{fontenc}

\DeclareRobustCommand{\VAN}[3]{#2}
\let\VANthebibliography\thebibliography
\def\thebibliography{\DeclareRobustCommand{\VAN}[3]{##3}\VANthebibliography}

\usepackage{graphicx}	
\usepackage{amsmath}	
\usepackage{dcolumn}
\usepackage{bm}
\usepackage{hyperref}
\usepackage{color}
\usepackage{xcolor, soul}
\sethlcolor{orange} 

\newcommand{\bna}{\bm \nabla}
\newcommand{\bal}{\bm \alpha}
\newcommand{\be}{\begin{equation}}
\newcommand{\ee}{\end{equation}}
\newcommand{\ba}{\begin{eqnarray}}
\newcommand{\ea}{\end{eqnarray}}

\newcommand{\en}{\nonumber\\}
\newcommand{\de}{\delta}

\newcommand{\nab}{\nabla}

\newcommand{\da}{\dagger}
\newcommand{\kk}{\mathbf{k}}
\newcommand{\xx}{\mathbf{x}}
\newcommand{\rr}{\mathbf{r}}
\newcommand{\uu}{\mathbf{u}}

\newcommand{\ab}{{\bf A}}
\newcommand{\ktil}{{\tilde{\kappa}}}

\title[Small-scale anisotropies and interlopers]{Interlopers speak out: studying the dark universe using small-scale lensing anisotropies}

\author[B. Dhanasingham et al.]{Birendra Dhanasingham $^{1}$\thanks{E-mail: birendradh@unm.edu}, Francis-Yan Cyr-Racine $^{1}$\thanks{E-mail: fycr@unm.edu}, Annika H. G. Peter$^{2, 3, 4}$, Andrew Benson$^{5}$,~and~
\newauthor Daniel Gilman$^{6}$
\\
$^{1}$Department of Physics and Astronomy, University of New Mexico, 210 Yale Blvd NE, Albuquerque, NM 87106, USA\\
$^{2}$Department of Physics, The Ohio State University, 191 W. Woodruff Ave., Columbus, OH 43210, USA\\
$^{3}$Center for Cosmology and Astroparticle Physics, The Ohio State University, 191 W. Woodruff Ave., Columbus, OH 43210, USA\\
$^{4}$Department of Astronomy, The Ohio State University, 140 W. 18th Ave., Columbus, OH 43210, USA\\
$^{5}$ Carnegie Observatories, 813 Santa Barbara Street, Pasadena, CA 91101, USA\\
$^{6}$ Department of Astronomy and Astrophysics, University of Toronto, 50 St. George Street, Toronto, ON M5S 3H4, Canada
}

\date{Accepted 2022 October 14. Received 2022 October 13; in original form 2022 March 31}

\pubyear{2022}

\begin{document}
\label{firstpage}
\pagerange{\pageref{firstpage}--\pageref{lastpage}}
\maketitle

\begin{abstract}
Strongly lensed systems are powerful probes of the distribution of dark matter on small scales. In this paper, we show that line-of-sight haloes between the source and the observers give rise to a distinct anisotropic signature in the two-point function of the effective lensing deflection field. We show in particular that the non-linear coupling between line-of-sight haloes and the main-lens plane imprints a characteristic quadrupole moment on this two-point function whose amplitude reflects the abundance of such haloes within the strongly lensed field. We discuss how, by taking ratios of different multipole moments, such observables could be made robust under the mass-sheet transform. We also demonstrate that future extremely large telescopes have the ability to detect the quadrupole moment due to this unique anisotropic signature under ideal conditions. Our approach opens the door to statistically distinguish the effect of line-of-sight haloes from that of the main-lens substructure on lensed images, hence allowing one to probe dark matter physics in a new way.
\end{abstract}



\section{Introduction}

Our understanding of the Universe is still incomplete without a grasp of the fundamental nature of dark matter. The Lambda cold dark matter ($\Lambda$CDM) model describes observations well on large cosmological scales, but questions remain about its success deep in the non-linear regime \citep[see e.g.,][]{Bullock:2017xww}. However, testing predictions on such small length scales, where key information about the elusive nature of dark matter may lie, is challenging. There are multiple astrophysical probes of dark matter distribution on sub-galactic scales. The most popular techniques are studying the kinematics of ultra-faint dwarf galaxies \citep{Walker_2011, Salucci_2012, Kirby_2013, Laporte_2013, Teyssier_2013, Ackermann_2014, Bonnivard_2015, Calabrese_2016, Lin_2016, Caldwell_2017, Hayashi_2021}, analysing stellar wakes to detect substructure locally \citep{Buschmann_2018}, investigating the potential gravitational perturbations caused by dark matter substructure on cold stellar tidal streams \citep{Banik:2018pjp, Banik:2019cza, Banik:2019smi, Bonaca_2018, Delos_2021},  and using pulsar timing measurements \citep{Siegel_2007, Clark_2016, Ramani_2020}. Recent analyses of the data collected by deep photometric surveys have made discoveries of a number of ultra-faint dwarf galaxies and increased the Milky Way satellite galaxy count, which thus provides tools to infer the substructure mass function and dark matter microphysics since the satellite galaxies are assumed to reside in subhaloes \citep{Vale_2006, Ackermann_2014, Bechtol_2015, Drlica-Wagner_2015, Drlica-Wagner_2020, Koposov_2015, Geha_2017, Nadler:2017dxq, Nadler:2019zrb, DES:2019ltu, Newton_2018, Cerny_2021, Mao_2021, Nadler_2021}. In addition to the forementioned methods, strong gravitational lensing provides a promising way to probe the distribution of dark matter on these sub-galactic scales and at higher redshifts than the Local Group of galaxies \citep{Mao:1998aa,Chiba:aa,Dalal_2002, Metcalf:ac,Kochanek_2003, Kochanek_2004,koopmans2005,vegetti2009a}.

In general, in a lensed image of an extended source created by a galaxy-scale strong lens system, the bulk of the distortions are typically caused by a massive elliptical galaxy ($M_{\rm h}\sim10^{13}{\rm M}_\odot$) situated roughly half-way between the observer and the source. However, a careful examination of these strongly lensed images shows subtle localized gravitational perturbations beyond the ``smooth'' mass distribution of the main-lens galaxy \citep{Bolton_2006, Bolton_2008, Gavazzi_2008, Auger_2009, Brownstein_2012, Shu_2016, oldham_2017, Cornachione_2018, Vernardos_2020}. These so-called lensing pertubers can either be substructure (i.e.,~satellites) of the main-lens galaxy \citep{Mao:1998aa,Chiba:aa} or line-of-sight haloes, which are also referred to as ``interlopers'' \citep{Li:2016afu}, spatially distributed between the observer and the source which happen to project very close to the lensed images  \citep{Keeton:2003aa,Xu:2011ru,Li:2016afu,Despali:2017ksx, Amorisco_2021}. Highly significant such perturbers have been detected in a handful of strong lenses \citep{vegetti2010a,vegetti2010b,vegetti2012,vegetti2014,Nierenberg_2014, Nierenberg_2017, Hezaveh2016}, and many more lenses have been analysed where no significant detection was found \citep{vegetti2014,Ritondale:2018cvp}. Taken together, these detections and non-detections have been used to constrain the subhalo mass function at the typical redshift of the main lenses \citep{vegetti2014,Hezaveh2016,Ritondale:2018cvp}.  Similarly, statistics of quasar flux ratio anomalies have also been used to directly constrain the dark matter halo mass function \citep{Gilman:2019vca,Gilman:2019nap, Gilman:2019bdm, Hsueh:2019ynk, Nadler:2021dft, Gilman_2021_pwr}.

An aspect of the problem that has become clear in the last half decade is the fact that the line-of-sight dark matter haloes make up a significant contribution to the lensing perturbers \citetext{see e.g., \citealp{Xu:2011ru, Li:2016afu, Despali:2017ksx,  Graus:2017rrr}; see also \citealp{Amorisco_2021}}. 
 This means that \emph{multiplane} gravitational lensing is an essential effect, especially for sources beyond redshift $z_{\rm src}\sim0.5$. This is why all recent gravitational imaging analyses of magnified arcs and Einstein rings always include the line-of-sight dark matter halo contribution in addition to the substructure contribution in the pixellated linear and localized potential corrections to the main lensing potential \citep{Vegetti_2018, Ritondale:2018cvp, Enzi:2020ieg}. Furthermore, all recent analyses of quasar flux-ratio anomalies \citep{Gilman:2019vca,Gilman:2019nap,Gilman:2019bdm,Hsueh:2019ynk, Nadler:2021dft} include the line-of-sight dark matter halo contribution, usually by generating entire line-of-sight populations of haloes and doing ray tracing through them to compute their impact on flux ratios. Doing so, however, requires the generation of millions of line-of-sight halo realizations for each lens. For quasar lenses, this is tractable as the source is described by a handful of parameters. For lenses with extended images (arcs and rings) for which the source might be described by hundreds of shapelet coefficients \citep{Refregier:2001fd, Refregier_shp_ii, Birrer_2015}, it is computationally very expensive to generate enough line-of-sight realizations to cover the full space of possible image perturbation \citetext{see \citealp{birrer2017} for an early attempt; see also \citealp{He_et_al_2020}}. To make further progress in probing the small-scale dark matter distribution with extended lensed images, it is thus important to explore new ways of characterizing lensing perturbations. 

Multiplane lensing is usually handled via a recursive method that requires solving a lens equation for each lens plane \citep{Blandford_1986, McCully_2014, Schneider_2014}. Therefore, handling these multiplane lensing effects is computationally expensive. It is thus critical to introduce a different approach, which we call ``effective multiplane gravitational lensing'', to investigate the collective effect of line-of-sight haloes and main-lens dark matter substructure, which reflect on a two-dimensional map from the source plane to the image plane. In this approach, the lens mapping between the source and image planes can be fully characterized by two “effective” lensing potentials encompassing the complete structure of the deflection field. Since line-of-sight haloes and main-lens substructure contribute differently to each effective potential, this approach has the ability to distinguish these two contributions from each other, hence potentially improving constraints on dark matter from strong gravitational lensing.

Using this approach, we point out, for the first time, that line-of-sight haloes imprint a distinct quadrupolar signature on the two-point function of these effective potentials. Such an anisotropic signature is the result of the breakdown of translation symmetry in the image plane due to the presence of the main-lens galaxy. In conjunction with the angular-averaged (monopole) two-point correlation function \citep[see e.g.,][]{Hezaveh_2014, Bayer_2018, Chatterjee:2018ast, DiazRivero:2017xkd,DiazRivero:2018oxk, Brennan:2018jhq, Cyr-Racine_2019, CaganSengul:2020nat}, this characteristic anisotropic signature could be used to statistically distinguish the line-of-sight contribution to lensing perturbations from that of main-lens substructure, hence opening the door to probing the redshift evolution of dark matter structure with strong lensing. Interestingly, this anisotropic feature might have already been seen in analyses of the strong lens system JVAS B1938+666. Indeed, \cite{Sengul_2021} claim that a low-mass dark matter perturber first discovered in \cite{vegetti2012} is a line-of-sight halo, a conclusion supported by the work of \cite{Despali:2017ksx}. The distinctive anisotropic signature that we introduce in this work can be seen in the convergence maps shown in \cite{Sengul_2021} and, to some extent, in \cite{vegetti2012}. In this latter case, the anisotropic feature is a little more difficult to see in the convergence correction map because of the coarse pixelization and the significant noise of the convergence reconstruction. Given these limitations (driven by data resolution and noise), their results nonetheless appear broadly consistent with the anisotropic signatures we study in this work.

The structure of this manuscript is as follows: First, in Section~\ref{sec:sec2} we introduce the basic concepts of effective multiplane lensing. In Section~\ref{sec:sec3}, we discuss the method of analysing small-scale ansiotropies on projected mass density maps of strong lens systems. In Section~\ref{sec:sec4}, we present the results of applying our formalism to simulations with mock data. In Section~\ref{sec:sec5}, we assess the sensitivity of the anisotropic signals to the existing and future observational technologies. Finally, we summarize our findings and conclude in Section~\ref{sec:Conc}.

\section{Effective Multiplane Lensing}\label{sec:sec2}

\begin{figure}
\begin{center}
\includegraphics[clip, trim=0cm 0.5cm 4.3cm 3cm, width=0.481\textwidth]{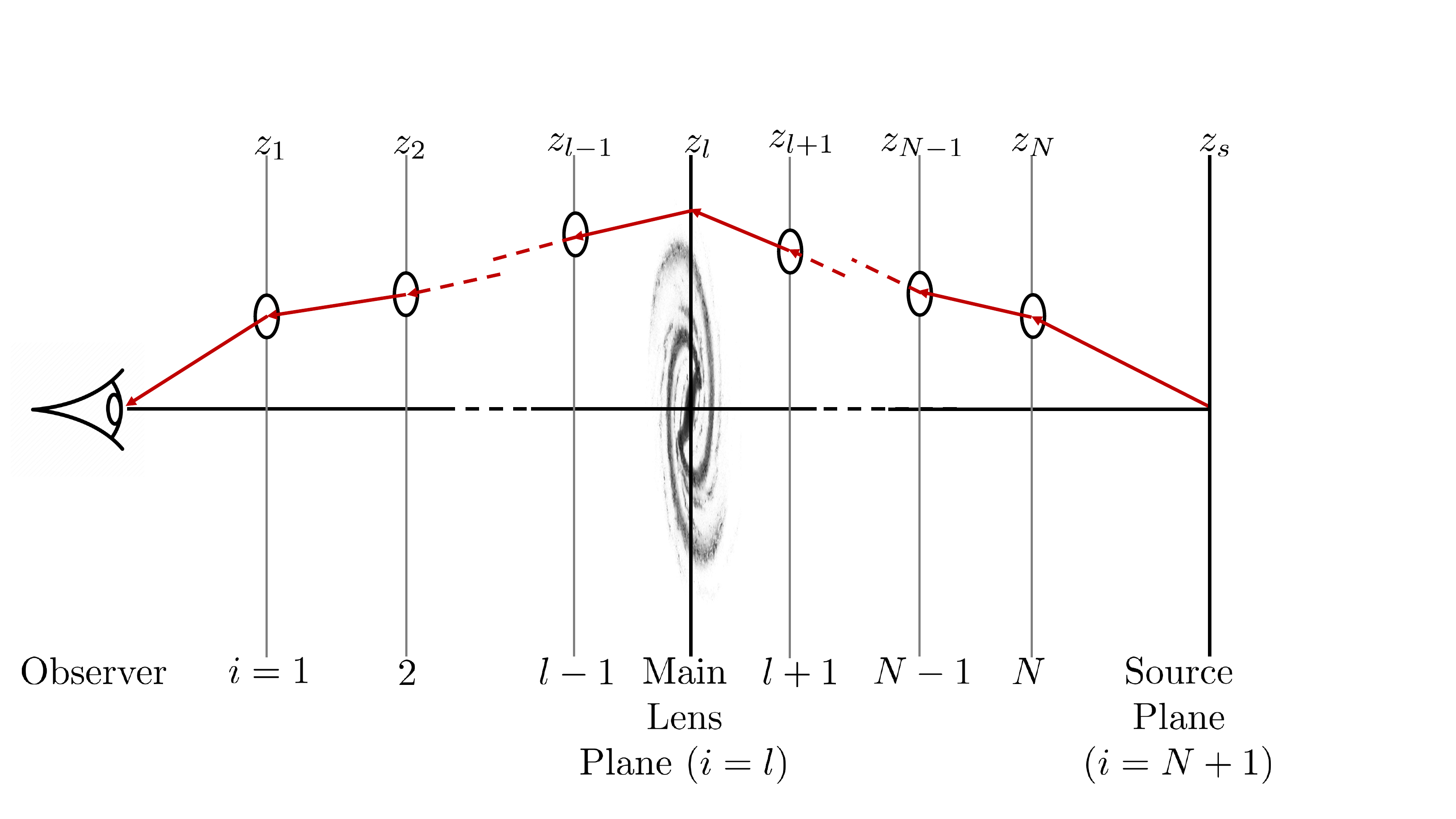}
\end{center}
\caption{\label{multiplane} A sketch to show a strong lensing system with line-of-sight haloes. The system has $N$ number of multiple lens planes, with the main-lens located on the $l^{\rm th}$ lens plane. Red arrows illustrate the path of a single light ray from the source to the observer. The redshifts of the multiplanes and the source plane ($z_i$) are also shown.}
\end{figure}

Consider a number, $N$, of thin lenses with redshifts $z_i$, $i=1,2,\dots,N$, indexed such that for $i<j$, $z_i < z_j$. The source redshift is  $z_{\rm s} (>z_N)$ and is in the $(N+1) {\rm th}$ plane (see Fig.~\ref{multiplane}). The angular diameter distances $D_i ,D_s, D_{is}$, and $D_{ij}$ ($i<j$) are defined as the  distance between the observer and the $i^{\rm th}$ lens plane, the distance between the observer and the source, the distance between the $i^{\rm th}$ lens and the source, and the distance between the $i^{\rm th}$ lens and the $j^{\rm th}$ lens plane, respectively.
 For multiplane gravitational lensing, the lens equation is given by

\be
\uu = \xx_1-\sum_{i=1}^N \bal_i (\xx_i),
\ee where $\uu \in {\mathbb{R}^2}$ stands for the angular coordinate in the source plane, and $\xx_1 \in {\mathbb{R}^2} $ is the image plane coordinate system. The angle $\bal_i (\xx_i)$ denotes the deflection of light from the lens $i$ at position $\xx_i$. The recursion relation to solve the lens equation for an intermediate lens plane is given by

\be 
\xx_j= \xx_1 - \sum_{i=1}^{j-1}\beta_{ij}\bal_i(\xx_i), \hspace{5mm} {\rm where} \hspace{5mm} \beta_{ij} \equiv \frac{D_{ij}D_s}{D_j D_{is}}. 
\ee

\noindent Note that $\xx_j$ depends on all the lens planes in front of $j^{\rm th}$ lens plane. Ray tracing begins with the angular position on the observer's sky ($\xx_1$) and continues until the source plane is reached \citep{McCully_2014}. In general, solving this array of lens equations at each lens plane is computationally intensive.

However, gravitational lensing is essentially a two-dimensional mapping from the source plane to the image plane. In the single-plane case, this non-linear map is a pure gradient of a scalar lensing potential. For multiplane lensing, this map is no longer a pure gradient due to the non-linear coupling between the consecutive lens planes. Therefore, it is useful to introduce the \emph{``effective multiplane lensing''} approach, which boils down the multiplane gravitational lensing problem to a single map between a source plane and an image plane. In this new approach, one scalar and one vector effective potential are introduced to capture all the effects of different lens planes. The general lens equation then takes the form

\be\label{eq:main_lens_eq}
\uu (\xx) = \xx - \bal_{\rm eff}(\xx),  
\ee where the "effective" deflection field $\bal_{\rm eff}(\xx)$ is

\be 
\bal_{\rm eff}(\xx) = \bna \phi_{\rm eff}(\xx) + \bna \times \ab_{\rm eff}(\xx).
\ee

\noindent Here, $\uu, \xx \in {\mathbb{R}^2}$ are the coordinates on the source plane and image plane, respectively. Also, $\phi_{\rm eff}$ is the effective scalar potential, and $\ab_{\rm eff}$ is the effective vector potential. The functions $\phi_{\rm eff}$ and $\ab_{\rm eff}$ encode the full non-linear multiplane lensing map from the source plane to the image plane. The decomposition of the effective deflection field into a divergence and a curl component is completely general and follows from Helmholtz's theorem, as long as the effective deflection field $\bal_{\rm eff}$ is twice continuously differentiable. By including the appropriate boundary terms, the effective scalar and vector potentials can be computed for any behavior of the deflection field at large distances from the centre of the lens. Note that $\ab_{\rm eff}$ points purely in the line-of-sight (or $\hat{\bf z}$) direction. Thus, introducing a scalar function $\zeta_{\rm eff}(\xx)$, such that $\ab_{\rm eff}(\xx) = \zeta_{\rm eff}(\xx) \hat{\bf z}$ is useful.

It is useful to introduce the divergence and curl of the effective deflection field \citep{Gilman:2019vca,CaganSengul:2020nat, Sengul_2021},

\be \label{Eq. kappa div}
\kappa_{\rm div} \equiv \frac{1}{2} \bna \cdot \bal_{\rm eff} - \kappa_0 =  \frac{1}{2} \nab^2 \phi_{\rm eff} - \kappa_0, 
\ee
and
\be \label{Eq. kappa curl}
\kappa_{\rm curl} \equiv \frac{1}{2} \bna \times \bal_{\rm eff} \cdot \hat{\bf z} =  -  \frac{1}{2} \nab^2 \zeta_{\rm eff},
\ee where $\kappa_0$ is the projected mass density of the main-lens model. This shows that both scalar potentials $\phi_{\rm eff}$ and $ \zeta_{\rm eff}$ satisfy a Poisson equation with source terms $2(\kappa_{\rm div} + \kappa_0)$ and $-2\kappa_{\rm curl}$, respectively. We shall refer to $\kappa_{\rm div}$ as the divergence component of the effective deflection field, while $\kappa_{\rm curl}$ will be its curl component. To isolate the collective convergence of the subhaloes and line-of-sight haloes, we subtract the macrolens (also known as the main-lens or the main deflector) convergence $\kappa_0$ in equation~\eqref{Eq. kappa div}. Since the effective deflection field in single-plane gravitational lensing is a pure gradient, $\kappa_{\rm curl}$ vanishes in this case.

To gain some intuition about the structure of the effective multiplane deflection field for a typical lens configuration, it is informative to numerically generate realistic realizations of line-of-sight haloes and main-lens subhaloes. We outline our procedure for generating such realizations in the next subsection.    

\subsection{Generating Convergence Maps}\label{sec:sec2.1}

We perform all the gravitational lensing computations using \textsc {lenstronomy} \citep{birrer2018lenstronomy, Birrer2021}, which is an open-source \textsc {python} software package. We also use \textsc {pyhalo} \citep{gilman2021strong} to generate main-lens substructure and line-of-sight halo realizations. Throughout this paper, we assume a flat $\rm \Lambda CDM$ cosmology based on the results of Planck 2018 \citep{planck_2018}. We also set the redshift of the main-lens at $z_{\rm macro}=0.5$, and that of the source at $z_{\rm s} = 1.0$. As the main-lens model, we use a power-law ellipsoid with an external shear. We set the Einstein radius to $\theta_{\rm E}=1.0$ arcsec, the eccentricity components to $(e_1, e_2)=(0.05, 0.08)$, and the logarithmic slope of the main-lens to $\gamma=2.078$ \citep{Auger_2010}. The components of the external shear tensor are set to $\gamma_1 = 0.01$ and $\gamma_2=-0.01$.

We generate multiple random realizations of CDM subhaloes modelled as truncated Navarro--Frenk--White (tNFW) profiles \citep{tNFW_2009}. The subhalo masses are rendered in the range of $10^6-10^{10}{\rm M_\odot}$ according to the power-law mass function given in equation~\eqref{Eq. 12} with the main halo mass of $M_{\rm halo} = 10^{13}{\rm M}_\odot$ and the logarithmic slope of $\alpha = -1.90$. This value is consistent with the power-law slope predicted by $N$-body simulations \citep{Springel_2008, Benson_2020}. We take the subhalo mass function to have the form

\be \label{Eq. 12}
\frac{{\rm d}^2N_{\rm sub}}{{\rm d}m{\rm d}A}=\frac{\Sigma_{\rm sub}}{m_0}\left(\frac{m}{m_0} \right)^\alpha\mathcal{F}(M_{\rm halo}, z),
\ee where $\Sigma_{\rm sub}=0.025 \,\, {\rm kpc}^{-2}$ is the normalization of the mass function, $m$ is the infall mass, and where the scaling function $\mathcal{F}(M_{\rm halo}, z)$ determining the evolution of the number density of subhaloes as a function of host halo mass $M_{\rm halo}$ and redshift $z$ is given by

\be \label{Eq. 13}
\log_{10}(\mathcal{F}) = k_1 \log_{10}\left(\frac{M_{\rm halo}}{10^{13}\rm M_\odot}\right)+k_2 \log_{10}(z+0.5) \,\,.
\ee Based on the dark matter halo realizations generated with the \textsc {galacticus} semi-analytical model \citep{BENSON_galacticus},   $k_1$ and $k_2$ are set to 0.88 and 1.7, respectively \citep{Gilman:2019bdm}.

The line-of-sight haloes are modelled as NFW profiles, and the number of line-of-sight haloes on each lens plane is chosen between the mass range of $10^6-10^{10}{\rm M_\odot}$ according to the re-scaled version \citep{Gilman:2019vca} of the Sheth--Tormen (ST) mass function \citep{Sheth_2001} given by

\be \label{Eq. 14}
\frac{{\rm d}^2N_{\rm LOS}}{{\rm d}m{\rm d}V} = \de_{\rm LOS}\left(1+\xi_{\rm 2halo}(r, M_{\rm halo}, z) \right)\left[\frac{{\rm d}^2N_{\rm}}{{\rm d}m{\rm d}V}\right]_{\rm ST}.
\ee

\noindent The scaling parameter $\de_{\rm LOS}$ captures the systematic shifts associated with the mean number of haloes predicted by the Sheth-Tormen mass function and the baryonic effects on small-scale structure formation \citep{Benson_2020, Gilman:2019bdm}. In our realizations this $\de_{\rm LOS}$ value is fixed to 1.0, unless we mentioned otherwise in the text. Here, $\xi_{\rm 2halo}$ is the two-halo term of the three-dimensional two-point correlation function, which captures correlated structure outside of the virial radius of the host dark matter halo of the main-lens \citep{Gilman:2019vca, gilman2021strong, Lazar_2021}, and thus alters the scale of amplitude of the line-of-sight halo mass function and background density. This term has the form

\be 
\xi_{\rm 2halo}(r, M_{\rm halo}, z) = b(M_{\rm halo}, z)\, \xi_{\rm lin}(r, z),
\ee where $b(M_{\rm halo}, z)$ is the halo bias around the lens halo with mass $M_{\rm halo}$ as computed in \cite{Sheth_torman_1999}, and $\xi_{\rm lin}(r, z)$ is the linear matter--matter correlation function at a distance $r$, which is determined using the linear power spectrum at redshift $z$.

\subsection{Convergence Map Anisotropies}

\begin{figure}
\begin{center} 
	\includegraphics[clip, width=0.485\textwidth]{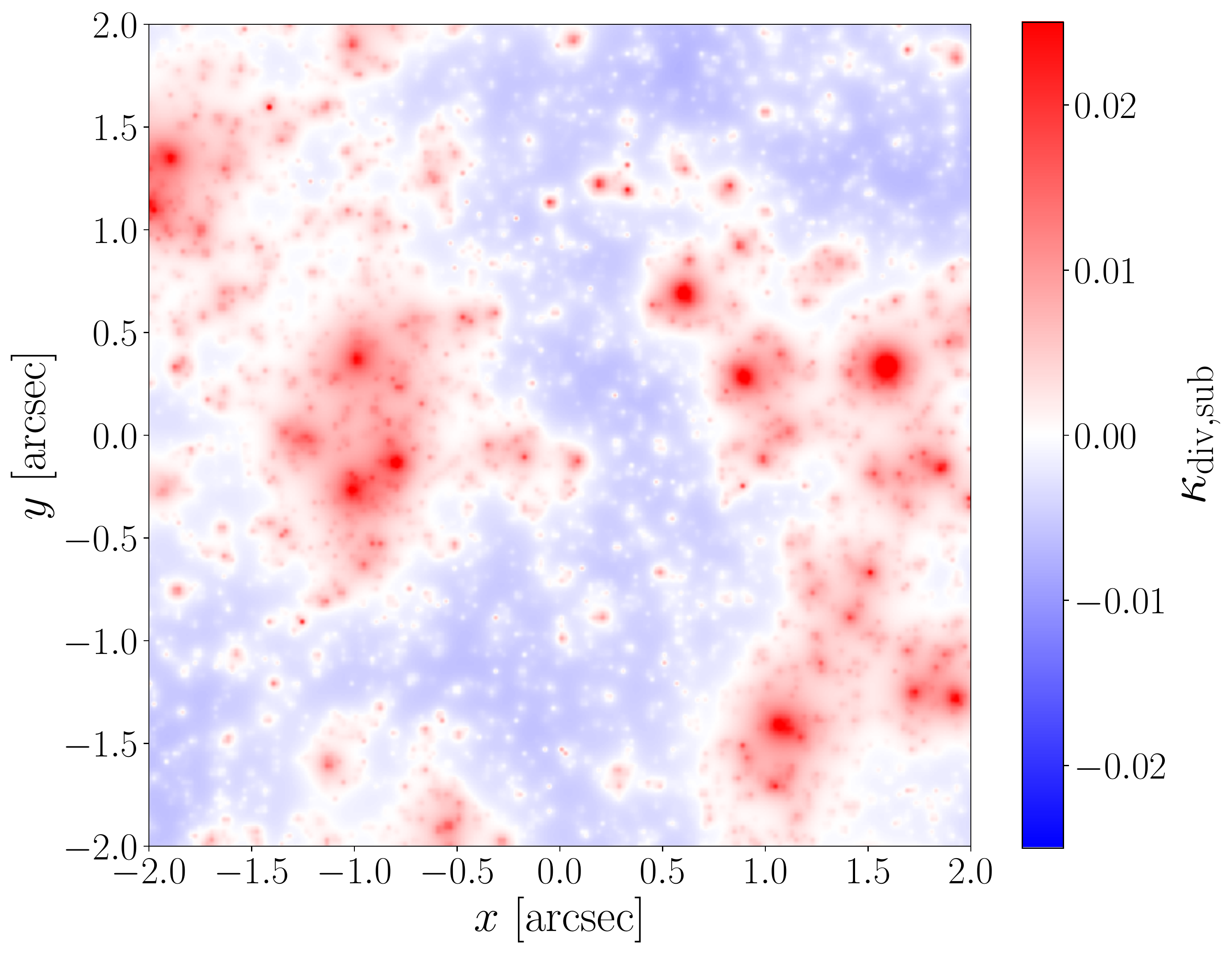} 
	\includegraphics[clip, width=0.485\textwidth]{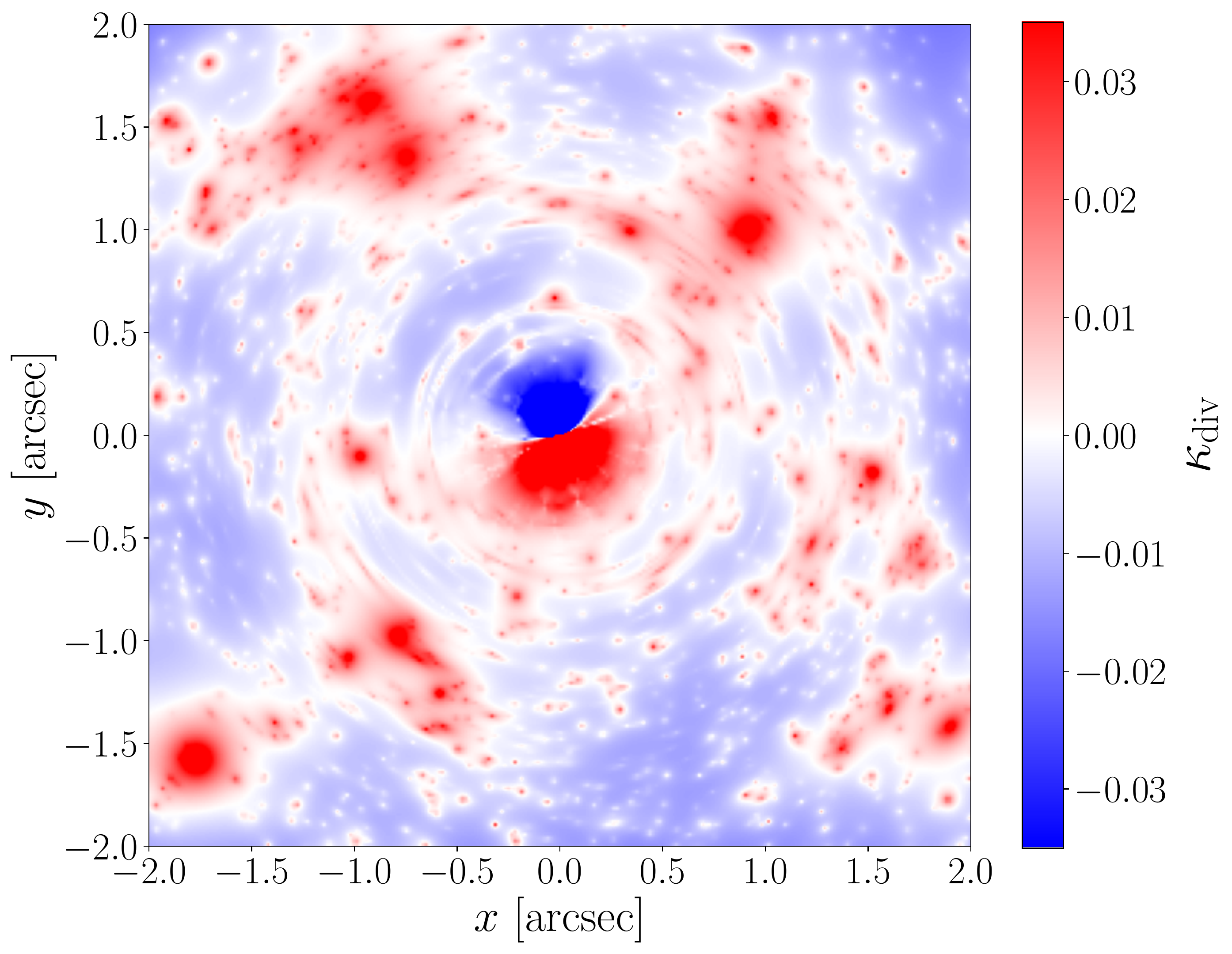}
	\includegraphics[clip, width=0.485
	\textwidth]{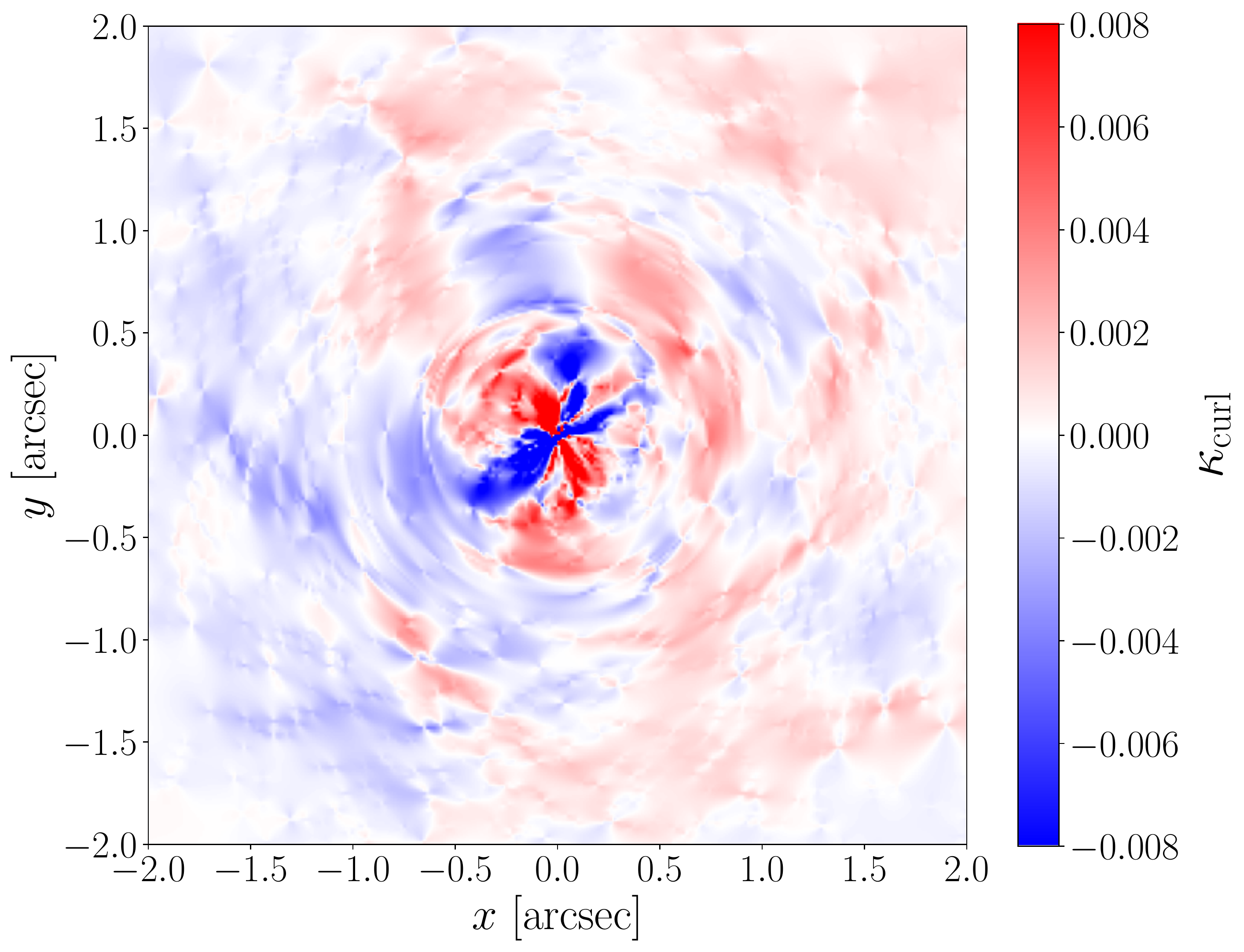}
\end{center}
\caption{\label{fig:kappa_sgc} The $\kappa_{\rm div}$ and $\kappa_{\rm curl}$ maps for a cold dark matter model considering both the single plane and multiplane gravitational lensing. The top panel shows the divergence of the single plane deflection of the main-lens substructure. The middle and the lower panels show the divergence and curl components of the multiplane convergence field for a full halo realization, including both the contribution from line-of-sight dark matter haloes and lens galaxy substructure. Note that $\kappa_{\rm curl}=0$ for the single plane gravitational lensing. }
\end{figure}

 The divergence and the curl of the effective deflection field given in equations~\eqref{Eq. kappa div} and \eqref{Eq. kappa curl} are shown in Fig.~\ref{fig:kappa_sgc} for a cold dark matter model, both for the single-plane and multiplane lensing cases. The top panel shows the $\kappa_{\rm div}$ component for the main-lens substructure only, which is fundamentally a single-plane computation. The middle and lower panels show the divergence and curl of the effective multiplane deflection field for a full CDM halo realization, i.e., this time the realization contains both the contributions from the line-of-sight haloes and the main-lens substructure. Note the different angular and symmetry structures in each case. In the top panel, the main-lens substructure leads to fluctuations that are largely statistically homogeneous and isotropic across the map, with individual perturbations being largely circular. In the middle panel, we see a clear break of homogeneity due to the presence of the main-lens, which selects a preferred centre on the map (here, $x=y=0$ for simplicity). In this case, the line-of-sight haloes between the source and the observer appear distorted (stretched) in the tangential direction, while the main-lens subhaloes still appear largely circular. This apparent anisotropy between the tangential and radial directions in the $\kappa_{\rm div}$ map, which only occurs for line-of-sight objects, immediately suggests a way to potentially \emph{distinguish} between main-lens substructure and line-of-sight haloes. Finally, the lower panel of Fig.~\ref{fig:kappa_sgc} shows $\kappa_{\rm curl}$, which is characterized by a quadrupolar pattern at the position of line-of-sight haloes, with the main-lens substructure making no contribution to this map. While there is obviously interesting structure in the curl component of the effective deflection field, we note that the signal there is nearly an order of magnitude smaller in strength. Thus we concentrate here on the anisotropies of the $\kappa_{\rm div}$ maps, and leave the study of the curl component, including its cross-correlation with $\kappa_{\rm div}$ maps, to future works.

\section{The two-point correlation function}\label{sec:sec3}

\begin{figure}
\begin{center}
\includegraphics[clip, trim=0.8cm 5.5cm 12.7cm 6cm, width=0.3\textwidth]{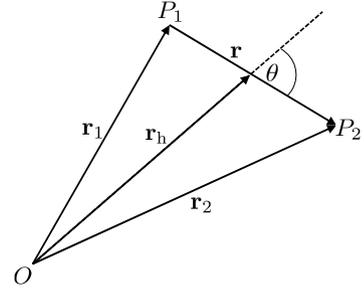}
\end{center}
\caption{\label{fig:2} A sketch to show the position vectors $\rr_1$ and $\rr_2$ of points $P_1$ and $P_2$, and the line-of-sight $\rr_{\rm h}$ to the mid-point of the two points. }
\end{figure}

As we saw above, in the presence of line-of-sight haloes, the resulting convergence field $\kappa_{\rm div}$ is no longer statistically isotropic. To study the additional information encoded in these anisotropies, we define the image-plane averaged two-point correlation function $\xi(\rr)$ as 

\be \label{eq:corr_func_def}
\xi(\rr)=\xi(\rr_2 - \rr_1)=\frac{1}{A}\int_A {\rm d}^2\rr_1 \,\, \kappa_{\rm div}(\rr_1) \,\, \kappa_{\rm div}(\rr +\rr_1),
\ee where $\rr_1$ and $\rr_2=\rr_1+\rr$ are the positions of points $P_1$ and $P_2$ from the centre of the convergence map ($O$) and $\rr$ is the vector connecting two points as shown in Fig.~\ref{fig:2}. Moreover, $A$ is the area of the image over which we perform the spatial average and $\rr_{\rm h}$ is the line-of-sight vector from the centre to the mid-point of the vector $\rr$ (see Fig.~\ref{fig:2}). The two-point correlation function can be written as the summation
\be 
\xi(\rr) = \xi_{\rm 1h}(\rr)+\xi_{\rm 2h}(\rr).
\ee In this case, the "one-halo" term ($\xi_{\rm 1h}(\rr)$) describes the contribution from a single dark matter halo, whereas the "two-halo" term ($\xi_{\rm 2h}(\rr)$) describes the contribution from pairs of distinct haloes. We focus on the one-halo term in this work because we are primarily interested in the intrinsic properties of dark matter haloes. This term dominates the two-point correlation function on scales smaller than the typical angular separation between lensing perturbers \citep{Cooray_halo_models, Zheng_2004, Jose_2017, DiazRivero:2017xkd, DiazRivero:2018oxk}. 

To characterize the anisotropies in the effective convergence maps, we decompose the two-point correlation function onto the orthonormal basis of Chebyshev polynomials $T_\ell(x)$ as

\be 
\xi(\rr)=\xi(r, \mu_\rr)=\sum_{\ell=0}^\infty \xi_\ell(r)T_\ell(\mu_\rr),
\ee where $\mu_\rr=\hat{\rr}\cdot\hat{\rr}_{\rm h} = \cos \theta$ and $\xi_\ell$ denotes the multipole of order $\ell$. The corresponding correlation multipoles are thus given by
\be 
\xi_\ell(r) = \frac{2-\de_{\ell 0}}{\pi}\int_{-1}^1 {\rm d} \mu_\rr \,\, \frac{\xi(r, \mu_\rr)T_\ell(\mu_\rr)}{\sqrt{1-\mu_\rr^2}},
\ee
 where $\de_{ij}$ is the Kronecker delta. We use Chebyshev polynomials instead of Legendre polynomials to decompose the two-point correlation function since our study is based on two-dimensional maps of projected mass densities, as opposed to the three-dimensional nature of anisotropic galaxy clustering \citep{Samushia_2014, Okumura_2015, Beutler_2017}.
 
The convergence power spectrum is the Fourier transform of the two-point correlation function $\xi(r,\mu_\rr)$ of the convergence map. The power spectrum can be computed and expanded in terms of Chebyshev polynomials as

\be 
P(k, \mu_\kk)=\int {\rm d}^2\rr \,\, e^{-i\kk \cdot \rr} \xi(r, \mu_\rr) = \sum_{\ell=0}^\infty P_\ell(k)T_\ell(\mu_\kk),
\ee where  $\mu_\kk = \hat{\kk}\cdot\hat{\rr}_{\rm h}$.  In Appendix \ref{App.A} we derive that the power spectrum multipoles are related to the correlation function multipoles by
\be \label{Eq._11}
P_\ell(k) = 2\pi(-i)^\ell \int {\rm d}r \,\, r \, J_\ell(kr) \xi_\ell(r),
\ee where $J_\ell(x)$ denotes the Bessel functions of the first kind. The inverse transform is given by
\be
\xi_\ell(r) = \frac{1}{2\pi}i^\ell \int {\rm d}k \,\, k \, J_\ell(kr) P_\ell(k).
\ee

\section{Analysis of simulated strong lensing systems}\label{sec:sec4}

In this section, we analyze the anisotropies appearing in the $\kappa_{\rm div}$ maps computed using equation~\eqref{Eq. kappa div} and generated according to the procedure described in Section~\ref{sec:sec2.1}. We focus here on standard CDM, and leave the study of the anisotropic correlation signal in different dark matter theories to future works.

For each numerical realization of line-of-sight haloes, we populate haloes in the redshift domain of interest according to the re-scaled version of the Sheth-Tormen mass function as we discussed in Section~\ref{sec:sec2.1}. Below, we will consider two distinct cases. In the first case, we populate line-of-sight haloes in between the observer and the main-lens. In the second case, we focus our attention only on the line-of-sight haloes behind the main-lens. Since lensing observations are primarily sensitive to deflection perturbations near the critical lines where lensed images form \citep{minor2016}, we apply an annular mask centred on the main deflector to focus our attention on the most relevant region of the image plane. Our procedure to compute the correlation function in the presence of this mask is shown in Appendix \ref{App.B}.

\subsection{General Lensing Invariance}

As discussed in many lensing-related works \citep[see e.g.,][]{Schneider_1995, Bradac_2004, Schneider_2013, Rexroth_2016, Birrer_2021, Cremonese_2021}, strong lensing systems have a built-in mathematical degeneracy under which a simultaneous transformation of the source and lens planes can leave several astrometric and photometric lensing observables invariant. Fundamentally, this general lensing invariance \citep[also known as the Source Position Transform;][]{Schneider:2014hsi,Unruh:2017yhi,Wertz_2018} occurs because we observe strong lenses in angular projection on the sky. Since angles are ratios of distances, it is always possible to simultaneously rescale such distances in a way that leave the observed angles invariant. Only observations of an \emph{absolute} scale, such as the intrinsic size of the lensed source or the relative time delay between images, can break this fundamental degeneracy. 

The scalar part of this transformation is the well-known mass-sheet degeneracy \citep{Falco_1985}. The mass-sheet transform (MST) isotropically maps the source plane coordinates as $\uu \rightarrow  \lambda\uu$, and thus transforms the effective deflection field, $\kappa_{\rm eff}=\frac{1}{2} \bna \cdot \bal_{\rm eff}$ as
\begin{align}
    \kappa'(\xx) & = \lambda\kappa_{\rm eff}(\xx)+(1-\lambda)\en
     & = \lambda\kappa_{\rm div}(\xx)+\kappa'_0(\xx),
\end{align} where the MST of the main-lens convergence is given by $\kappa_0'(\xx)=\lambda\kappa_0(\xx)+(1-\lambda)$. Then we may write down the transformation of the $\kappa_{\rm div}$ field under a MST  as $\kappa'_{\rm div}(\xx)=\kappa'(\xx)-\kappa'_0(\xx)=\lambda\kappa_{\rm div}(\xx)$. This transform scales the two-point correlation function, $\xi(\rr) \rightarrow \lambda^2\xi(\rr)$, and can thus bias its amplitude. This degeneracy appears to have been missed in previous works on this topic. To eliminate the effects of this degeneracy, we normalize the two-point correlation function with the zero-shifted autocorrelation (similar to the zero-lag autocorrelation in signal processing) of the mean subtracted $\kappa_{\rm div}$ fields, as discussed in Appendix \ref{App.B}. This effectively sets the amplitude of the two-point function to unity at the smallest scales. The relevant quantities invariant under the MST are thus the relative amplitudes and shapes of the different correlation function multipoles, on which we focus below. 

\subsection{Normalized Correlation Multipoles}

\begin{figure*}
\begin{center}
	\includegraphics[width=0.375\textwidth]{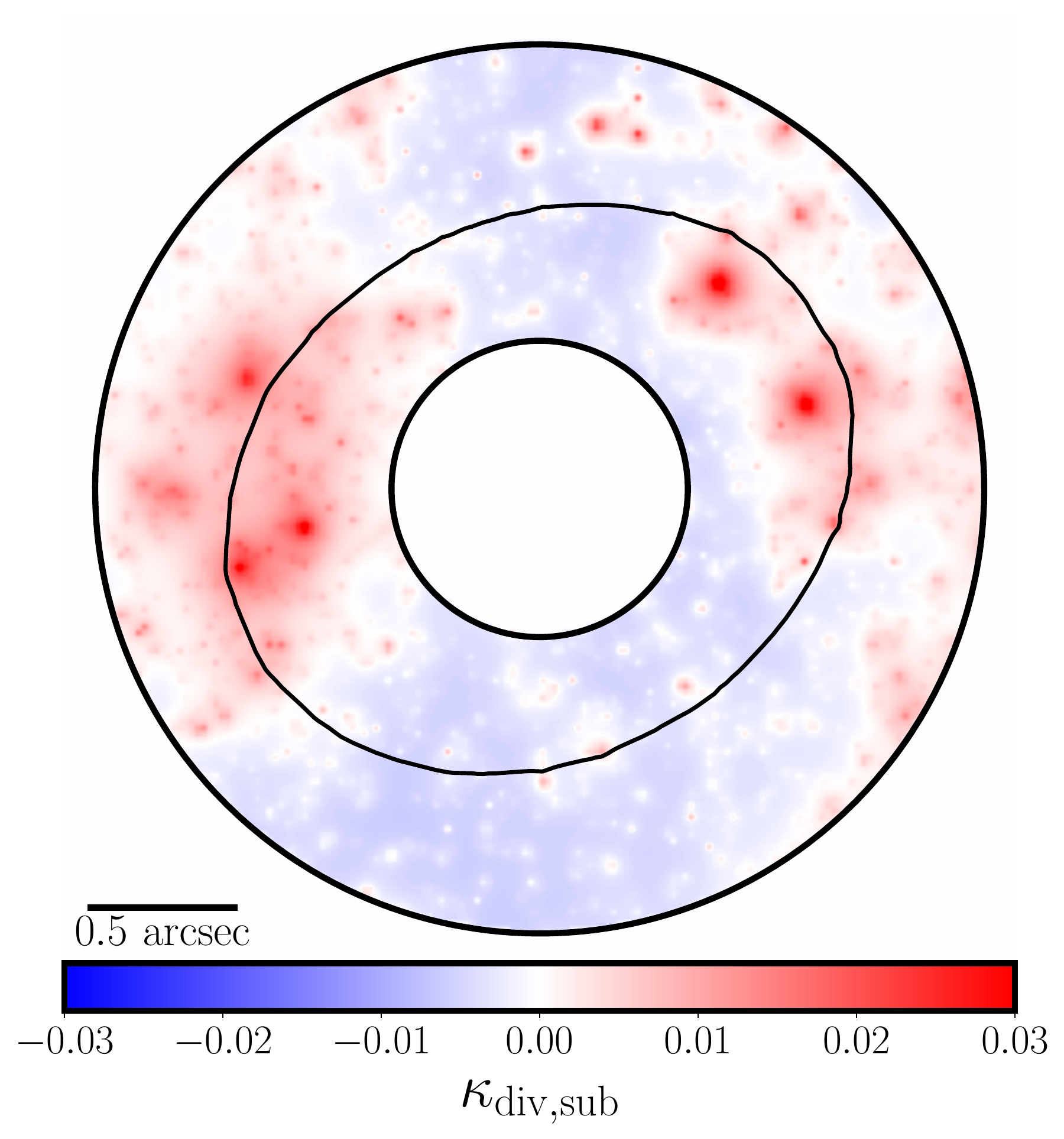}
	\includegraphics[width=0.53\textwidth]{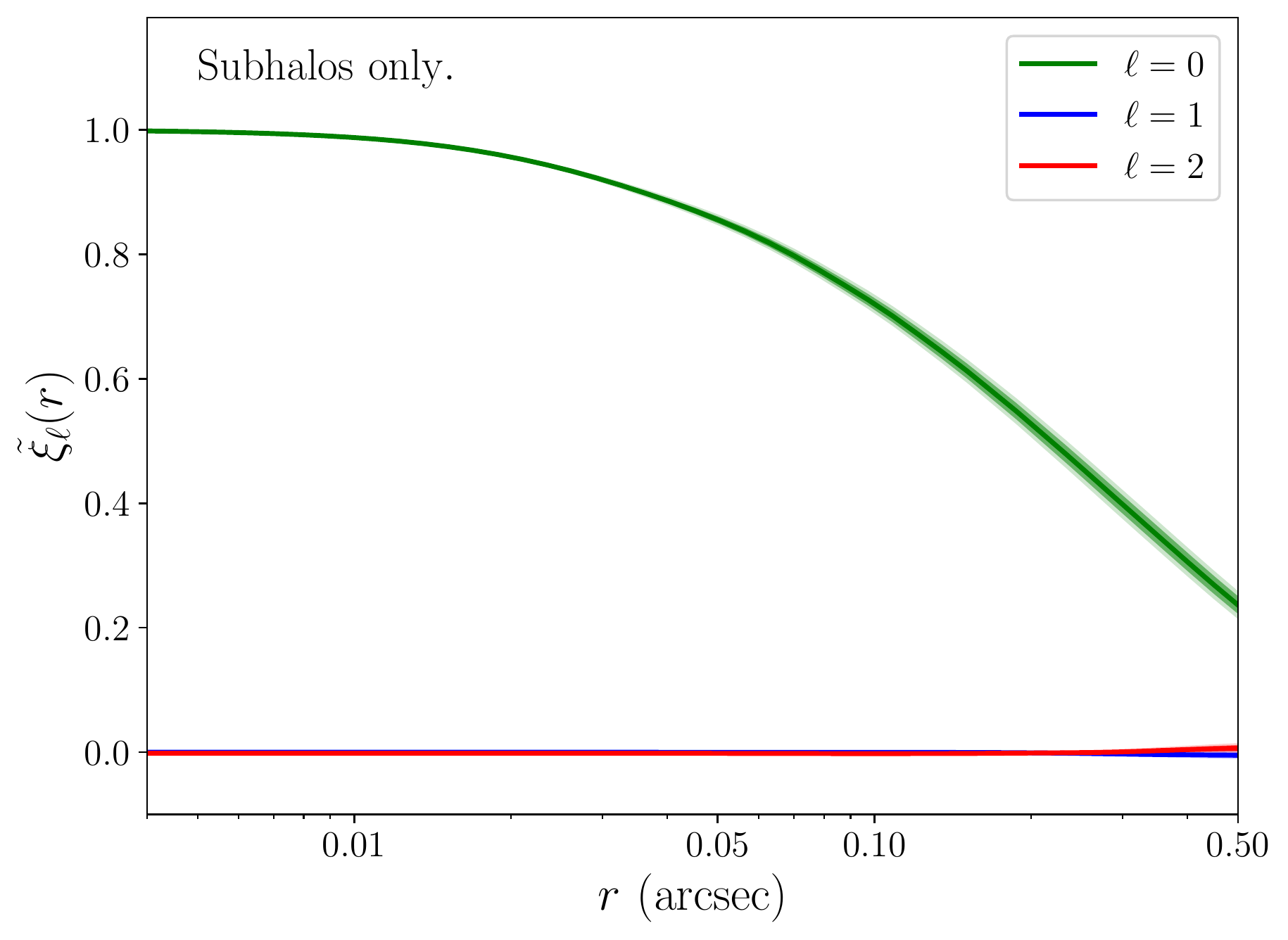}
\end{center}
\begin{center}
	\includegraphics[width=0.38\textwidth]{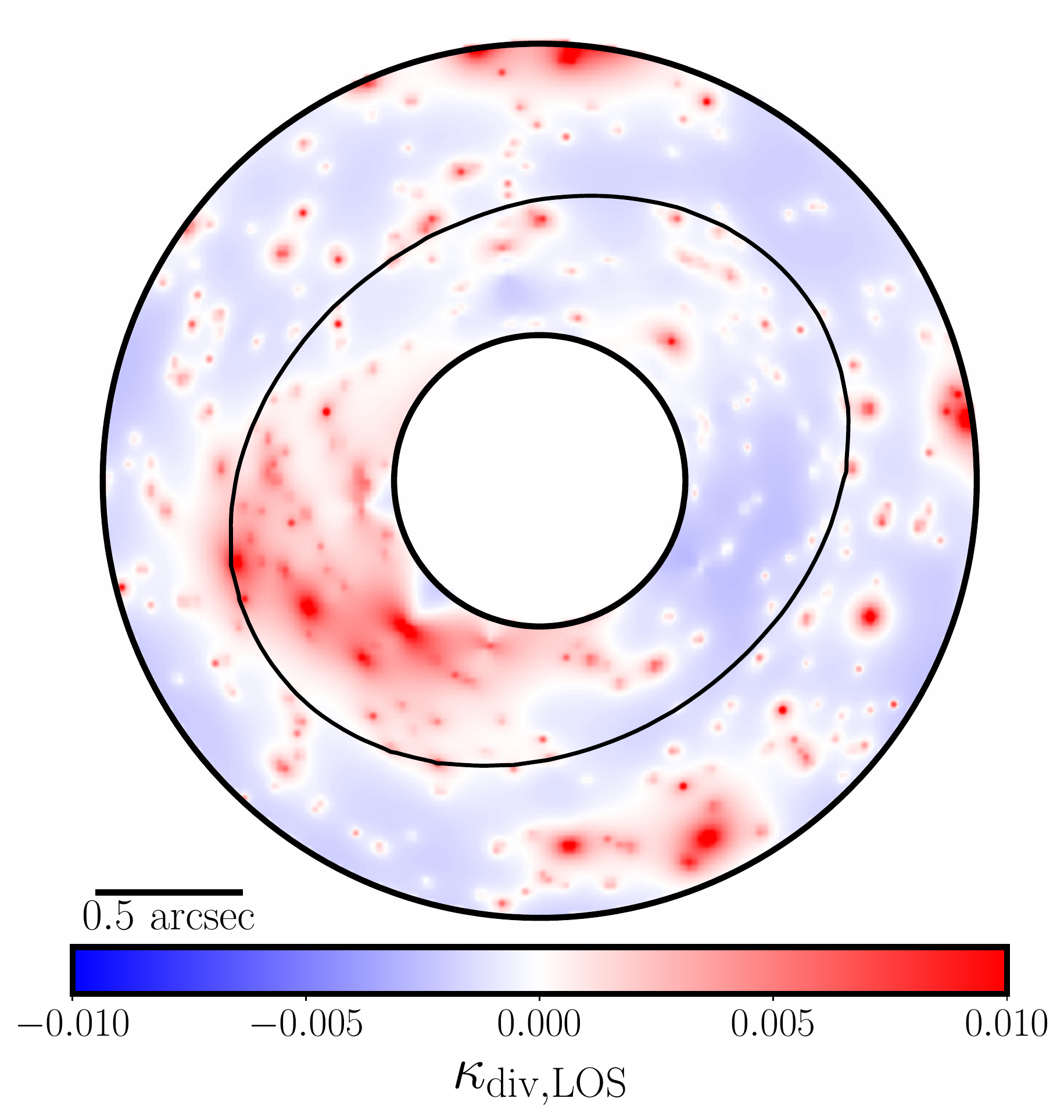}
	\includegraphics[width=0.55\textwidth]{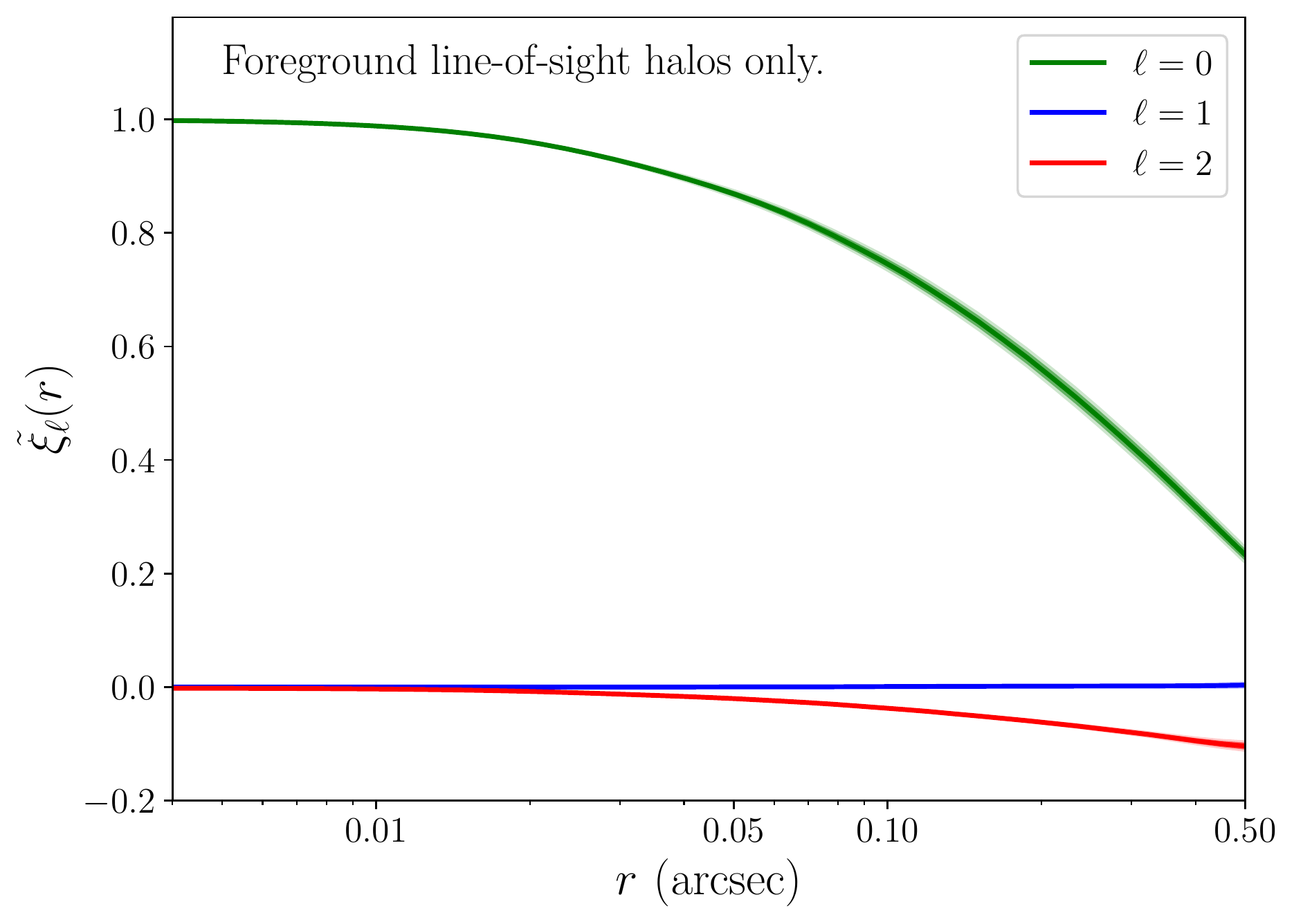}
\end{center}
\begin{center}
	\includegraphics[width=0.38\textwidth]{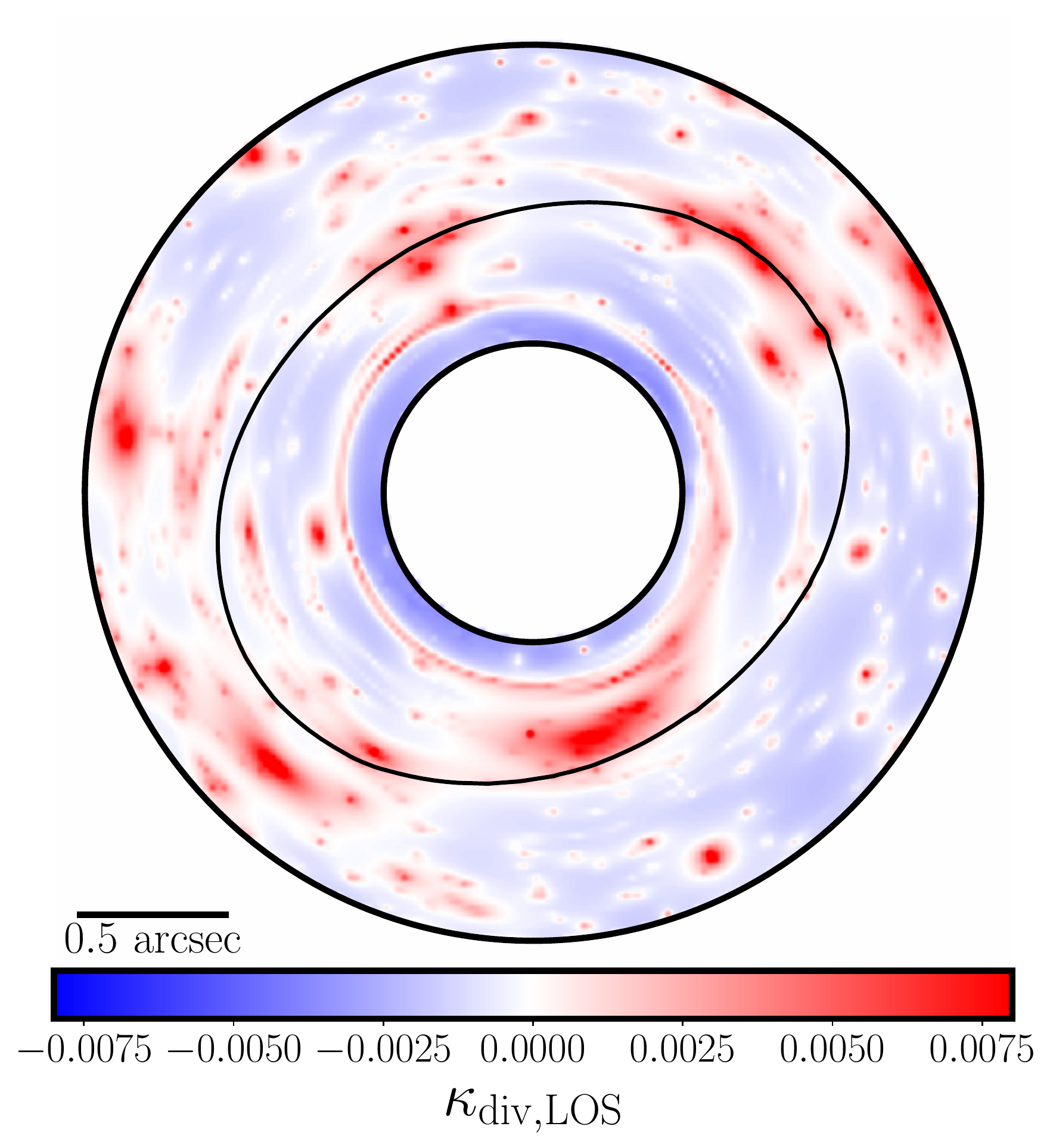}
	\includegraphics[width=0.55\textwidth]{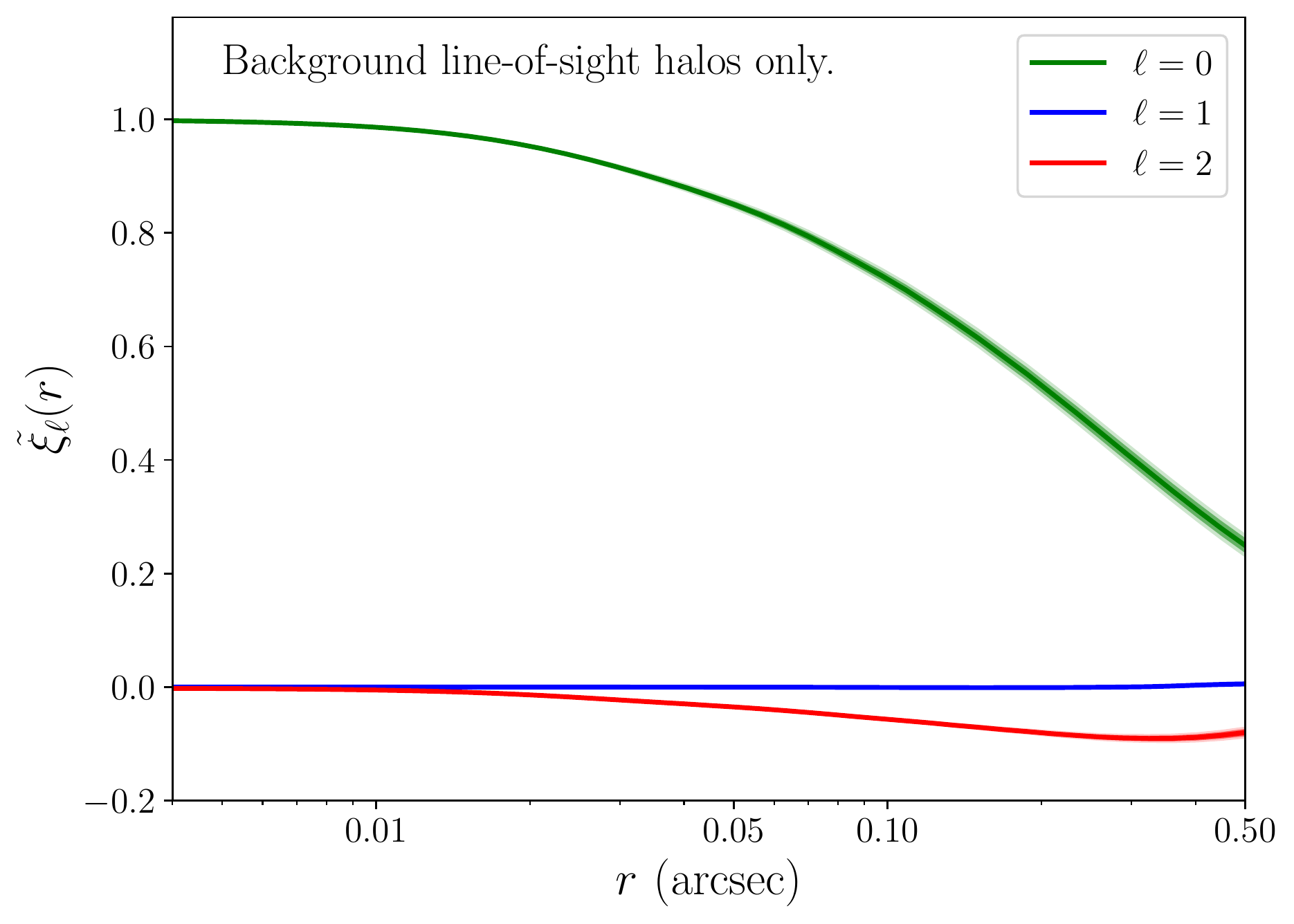}
\end{center}
\caption{\label{fig:corr_multi}Multipole moments of the masked-normalized two-point correlation function ($\tilde{\xi}_\ell(r)$) of the masked $\kappa_{\rm div}$ field. The solid black lines inside the annuli show the critical lines. The top panel shows the multipole moments for a substructure-only realization. Note that the properties of the subhaloes are fully captured by the non-zero monopole, ($\ell=0$). The middle panel (haloes in front of the main-lens) and the lower panel (haloes behind the main-lens) show the multipole moments for two line-of-sight halo only realizations. In both cases, line-of-sight haloes have a fairly large monopole ($\ell = 0$), and a quadrupole ($\ell = 2$). These non-zero quadrupoles capture the statistical properties of line-of-sight haloes.  }
\end{figure*}

\begin{figure} 
\begin{center}
\includegraphics[clip, trim=0.25cm 0.1cm 0.05cm 0.08cm, width=0.48\textwidth]{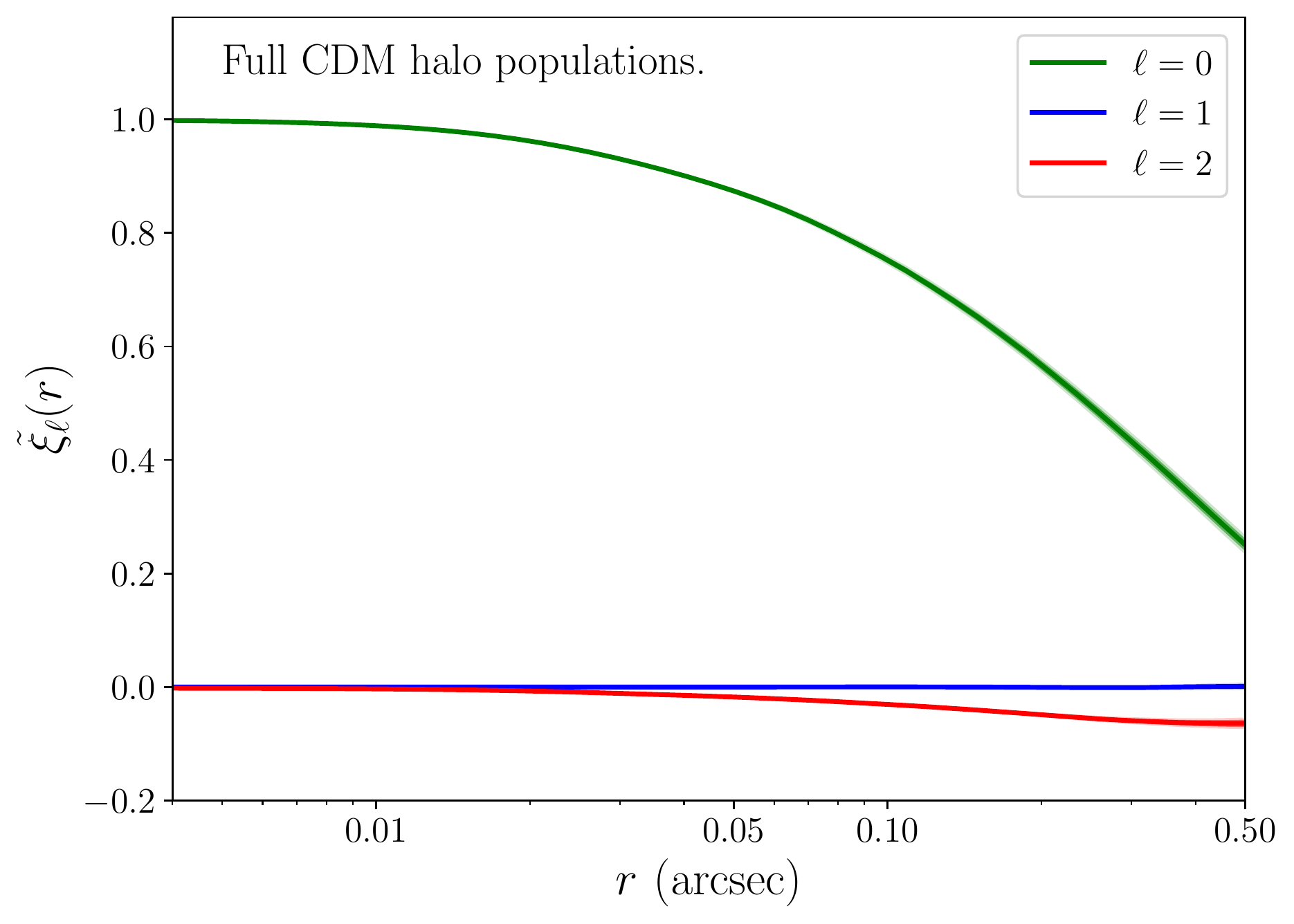}
\end{center}
\caption{\label{full_realization} The multipole moments of the masked-normalized two-point function ($\tilde{\xi}_\ell(r)$) of the $\kappa_{\rm div}$ field considering an average of 100 full CDM halo realizations that included both the substructure and line-of-sight haloes. The values of $\de_{\rm LOS}$ and $\Sigma_{\rm sub}$ are set to $1.0$ and $0.025 \,\, {\rm kpc}^{-2}$, respectively. Notice that the monopole ($\ell = 0$) captures the statistical properties of both the substructure and the line-of-sight haloes, while the quadrupole ($\ell = 2$) describes the properties of the line-of-sight haloes.}
\end{figure}

The multipole moments of the masked and normalized two-point correlation function $\tilde{\xi}_\ell(r)$, computed using the method discussed in Appendix~\ref{App.B}, are shown in the right-hand column of Fig.~\ref{fig:corr_multi}, while the left-hand column displays corresponding examples of $\kappa_{\rm div}$ maps. We apply annular masks to these maps with inner and outer radii of 0.5 and 1.5 arcsec, respectively, and keep the mean value of $\kappa_{\rm div}$ constant within the annuli. We focus on three distinct cases as labelled in the figure and generate 100 CDM realizations in each case. Because we are primarily interested in the behavior of the one-halo term of the two-point correlation function ($\xi_{\rm 1h}(\rr)$), the maximum radial distance we consider here is $r=0.5$ arcsec. As shown in \cite{DiazRivero:2017xkd}, the two-halo term of the monopole can only become significant for $r > r_{\rm t, max}$, where $r_{\rm t, max}$ is the truncation radius of the most massive subhalo. In the CDM halo realizations we consider in this study, the maximum truncation radius of subhaloes is $r_{\rm t,max}\simeq 0.35$ arcsec. Therefore, focusing on $r\leq 0.5$ arcsec ensures that we are primarily sensitive to the one-halo term for the monopole.

The top panel in Fig.~\ref{fig:corr_multi} shows the correlation multipoles for a pure subhalo population. The solid lines show the mean and shaded areas represent 68\%  and 90\% credible intervals. Due to the normalization of the correlation function, these confidence intervals are narrow and hard to resolve. The non-zero monopole ($\ell=0$) for this population contains most statistical properties of the main-lens substructure. A very small positive quadrupole is visible at the largest scales shown in the figure, which is caused by the two-halo term becoming sizable there. The multipole moments for foreground line-of-sight haloes (shown in the middle panel) and haloes behind the main-lens (shown in the lower panel) both have sizable negative quadrupole ($\ell=2$) moments as well as large monopoles. These quadrupole moments reflect the anisotropy between the radial and tangential directions in the effective deflection field caused by the impact of the main-lens galaxy on the deflection perturbation from line-of-sight haloes. 

Unlike the dark matter clumps located in front of the macrolens, the projected mass densities of the background line-of-sight haloes are distorted more in the tangential direction due to the massive lensing potential of the main-lens galaxy and its substructure. Therefore, they show a slightly larger (in absolute value) one-halo quadrupole moment compared to the quadrupole moment created in the presence of only foreground line-of-sight haloes. However, the more significant distortion of background line-of-sight haloes also gives rise to a larger positive two-halo contribution to the quadrupole moment in this case, resulting in a partial cancellation between the one-halo and two-halo contribution at the largest scales. This explains why the quadrupole turns over around $r\sim0.35$ arcsec in the lowest right panel of Fig.~\ref{fig:corr_multi}. In Fig.~\ref{full_realization}, a more realistic situation in which both the entire line-of-sight population and the main-lens substructure are present (similar to the middle panel of Fig.~\ref{fig:kappa_sgc}) yields qualitatively similar results with a monopole ($\ell=0$) describing the properties of both types of perturbers and a quadrupole ($\ell=2$) expressing the properties of line-of-sight perturbers and the macromodel (see Section~\ref{macro_impact}). Again, the slight turnover of the quadrupole moment at $r>0.4$ arcsec is caused by the competition between the one-halo and two-halo terms at the largest scales. Detailed measurements of the scale dependence of the quadrupole moment could thus tell us important information about both the density profiles of line-of-sight haloes and their clustering properties.

\subsection{Halo Abundances and Quadrupole-to-Monopole Ratio}\label{sec:sec4.2}

\begin{figure} 
\begin{center}
\includegraphics[clip, trim=0.25cm 0.1cm 0.05cm 0.05cm, width=0.48\textwidth]{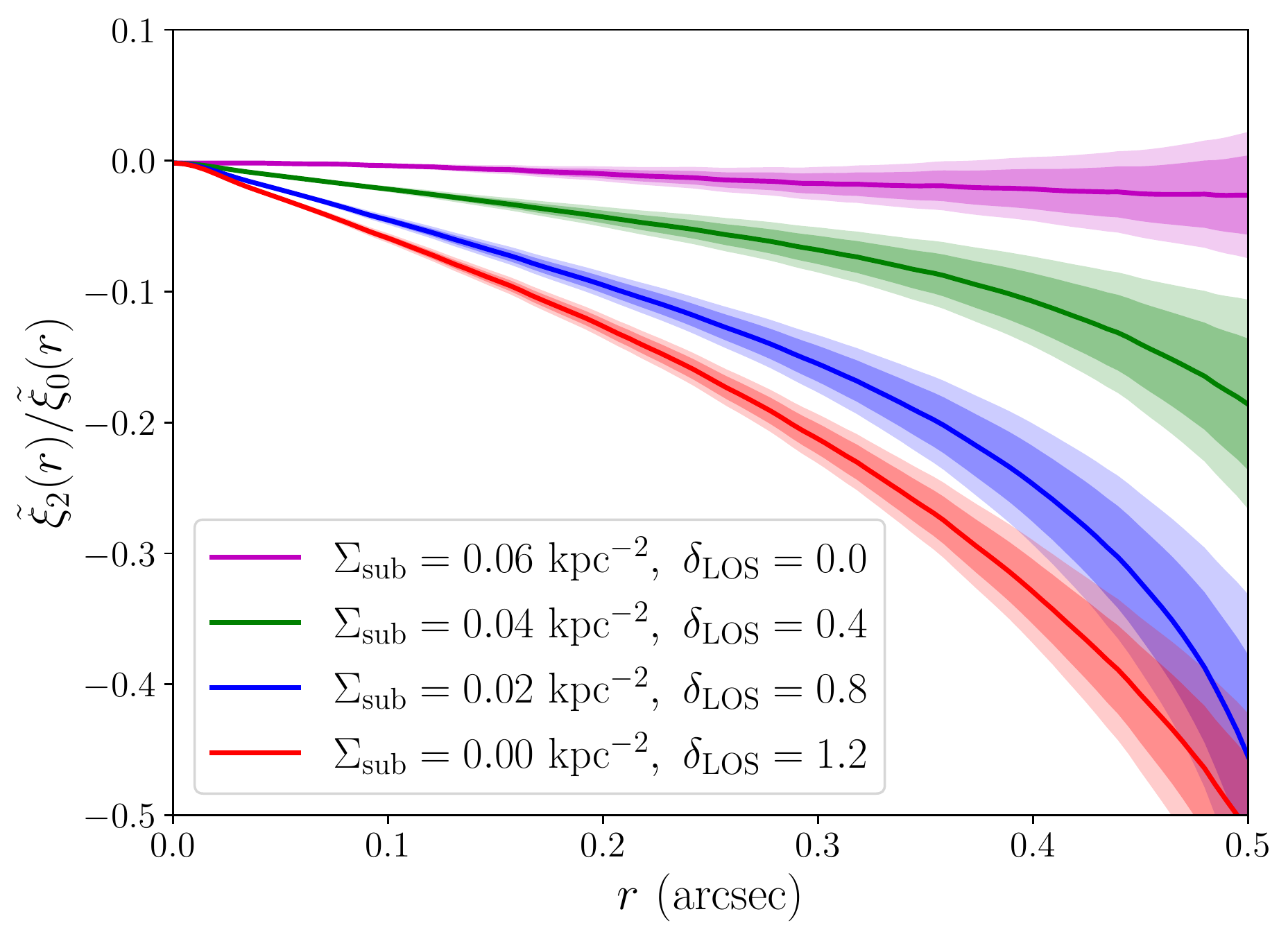}
\includegraphics[clip, trim=0.25cm 0.1cm 0.05cm 0.05cm, width=0.48\textwidth]{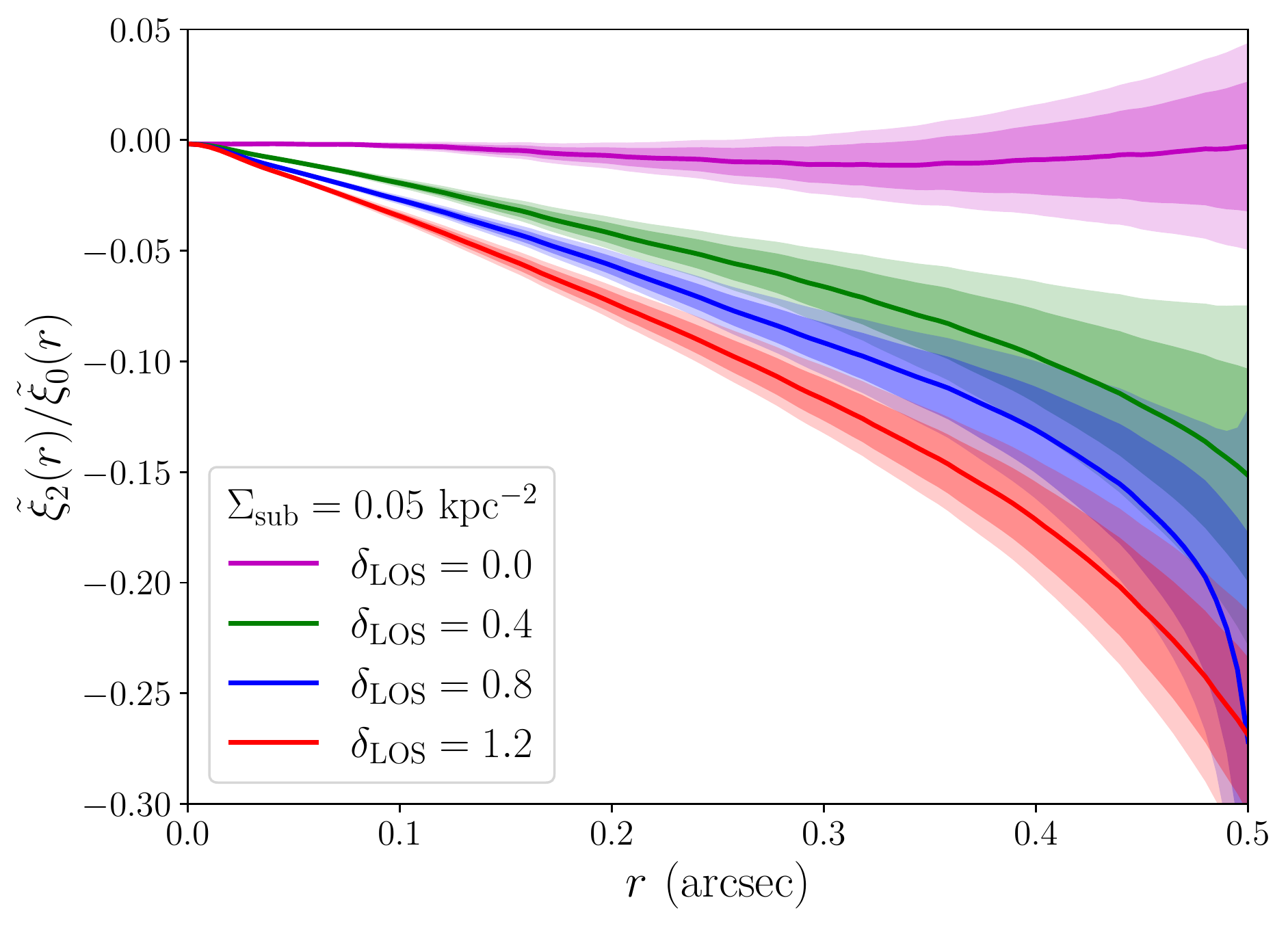}
\end{center}
\caption{\label{fig:abunadances} The quadrupole to monopole ratio for the effective projected mass density as a function of $\Sigma_{\rm sub}$ and $\de_{\rm LOS}$ (top panel), and as a function of $\de_{\rm LOS}$ (bottom panel). We set the value of  $\Sigma_{\rm sub}$ to a constant ($0.05\,\,{\rm kpc}^{-2}$) in the bottom panel. This ratio increases progressively with the line-of-sight dark matter halo contribution, and is almost zero for just subhaloes. }
\end{figure}

As discussed above, both main-lens substructure and line-of-sight haloes contribute to the correlation function monopole, while only line-of-sight haloes are responsible for a non-zero quadrupole moment, via their non-linear coupling with the main-lens plane. To isolate the \emph{relative} contribution of substructure and line-of-sight haloes to the correlation function, it is thus interesting to look at the quadrupole-to-monopole ratio as a function of scale. Importantly, this ratio is invariant under the MST. 

Figure \ref{fig:abunadances} depicts how the quadrupole-to-monopole ratio changes as a function of $\Sigma_{\rm sub}$ and $\de_{\rm LOS}$ (top panel), and as a function of $\de_{\rm LOS}$ for a fixed value of $\Sigma_{\rm sub}=0.05\,\,{\rm kpc}^{-2}$ (bottom panel). These plots have been made with 100 different realizations for each pair of $\Sigma_{\rm sub}$ and $\de_{\rm LOS}$. In each case, we keep the mean value of $\kappa_{\rm div}$ inside the annular mask fixed. The solid lines represent the mean, and the shaded regions show the $68\%$ and $90\%$ confidence intervals for the 100 realizations. As anticipated, the quadrupole to monopole ratio is nearly zero for substructure only realizations, and increases as the line-of-sight halo abundance increases (parametrized via $\de_{\rm LOS}$). These plots demonstrate that the quadrupole signal is indeed coming from the line-of-sight haloes. With a careful study using real observational data, this ratio of correlation function multipoles may reveal useful clues about the relative abundance of subhaloes and line-of-sight haloes. 

\subsection{Impact of Macrolens Model}\label{macro_impact}

Since the anisotropies of the two-point correlation function are the result of the breaking of translation symmetry in the image plane due to the presence of the macrolens, it is important to study the impact of the macrolens model on the quadrupole moment. Perhaps the first question that comes to mind is whether the correlation function quadrupole is somewhat degenerate with the external gravitational shear field caused by matter in the neighbourhood of the main galaxy. After all, the external shear has a (global) quadrupolar structure. However, since we define multipole moments with reference to the radial and tangential directions of the critical curves, the quadrupole we describe here is localized, with the contribution from the external shear largely averaging out when performing the integral in equation~\eqref{eq:corr_func_def}. As we discussed in Appendix~\ref{App.C}, this results in the external shear having a negligible impact on the quadrupole and other moments of the correlation function. Moreover, the shear-ellipticity degeneracy also has a negligible effect.

On the other hand, the logarithmic slope of the elliptical power-law mass profile we use for the macrolens has a significant influence on the amplitude of the quadrupole moment. This parameter affects the non-linear coupling between lens planes by changing the central density and enclosed mass within the radius $r$ of the main-lens plane \citep{Lyskova_2018}. In Appendix~\ref{App.C}, we show that changing the logarithmic slope of the main-lens density profile does modify the exact amplitude of the quadrupole moment, although the overall qualitative picture remains the same. Unlike the logarithmic slope, the main-lens ellipticities have small effect. This suggests that a good knowledge of the macrolens mass profile is necessary to correctly interpret the quadrupole moment of the two-point function in terms of the abundance and properties of line-of-sight haloes. Fortunately, the macrolens parameters are usually the easiest to constrain from strong lensing observations. 

\section{Sensitivity to the Quadrupole Moment}\label{sec:sec5}
In this section, we assess the ability of existing space-based telescopes and future ground-based observatories to detect the quadrupole and monopole power spectra generated by subhaloes and line-of-sight haloes. To do so, we estimate the error in the binned monopole and quadrupole power spectra using a Fisher formalism. We begin by reviewing the necessary statistical formalism described in \cite{Cyr-Racine_2019} and \cite{Hezaveh_2014}, adapting it to the multiplane gravitational lensing case. We refer the reader to \cite{Cyr-Racine_2019} for an in-depth description of the full likelihood analysis. 

\subsection{Image Residuals and Decomposition}

To extract information about the subhalo and line-of-sight halo population from observed strongly lensed images of an extended source, it is in general necessary to perform detailed lens modelling using very flexible models for both the source and the macrolens structure \citep[see e.g.,][]{koopmans2005, vegetti2012, vegetti_2014, birrer2018lenstronomy, Birrer2021, Vernardos_2022}. This procedure then returns the best-fitting single-plane deflection field $\bal_{\rm eff, bf}(\xx)$ and the corresponding source brightness profile $S({\bf u})$. Since we are interested here in determining whether the anisotropic signal contained in the quadrupole moment of the correlation function \emph{is at all detectable under the best of circumstances} for a typical lens configuration, we fix the macrolens and extended source structure to their true values in the following analysis. The true and best-fitting effective deflection fields have the following shape:
\be 
\bal_{\rm eff, true} (\xx) = \bna \phi_{\rm o}(\xx) + \bna \phi_{\rm eff, sub+LOS}(\xx) +\bna \times \ab_{\rm eff, LOS}(\xx)
\ee and 
\be 
\bal_{\rm eff, bf}(\xx) = \bna\phi_{\rm o}(\xx) ,
\ee respectively. Note that $\phi_{\rm eff}$ has contributions from the main-lens, its substructure, and line-of-sight haloes, while $\ab_{\rm eff}$ describes the non-linear coupling between the different lens planes. Assuming that the source brightness profile, $S$, and the best-fitting single-plane deflection field, $\bal_{\rm eff, bf}(\xx)$, provide a reasonable fit to the observed lens, the image residuals between the best-fitting image and the true lensed image can be expressed as

\begin{align} \label{Eq. 18_}
    \de O_{\rm sub+LOS}(\xx) &= S[\xx - \bal_{\rm eff, true} (\xx)] -S[\xx- \bal_{\rm eff, bf}(\xx)]\en
     & \approx -\bna_\uu S(\uu)|_{\uu = \xx - \bna \phi_{\rm o}(\xx)}\cdot[\bna \phi_{\rm eff, sub+LOS}(\xx) \en
     & \hspace{3.5cm}+\bna \times \ab_{\rm eff, LOS}(\xx)] \en
     & = \de O_{\rm div} + \de O_{\rm curl},
\end{align} where $\bna_\uu$ indicates the source-plane gradient. Note that the right-hand side of equation~\eqref{Eq. 18_} is to be convolved with the point spread function (PSF) to take into account the response of the optical imaging system and potential astronomical (or atmospheric) seeing. To avoid clutter, we do not write down this convolution step explicitly here, but it is always included in our numerical computations. In the second step of equation~\eqref{Eq. 18_}, we use the Taylor series as the perturbation series to expand the true lensed source term since the distortions created by perturbers are generally small. We use the notations $\de O_{\rm div}$ and $\de O_{\rm curl}$ to denote the residuals caused by the divergence and the curl part of the effective deflection field, respectively. In reality, both the effects of perturbers and instrumental noise account for the observed residuals $\de O_{\rm obs}(\xx)$ \citep{Cyr-Racine_2019} and as well as having the form
\be \label{Eq.19}
\de O_{\rm obs}(\xx) = \de O_{\rm sub+LOS}(\xx) + N(\xx),
\ee where $N(\xx)$ denotes the instrumental noise. 
 
The image residuals given in equation~\eqref{Eq. 18_} may reveal the important information associated with the small-scale lensing anisotropies in the projected mass density fields of dark matter haloes. Here, we focus on the image residuals caused by the effective potential field $\phi_{\rm eff, sub+LOS}(\xx)$, as the curl contribution is always subdominant. We decompose the residuals $\de O_{\rm div}$ in terms of a set of orthonormal basis functions. As the optimal choice of basis functions, we choose a Fourier basis $\bna \varphi_{m}$ to expand the deflection field $\bna \phi_{\rm eff, sub+LOS}(\xx)$. From here onwards, we use the term $\bna \phi_{\rm div}(\xx)$ to denote the effective deflection field, $\bna \phi_{\rm eff, sub+LOS}(\xx)$, and expand as

\be 
\bna \phi_{\rm div}(\xx) = \sum_{m=1}^{N_{\rm modes}} \mathcal{B}_m \bna \varphi_{m}(\xx),
\ee

\noindent where $\mathcal{B}_m$ are the mode amplitudes of the $\bna \phi_{\rm div}(\xx)$ field. Here, index $m$ denotes the doublet $(m_x, m_y)$ indicating the indices in Cartesian coordinates. These two-dimensional Fourier basis functions are orthonormal in the sense that

\be \label{Eq.21}
\frac{1}{A}\int_A {\rm d}^2\xx \,\, \bna \varphi_{m}(\xx) \cdot \bna \varphi^*_{m'}(\xx) = \de_{mm'},
\ee where $\de_{mm'}$ is the Kronecker delta and $A$ is the area of the strong lensing region where we perform the basis function expansion. From the orthonormality condition given in equation~\eqref{Eq.21}, the mode amplitudes of the Fourier basis functions are given by

\begin{align}
    \mathcal{B}_m = \frac{1}{A}\int_A {\rm d}^2\xx \,\, \bna\varphi_m^*(\xx)\cdot \bna \phi_{\rm div}(\xx).
\end{align} 
\noindent The discrete orthonormal Fourier basis functions for $\varphi_m(\xx)$ are defined in such a way that
\be \label{Eq.23_}
\varphi_m(\xx) =
  \begin{cases}
    \frac{e^{i {\bf k}_m\cdot \xx}}{k_m}       & \quad \text{if } \xx \in A,\\
   0  & \quad  \text{otherwise}.
  \end{cases}
\ee Given the fact that these basis functions are zero on the image boundaries, one can show that

\be \label{Eq.24}
\mathcal{B}_m = -\frac{2}{A}\int_A {\rm d}^2\xx \,\, \varphi_m^*(\xx) \kappa_{\rm div}(\xx),
\ee by using integration by parts and $\kappa_{\rm div}$ given in equation~\eqref{Eq. kappa div}. Now one can write down the residuals as

\be
\de O_{\rm div}(\xx) = \sum_{m=1}^{N_{\rm modes}} \mathcal{B}_m \mathcal{W}_{m}(\xx),
\ee where

\be 
\mathcal{W}_{m}(\xx) = -\bna_\uu S(\uu)|_{\uu = \xx - \bna \phi_{\rm o}(\xx)}\cdot\bna \varphi_{m}(\xx).
\ee The gradient of the source projected onto the $m^{\rm th}$ mode of the orthonormal basis function $\bna \varphi_m$ and convolved with the PSF yields the $\mathcal{W}_{m}(\xx)$ kernel.

It is convenient to put all the information we discussed above associated with lensing residuals into vector and matrix notation. The observed residuals ($\de O_{\rm obs}$) and perturber-induced lensing residuals ($\de O_{\rm div}$) are combined into two vectors $\de {\bf O}_{\rm obs}$ and $\de {\bf O}_{\rm div}$ of length $N_{\rm pix}$, respectively. Here $N_{\rm pix}$ is the number of pixels in the image which is given by $N_{\rm pix} = N_{\rm pix,x}\times N_{\rm pix,y}$, where $N_{\rm pix,x}$ and $N_{\rm pix,y}$ are the number of pixels along the $x$ and $y$ axes of the image, respectively. The block matrix $\bf W$ of size $N_{\rm pix} \times N_{\rm modes}$ contains $N_{\rm modes}$ number of $\mathcal{W}_{m}$ kernels of length $N_{\rm pix}$ when the kernels are changed from mode space to pixel space. Finally, storing mode amplitudes $\mathcal{B}_m$ in a vector ${\bf b} \equiv \{ \mathcal{B}_m\}$ of length $N_{\rm modes}$ results in

\be
\de {\bf O}_{\rm div} \equiv {\bf Wb}.
\ee

 Assuming that the residuals $\de O_{\rm curl}$ are small in comparison to the residuals $\de O_{\rm div}$ and gathering the instrumental noise into a vector ${\bf N}$ of length $N_{\rm pix}$, we can write equation~\eqref{Eq.19} as
\be \label{Eq.28}
\de {\bf O}_{\rm obs} \approx \de {\bf O}_{\rm div} + {\bf N}.
\ee

Now, assume that the signal ($\de O_{\rm div}(\xx)$) and noise ($N(\xx)$) are uncorrelated. Using equation~\eqref{Eq.28} the full covariance matrix of shape $N_{\rm pix} \times N_{\rm pix}$ can be thus derived as

\be \label{Eq.29ii}
{\bf C}\equiv {\bf C}_{S}+{\bf C}_{N},
\ee where ${\bf C}_{S}$ and ${\bf C}_{N}$ are the signal and noise covariance matrices, respectively. The signal covariance matrix ${\bf C}_{\rm S}$ is given by
\begin{align}
    {\bf C}_{S,ij} & \equiv \langle \de {\bf O}_{{\rm div},i} \,\,\de {\bf O}_{{\rm div},j}^\da\rangle = ({\bf WC}_{\rm div}{\bf W}^\da)_{ij}\en
    & = \sum_{m=1}^{N_{\rm modes}}\sum_{m'=1}^{N_{\rm modes}} ({\bf W}_m)_i{\bf C}_{{\rm div},mm'}({\bf W}^\da_{m'})_j \,\,,
\end{align} where the variance of the mode amplitudes ${\bf C}_{\rm div}$ has the form

\begin{align} \label{Eq.31_}
    {\bf C}_{{\rm div},mm'}& \equiv \langle {\bf b}_{m}\,\,{\bf b}^\da_{m'} \rangle= \langle\mathcal{B}_m \,\, \mathcal{B}_{m'}^*\rangle \en
    & = \sum_\ell{\bf C}^{(\ell)}_{{\rm div},mm'},
\end{align} and the ensemble average $\langle \dots \rangle$ is taken over multiple realizations, and subscripts $i$ and $j$ denote the pixel. In Appendix~\ref{App.D}, an arbitrary element in the covariance matrix of the $\mathcal{B}_m$ coefficients for the $\ell ^{\rm th}$ multipole is then derived as

\begin{align} \label{Eq. 24}
   {\bf C}^{(\ell)}_{{\rm div},mm'}  & = \frac{8\pi(-i)^\ell}{A^2k_mk_{m'}} P_\ell\left(\frac{k''}{2}\right )\, T_\ell(\cos \theta) \int {\rm d}r_{\rm h} \, r_{\rm h} \, J_\ell(k'r_{\rm h}),
\end{align} where

\begin{align}
   k' & = |\kk_m-\kk_{m'}| = \sqrt{k_m^2 + k_{m'}^2 - 2k_mk_{m'}\cos(\theta_m-\theta_{m'})} \,\,\, ,
\end{align}   

\begin{align}
   k'' & = |\kk_m+\kk_{m'}| = \sqrt{k_m^2 + k_{m'}^2 + 2k_mk_{m'}\cos(\theta_m-\theta_{m'})} \,\,\, ,
\end{align} and
\begin{align}
   \cos \theta & = \frac{k_m^2-k_{m'}^2}{\sqrt{k_m^4 + k_{m'}^4 - 2k_m^2k_{m'}^2\cos[2(\theta_m-\theta_{m'})]}} \,\,\, .
\end{align} Here, $\theta_m$ and $\theta_{m'}$ are the angles made by the vectors $\kk_{m}$ and $\kk_{m'}$ with the $x$ -axis, respectively.

Here we assume a simple noise model in which the error in each pixel is contributed by a Gaussian background term and a Poisson noise term based on the photon counts. Each of these two contributions is assumed to be uncorrelated. Ensemble averaging over noise realizations then gives the following noise covariance matrix \citep{Birrer_2015, birrer2018lenstronomy}:

\begin{align}
    {\bf C}_{N,ij} & = \langle {\bf N}_i \,\,  {\bf N}_j^{\rm T} \rangle_N \en
    & = \de_{ij}\left(\sigma^2_{{\rm bkg},i} + \frac{d_{{\rm model},i}}{f_{i}} \right),
\end{align} where $\sigma_{{\rm bkg},i}$ is the estimated background rms noise from the data image, $d_{{\rm model},i}$ is the best-fitting model at the $i^{\rm th}$ pixel, and $f_i$ is the count statistics of pixel $i$. For the Poisson noise term, CCD gain and exposure time are incorporated into the $f_i$ factor.

\begin{figure*}
\begin{center}
	\includegraphics[clip, trim=0.2cm 3.0cm 0.1cm 3.2cm, width=0.9\textwidth]{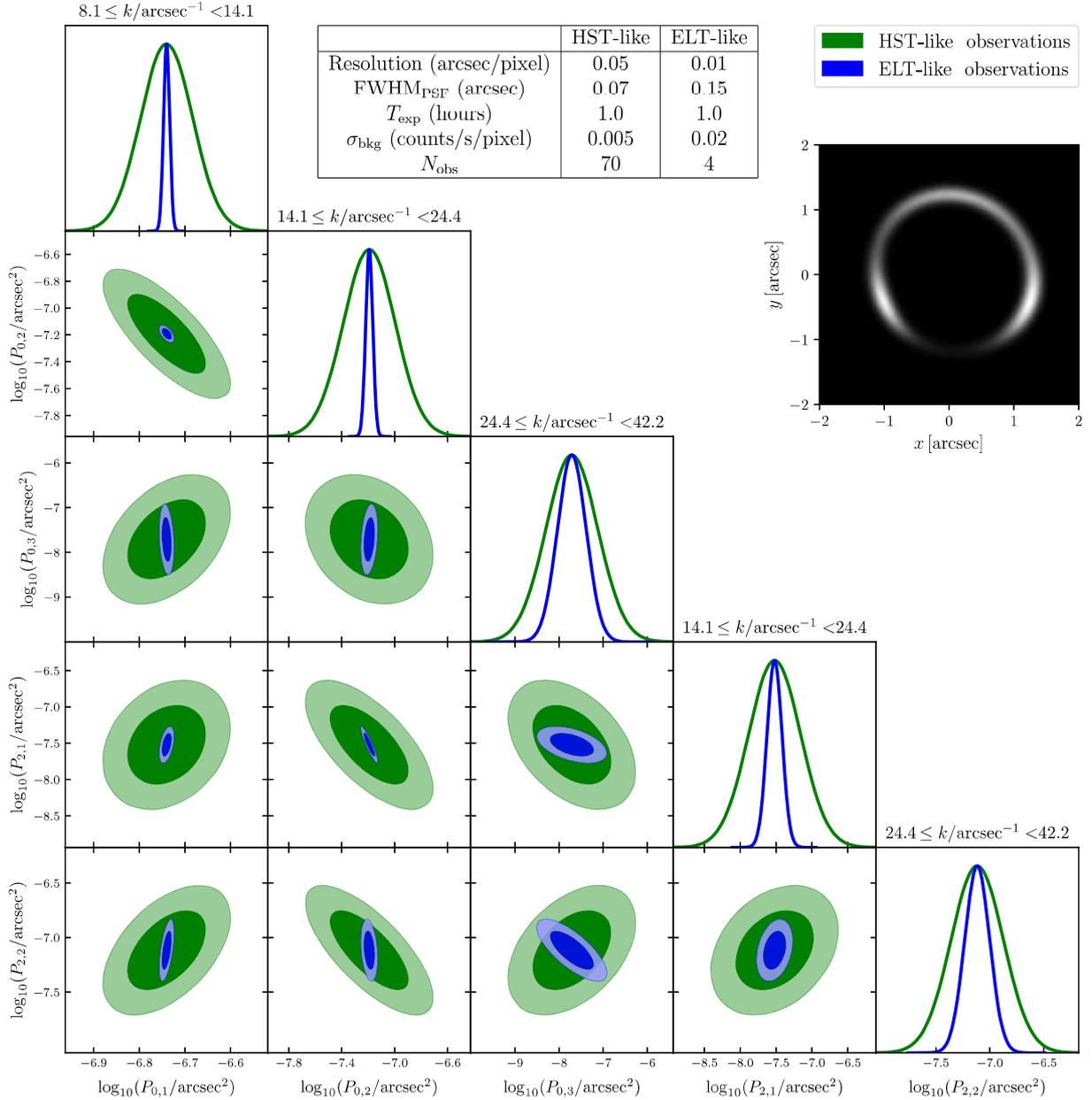}
\end{center}	
\caption{\label{fig:Fisher_triangle}The Fisher forecast results for the monopole power spectrum with $N_{{\rm bins},0}=3$ and quadrupole power spectrum with $N_{{\rm bins},2}=2$. The covariance matrices for \textit{HST}-like and ELT-like observations are taken using the inverse of the Fisher matrices, and the dark (light) shaded areas show the 68\% (95\%) credible regions. The simulated image of the Einstein ring and a table containing the relevant details of the observations are also given. Here, the surface brightness distribution of the main-lens is removed from the inset image of the Einstein ring. Note that the ELT-like observations display greater sensitivity to the monopole and quadrupole power spectra than the \textit{HST}-like observations.}
\end{figure*}

Making the standard simplifying assumption that the observed residuals are drawn from a Gaussian distribution with variance given by the sum of the signal and noise covariance matrices (equation~\eqref{Eq.29ii}), we can write down the likelihood marginalized over the mode amplitudes $\mathcal{B}_m$ as 

\be
{\mathcal L} \propto \frac{e^{-{\frac{1}{2} \de{\bf O}_{\rm obs}^{\rm T} {\bf C}^{-1}\de {\bf O}_{\rm obs}}}}{\sqrt{|{\bf C}|}} \,\,.
\ee 
Since our goal here is to determine whether the correlation function anisotropies are at all detectable under the best of circumstances, we focus here on performing Fisher forecasts.  We leave the exploration of this likelihood via Markov Chain Monte Carlo techniques to future works.

\begin{figure}
\begin{center}
	\includegraphics[clip, trim=0.25cm 0.25cm 0.1cm 0.2cm, width=0.48\textwidth]{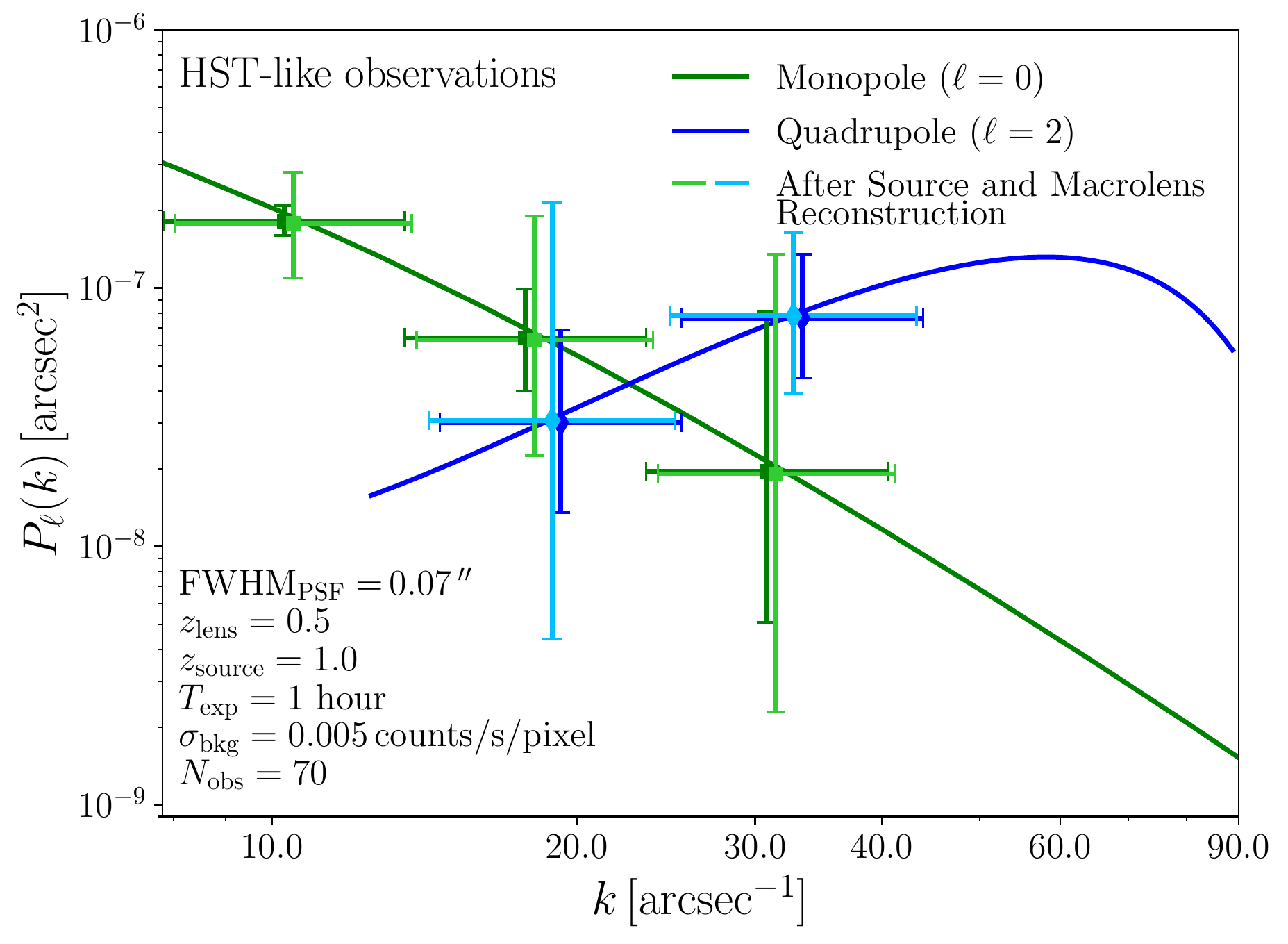}
\end{center}	
\begin{center}
	\includegraphics[clip, trim=0.25cm 0.25cm 0.1cm 0.2cm, width=0.48\textwidth]{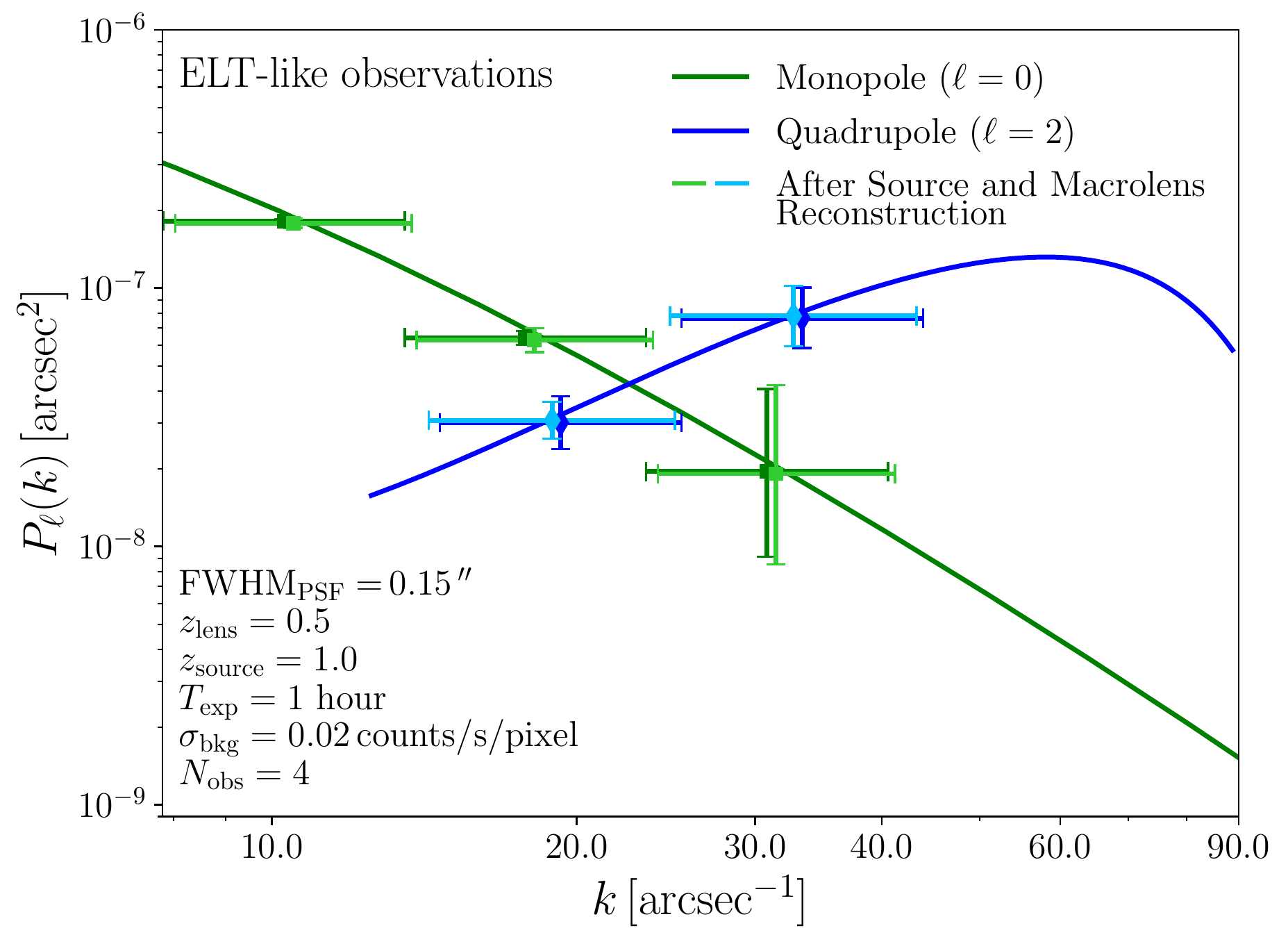}
\end{center}
\caption{\label{fig:Fisher_pk}The resulting error bars from the Fisher analysis on the fiducial monopole (green) and quadrupole (blue) power spectra for \textit{HST}-like observations (top panel) and ELT-like observations (lower panel). The error bars on the monopole (lighter green) and quadrupole (lighter blue) power spectra after the source and main-lens reconstruction are also shown. Note that the vertical error bars show the $1-\sigma$ confidence intervals. For clarity, the centres of the wavenumber bins of the power spectra are slightly shifted horizontally.}
\end{figure}

\subsection{Fisher Forecast}
To quantify the detectability of the two-point correlation function anisotropies induced by line-of-sight structure, we use as our key parameters the natural logarithm of the binned amplitudes of the monopole and quadrupole power spectra. We use the power spectrum multipoles rather than their correlation function equivalent since this drastically simplifies the computation of the Fisher matrix, as we will see below. 
Denoting these binned amplitudes as $\{\ln P_{0,i} \}_{i=1,\dots,N_{{\rm bins},0}}$ for the monopole and $\{\ln P_{2,j} \}_{j=1,\dots,N_{{\rm bins},2}}$ for the quadrupole (where $N_{{\rm bins},0}$ and $N_{{\rm bins},2}$ are the number of bins used for the monopole and quadrupole power spectra, respectively), the Fisher matrix can be written as

\begin{align}
    F_{ij} & \equiv - \left<\frac{\partial^2 \ln{\mathcal{L}}}{\partial\ln P_{\ell_1, i}\,\,\partial\ln P_{\ell_2, j}}\right>\en
    & = \frac{P_{\ell_1,i}\,\,P_{\ell_2,j}}{2}{\rm Tr}\left[{\mathbf\Gamma}\frac{\partial{\bf C}_{\rm div}}{\partial P_{\ell_1,i}} {\mathbf\Gamma}\frac{\partial{\bf C}_{\rm div}}{\partial P_{\ell_2,j}} \right],
\end{align} where ${\mathbf \Gamma} \equiv {\bf W}^\da {\bf C}^{-1}{\bf W}$ and $\ell_1,\, \ell_2 = 0,\,2$. The term $P_{\ell,i}$ denotes the amplitude of the $i^{\rm th}$ bin of the power spectrum multipole $P_{\ell}$. The total-covariance matrix of mode amplitudes ${\bf C}_{\rm div}$ represents the combined effect of the monopole and quadrupole signals in this case and admits the form
\be 
{\bf C}_{\rm div} = {\bf C}_{\rm div}^{(0)} + {\bf C}_{\rm div}^{(2)},
\ee
where the different multipole components can be computed via equation~\eqref{Eq. 24}.

In this forecast, we consider a fiducial lensed image made of a nearly complete Einstein ring that is perturbed with a realistic CDM halo realization containing both the substructure and line-of-sight haloes as shown in the middle panel of Fig.~\ref{fig:kappa_sgc}. As our fiducial lensing perturber model, we adopt the power spectrum monopole and quadrupole obtained by applying the Hankel transform given in equation~\eqref{Eq._11} to the unnormalized correlation multipoles for the divergence of the deflection field. We use three and two wave number bins uniformly spaced in $\log _{10}(k)$ for the monopole and quadrupole power spectra, respectively. Since we are interested here in detectability, we fix the source and main-lens parameters to their true values in our forecasts.

Figure~\ref{fig:Fisher_triangle} shows the parameter covariance for the simulated observation of the Einstein ring under two observational scenarios at optical wavelengths. We consider an observation similar to what is feasible with the \textit{Hubble Space Telescope} (\textit{HST}) and another observation achievable with a ground-based extremely large telescope (ELT) with advanced adaptive optics. The \textit{HST}-like image is made on a $80\times 80$ pixel grid with a resolution of 0.05 arcsec/pixel, while the ELT-like image with $400\times 400$ pixels has a 5 times higher resolution than \textit{HST}. The quality of the observation is decided by multiple factors, including the exposure time ($T_{\rm exp}$) per observation, the number of observations ($N_{\rm obs}$), the magnitude of the source, etc. We assume 70 \textit{HST}-like observations of a source with an unlensed AB magnitude of $m_{\rm AB}=18$ using the F 555 W filter with the WFC3 UVIS channel at the \textit{HST} for $T_{\rm exp}=1$ h. The full width at half-maximum of the PSF (${\rm FWHM}_{\rm PSF}$) is set to 0.07 arcsec. The choice of the filter and the camera is based on the previous observations of strong gravitational lenses \citep[see e.g.,][]{Pawase_2014, Diaz-Sanchez_2021}, and we leave a comprehensive study of sensitivity using different filters to future work. We also consider four observations of the same source with an ELT, but the quality of the observations are limited by its Moffat PSF \citep{moffat_1969} which has a FWHM of 0.15 arcsec. We also apply Gaussian background noise with rms values of $0.005\,{\rm counts/s/pixel}$ and $0.02\,{\rm counts/s/pixel}$ for the simulated \textit{HST} and ELT observations, respectively.

Overall, Fig.~\ref{fig:Fisher_triangle} shows that both the power spectrum monopole and quadrupole could be jointly detected under optimistic observational conditions. While it appears that the resources needed to make such measurement with \textit{HST} are likely enormous, the larger light-collecting area of an ELT would make detecting the power spectrum quadrupole possible with much more modest observational resources. We also see interesting correlation between the different binned amplitudes of the power spectrum multipoles. 

Fig.~\ref{fig:Fisher_pk} shows a different perspective on our Fisher forecast. Here we show the error bars (computed using the diagonal elements of the Fisher matrix) on the monopole and the quadrupole power spectra for our two different observational scenarios, overplotting them on the fiducial power spectrum model used in our analysis. This illustrates how the monopole power spectrum has most of its power on larger scales (small $k$), while the quadrupole is more important on smaller length scales (larger $k$). Again, due to a higher resolution and the larger light-gathering power of the instrument, ELT observations have an edge over \textit{HST} and show better sensitivity to the power spectra, even though the former is limited by the approximately 2 times higher FWHM of the PSF and the 4 times higher background noise contribution compared to the \textit{HST}. In comparison to the two lowest wave number bins of the monopole power spectrum obtained using mock ELT observations, the amplitude of the monopole power spectra in the third bin shows a significant drop in sensitivity due to the restrictions imposed by the FWHM of the PSF ($k_{\rm PSF}\approx 21\, {\rm arcsec}^{-1}$) and the source light profile ($k_{\rm src}\approx 22\, {\rm arcsec}^{-1}$). 

To account for potential main-lens and source mismodelling, we also perform main-lens and source reconstruction on perturbed images and use these updated lens and source parameters within our Fisher framework. Our reconstruction uses a shapelet basis set for the source surface brightness profile \citep{Refregier:2001fd, Refregier_shp_ii, Birrer_2015}. The lensed images are well reconstructed and the features of the source brightness profile are reproduced with $\sim1$\% errors. The error bars on the monopole (lighter green) and quadrupole (lighter blue) power spectra after this reconstruction are shown in Fig.~\ref{fig:Fisher_pk}. We note that the effect of the lens and source modelling is minimal for the source and lens models considered.
In summary, we see that the correlation function anisotropies induced by line-of-sight structure are in principle detectable with relatively modest observational resources on an ELT-class ground-based telescope. Of course, more complex source and main-lens structure would make these forecasted error bars somewhat larger. We leave such more realistic forecasts to future works. 

\section{Conclusions}\label{sec:Conc}

In this paper, we have introduced the concept of "effective multiplane gravitational lensing" and identified that the dark matter haloes along the line-of-sight between the observer and the source create a recognizable anisotropic signature in the convergence maps obtained using the divergence of the effective deflection field and its two-point correlation function. Unlike the subhaloes, these distinct anisotropic patterns generated by the line-of-sight add a non-vanishing quadrupole moment to the two-point correlation multipoles. We have derived the necessary formalism to extract and study this quadrupole moment from observations of strongly lensed extended sources, and we have also shown that the amplitude of this quadupole reflects the abundance of line-of-sight haloes in a strong gravitational lens system. Observations of this quadrupole could potentially reveal crucial information about the redshift evolution of the dark matter halo mass function. At a minimum, a detection of this characteristic anisotropic signature could serve as an important consistency check of the standard CDM picture on small scales. Finally, we have evaluated the sensitivity of \textit{HST}-like observations and upcoming ELT-like observations to the monopole and quadrupole signals under the best of circumstances. The results of our Fisher analysis indicate that current space-based observations could have the right sensitivity to the monopole and quadrupole power spectra, but at the price of very expensive observational resources. On the other hand, an ELT could achieve better sensitivity to the anisotropic signatures with much more reasonable observational resources due to its higher resolution and large light-collecting area. However, both the source and macrolens mismodelling would contribute to the residuals and increase the size of the error bars.

In this paper, we propose a new method to statistically study the microphysics of dark matter using small-scale anisotropies. In future works, we will investigate how different types of dark matter physics affect the shape and amplitude of the monopole and quadrupole moments. Moreover, to isolate the line-of-sight halo effects and thus advance our knowledge of the redshift distribution of dark matter haloes, we suggest studying the cross-correlation between the $\kappa_{\rm div}$ and $\kappa_{\rm curl}$ maps. Finally, to assess the sensitivity of realistic current and future observations to the anisotropic signal from line-of-sight haloes, performing full lens modelling and source reconstruction while also inferring the power spectrum multipoles will be necessary. A combined study of these non-zero multipoles could, in principle, unveil new and crucial information about the redshift distribution and inner properties of dark matter clumps affecting lensed images, thus providing sensitivity to the time evolution of the halo mass function, and to dark matter physics.

\section*{Acknowledgements}

We would like to express our gratitude to the anonymous referee for providing insightful comments and suggestions on this manuscript for improvements. We would also like to thank Tansu Daylan, Xiaolong Du, Carton Zeng, Anthony Pullen, Ekapob Kulchoakrungsun, and Charles Mace for useful discussions. This work was supported in part by the National Aeronautics and Space Administration Astrophysics Theory Program under grant 80NSSC18K1014. We also would like to thank the University of New Mexico Center for Advanced Research Computing, supported in part by the National Science Foundation, for providing the high-performance computing resources used in this work.

The work in this manuscript made partial use of the \textsc {python} packages \textsc {numpy} \citep{Numpy}, \textsc {numba} \citep{numba}, \textsc {matplotlib} \citep{matplotlib}, \textsc {scipy} \citep{SciPy}, \textsc {getdist} \citep{getdist}, \textsc {astropy} \citep{astropy}, \textsc {mcfit} \citep{mcfit}, and \textsc {h5py} \citep{h5py}. We are appreciative of the great work done by the developers of these software.

\section*{Data Availability}

All the data generated in this research, as well as the codes used, will be available upon reasonable request to the corresponding authors.




\bibliographystyle{mnras}
\bibliography{references} 

\begin{thebibliography}{}
\makeatletter
\relax
\def\mn@urlcharsother{\let\do\@makeother \do\$\do\&\do\#\do\^\do\_\do\%\do\~}
\def\mn@doi{\begingroup\mn@urlcharsother \@ifnextchar [ {\mn@doi@}
  {\mn@doi@[]}}
\def\mn@doi@[#1]#2{\def\@tempa{#1}\ifx\@tempa\@empty \href
  {http://dx.doi.org/#2} {doi:#2}\else \href {http://dx.doi.org/#2} {#1}\fi
  \endgroup}
\def\mn@eprint#1#2{\mn@eprint@#1:#2::\@nil}
\def\mn@eprint@arXiv#1{\href {http://arxiv.org/abs/#1} {{\tt arXiv:#1}}}
\def\mn@eprint@dblp#1{\href {http://dblp.uni-trier.de/rec/bibtex/#1.xml}
  {dblp:#1}}
\def\mn@eprint@#1:#2:#3:#4\@nil{\def\@tempa {#1}\def\@tempb {#2}\def\@tempc
  {#3}\ifx \@tempc \@empty \let \@tempc \@tempb \let \@tempb \@tempa \fi \ifx
  \@tempb \@empty \def\@tempb {arXiv}\fi \@ifundefined
  {mn@eprint@\@tempb}{\@tempb:\@tempc}{\expandafter \expandafter \csname
  mn@eprint@\@tempb\endcsname \expandafter{\@tempc}}}

\bibitem[\protect\citeauthoryear{Abramowitz \& Stegun}{Abramowitz \&
  Stegun}{1964}]{abramowitz+stegun}
Abramowitz M.,  Stegun I.~A.,  1964, Handbook of Mathematical Functions with
  Formulas, Graphs, and Mathematical Tables.
Dover, New York

\bibitem[\protect\citeauthoryear{{Ackermann} et~al.,}{{Ackermann}
  et~al.}{2014}]{Ackermann_2014}
{Ackermann} M.,  et~al., 2014, \mn@doi [\prd] {10.1103/PhysRevD.89.042001},
  \href {https://ui.adsabs.harvard.edu/abs/2014PhRvD..89d2001A} {89, 042001}

\bibitem[\protect\citeauthoryear{{Amorisco} et~al.,}{{Amorisco}
  et~al.}{2022}]{Amorisco_2021}
{Amorisco} N.~C.,  et~al., 2022, \mn@doi [\mnras] {10.1093/mnras/stab3527},
  \href {https://ui.adsabs.harvard.edu/abs/2022MNRAS.510.2464A} {510, 2464}

\bibitem[\protect\citeauthoryear{{Astropy Collaboration} et~al.,}{{Astropy
  Collaboration} et~al.}{2018}]{astropy}
{Astropy Collaboration} et~al., 2018, \mn@doi [\aj] {10.3847/1538-3881/aabc4f},
  \href {https://ui.adsabs.harvard.edu/abs/2018AJ....156..123A} {156, 123}

\bibitem[\protect\citeauthoryear{{Auger}, {Treu}, {Bolton}, {Gavazzi},
  {Koopmans}, {Marshall}, {Bundy}  \& {Moustakas}}{{Auger}
  et~al.}{2009}]{Auger_2009}
{Auger} M.~W.,  {Treu} T.,  {Bolton} A.~S.,  {Gavazzi} R.,  {Koopmans}
  L.~V.~E.,  {Marshall} P.~J.,  {Bundy} K.,   {Moustakas} L.~A.,  2009, \mn@doi
  [\apj] {10.1088/0004-637X/705/2/1099}, \href
  {https://ui.adsabs.harvard.edu/abs/2009ApJ...705.1099A} {705, 1099}

\bibitem[\protect\citeauthoryear{{Auger}, {Treu}, {Bolton}, {Gavazzi},
  {Koopmans}, {Marshall}, {Moustakas}  \& {Burles}}{{Auger}
  et~al.}{2010}]{Auger_2010}
{Auger} M.~W.,  {Treu} T.,  {Bolton} A.~S.,  {Gavazzi} R.,  {Koopmans}
  L.~V.~E.,  {Marshall} P.~J.,  {Moustakas} L.~A.,   {Burles} S.,  2010,
  \mn@doi [\apj] {10.1088/0004-637X/724/1/511}, \href
  {https://ui.adsabs.harvard.edu/abs/2010ApJ...724..511A} {724, 511}

\bibitem[\protect\citeauthoryear{{Baltz}, {Marshall}  \& {Oguri}}{{Baltz}
  et~al.}{2009}]{tNFW_2009}
{Baltz} E.~A.,  {Marshall} P.,   {Oguri} M.,  2009, \mn@doi [J. Cosmol.
  Astropart. Phys.] {10.1088/1475-7516/2009/01/015}, \href
  {https://ui.adsabs.harvard.edu/abs/2009JCAP...01..015B} {2009, 015}

\bibitem[\protect\citeauthoryear{Banik, Bertone, Bovy  \& Bozorgnia}{Banik
  et~al.}{2018}]{Banik:2018pjp}
Banik N.,  Bertone G.,  Bovy J.,   Bozorgnia N.,  2018, \mn@doi [J. Cosmol.
  Astropart. Phys.] {10.1088/1475-7516/2018/07/061}, 2018, 061

\bibitem[\protect\citeauthoryear{Banik, Bovy, Bertone, Erkal  \& de Boer}{Banik
  et~al.}{2021a}]{Banik:2019cza}
Banik N.,  Bovy J.,  Bertone G.,  Erkal D.,   de Boer T. J.~L.,  2021a, \mn@doi
  [\mnras] {10.1093/mnras/stab210}, 502, 2364

\bibitem[\protect\citeauthoryear{{Banik}, {Bovy}, {Bertone}, {Erkal}  \& {de
  Boer}}{{Banik} et~al.}{2021b}]{Banik:2019smi}
{Banik} N.,  {Bovy} J.,  {Bertone} G.,  {Erkal} D.,   {de Boer} T.~J.~L.,
  2021b, \mn@doi [J. Cosmol. Astropart. Phys.] {10.1088/1475-7516/2021/10/043},
  \href {https://ui.adsabs.harvard.edu/abs/2021JCAP...10..043B} {2021, 043}

\bibitem[\protect\citeauthoryear{{Bayer}, {Chatterjee}, {Koopmans}, {Vegetti},
  {McKean}, {Treu}  \& {Fassnacht}}{{Bayer} et~al.}{2018}]{Bayer_2018}
{Bayer} D.,  {Chatterjee} S.,  {Koopmans} L.~V.~E.,  {Vegetti} S.,  {McKean}
  J.~P.,  {Treu} T.,   {Fassnacht} C.~D.,  2018, arXiv e-prints, \href
  {https://ui.adsabs.harvard.edu/abs/2018arXiv180305952B} {p. arXiv:1803.05952}

\bibitem[\protect\citeauthoryear{{Bechtol} et~al.,}{{Bechtol}
  et~al.}{2015}]{Bechtol_2015}
{Bechtol} K.,  et~al., 2015, \mn@doi [\apj] {10.1088/0004-637X/807/1/50}, \href
  {https://ui.adsabs.harvard.edu/abs/2015ApJ...807...50B} {807, 50}

\bibitem[\protect\citeauthoryear{{Benson}}{{Benson}}{2012}]{BENSON_galacticus}
{Benson} A.~J.,  2012, \mn@doi [\na] {10.1016/j.newast.2011.07.004}, \href
  {https://ui.adsabs.harvard.edu/abs/2012NewA...17..175B} {17, 175}

\bibitem[\protect\citeauthoryear{{Benson}}{{Benson}}{2020}]{Benson_2020}
{Benson} A.~J.,  2020, \mn@doi [\mnras] {10.1093/mnras/staa341}, \href
  {https://ui.adsabs.harvard.edu/abs/2020MNRAS.493.1268B} {493, 1268}

\bibitem[\protect\citeauthoryear{{Beutler} et~al.,}{{Beutler}
  et~al.}{2017}]{Beutler_2017}
{Beutler} F.,  et~al., 2017, \mn@doi [\mnras] {10.1093/mnras/stw3298}, \href
  {https://ui.adsabs.harvard.edu/abs/2017MNRAS.466.2242B} {466, 2242}

\bibitem[\protect\citeauthoryear{{Birrer}}{{Birrer}}{2021}]{Birrer_2021}
{Birrer} S.,  2021, \mn@doi [\apj] {10.3847/1538-4357/ac1108}, \href
  {https://ui.adsabs.harvard.edu/abs/2021ApJ...919...38B} {919, 38}

\bibitem[\protect\citeauthoryear{{Birrer} \& {Amara}}{{Birrer} \&
  {Amara}}{2018}]{birrer2018lenstronomy}
{Birrer} S.,  {Amara} A.,  2018, \mn@doi [Physics of the Dark Universe]
  {10.1016/j.dark.2018.11.002}, \href
  {https://ui.adsabs.harvard.edu/abs/2018PDU....22..189B} {22, 189}

\bibitem[\protect\citeauthoryear{{Birrer}, {Amara}  \& {Refregier}}{{Birrer}
  et~al.}{2015}]{Birrer_2015}
{Birrer} S.,  {Amara} A.,   {Refregier} A.,  2015, \mn@doi [\apj]
  {10.1088/0004-637X/813/2/102}, \href
  {https://ui.adsabs.harvard.edu/abs/2015ApJ...813..102B} {813, 102}

\bibitem[\protect\citeauthoryear{{Birrer}, {Welschen}, {Amara}  \&
  {Refregier}}{{Birrer} et~al.}{2017}]{birrer2017}
{Birrer} S.,  {Welschen} C.,  {Amara} A.,   {Refregier} A.,  2017, \mn@doi [J.
  Cosmol. Astropart. Phys.] {10.1088/1475-7516/2017/04/049}, \href
  {https://ui.adsabs.harvard.edu/abs/2017JCAP...04..049B} {2017, 049}

\bibitem[\protect\citeauthoryear{Birrer et~al.,}{Birrer
  et~al.}{2021}]{Birrer2021}
Birrer S.,  et~al., 2021, \mn@doi [J. of Open Source Softw.]
  {10.21105/joss.03283}, 6, 3283

\bibitem[\protect\citeauthoryear{{Blandford} \& {Narayan}}{{Blandford} \&
  {Narayan}}{1986}]{Blandford_1986}
{Blandford} R.,  {Narayan} R.,  1986, \mn@doi [\apj] {10.1086/164709}, \href
  {https://ui.adsabs.harvard.edu/abs/1986ApJ...310..568B} {310, 568}

\bibitem[\protect\citeauthoryear{{Bolton}, {Burles}, {Koopmans}, {Treu}  \&
  {Moustakas}}{{Bolton} et~al.}{2006}]{Bolton_2006}
{Bolton} A.~S.,  {Burles} S.,  {Koopmans} L. V.~E.,  {Treu} T.,   {Moustakas}
  L.~A.,  2006, \mn@doi [\apj] {10.1086/498884}, \href
  {https://ui.adsabs.harvard.edu/abs/2006ApJ...638..703B} {638, 703}

\bibitem[\protect\citeauthoryear{{Bolton}, {Burles}, {Koopmans}, {Treu},
  {Gavazzi}, {Moustakas}, {Wayth}  \& {Schlegel}}{{Bolton}
  et~al.}{2008}]{Bolton_2008}
{Bolton} A.~S.,  {Burles} S.,  {Koopmans} L. V.~E.,  {Treu} T.,  {Gavazzi} R.,
  {Moustakas} L.~A.,  {Wayth} R.,   {Schlegel} D.~J.,  2008, \mn@doi [\apj]
  {10.1086/589327}, \href
  {https://ui.adsabs.harvard.edu/abs/2008ApJ...682..964B} {682, 964}

\bibitem[\protect\citeauthoryear{{Bonaca}, {Hogg}, {Price-Whelan}  \&
  {Conroy}}{{Bonaca} et~al.}{2019}]{Bonaca_2018}
{Bonaca} A.,  {Hogg} D.~W.,  {Price-Whelan} A.~M.,   {Conroy} C.,  2019,
  \mn@doi [\apj] {10.3847/1538-4357/ab2873}, \href
  {https://ui.adsabs.harvard.edu/abs/2019ApJ...880...38B} {880, 38}

\bibitem[\protect\citeauthoryear{{Bonnivard} et~al.,}{{Bonnivard}
  et~al.}{2015}]{Bonnivard_2015}
{Bonnivard} V.,  et~al., 2015, \mn@doi [\mnras] {10.1093/mnras/stv1601}, \href
  {https://ui.adsabs.harvard.edu/abs/2015MNRAS.453..849B} {453, 849}

\bibitem[\protect\citeauthoryear{{Brada{\v{c}}}, {Lombardi}  \&
  {Schneider}}{{Brada{\v{c}}} et~al.}{2004}]{Bradac_2004}
{Brada{\v{c}}} M.,  {Lombardi} M.,   {Schneider} P.,  2004, \mn@doi [\aap]
  {10.1051/0004-6361:20035744}, \href
  {https://ui.adsabs.harvard.edu/abs/2004A&A...424...13B} {424, 13}

\bibitem[\protect\citeauthoryear{Brennan, Benson, Cyr-Racine, Keeton, Moustakas
   \& Pullen}{Brennan et~al.}{2019}]{Brennan:2018jhq}
Brennan S.,  Benson A.~J.,  Cyr-Racine F.-Y.,  Keeton C.~R.,  Moustakas L.~A.,
   Pullen A.~R.,  2019, \mn@doi [\mnras] {10.1093/mnras/stz1607}, 488, 5085

\bibitem[\protect\citeauthoryear{{Brownstein} et~al.,}{{Brownstein}
  et~al.}{2012}]{Brownstein_2012}
{Brownstein} J.~R.,  et~al., 2012, \mn@doi [\apj] {10.1088/0004-637X/744/1/41},
  \href {https://ui.adsabs.harvard.edu/abs/2012ApJ...744...41B} {744, 41}

\bibitem[\protect\citeauthoryear{Bullock \& Boylan-Kolchin}{Bullock \&
  Boylan-Kolchin}{2017}]{Bullock:2017xww}
Bullock J.~S.,  Boylan-Kolchin M.,  2017, \mn@doi [Ann. Rev. Astron.
  Astrophys.] {10.1146/annurev-astro-091916-055313}, 55, 343

\bibitem[\protect\citeauthoryear{{Buschmann}, {Kopp}, {Safdi}  \&
  {Wu}}{{Buschmann} et~al.}{2018}]{Buschmann_2018}
{Buschmann} M.,  {Kopp} J.,  {Safdi} B.~R.,   {Wu} C.-L.,  2018, \mn@doi [\prl]
  {10.1103/PhysRevLett.120.211101}, \href
  {https://ui.adsabs.harvard.edu/abs/2018PhRvL.120u1101B} {120, 211101}

\bibitem[\protect\citeauthoryear{{Calabrese} \& {Spergel}}{{Calabrese} \&
  {Spergel}}{2016}]{Calabrese_2016}
{Calabrese} E.,  {Spergel} D.~N.,  2016, \mn@doi [\mnras]
  {10.1093/mnras/stw1256}, \href
  {https://ui.adsabs.harvard.edu/abs/2016MNRAS.460.4397C} {460, 4397}

\bibitem[\protect\citeauthoryear{{Caldwell} et~al.,}{{Caldwell}
  et~al.}{2017}]{Caldwell_2017}
{Caldwell} N.,  et~al., 2017, \mn@doi [\apj] {10.3847/1538-4357/aa688e}, \href
  {https://ui.adsabs.harvard.edu/abs/2017ApJ...839...20C} {839, 20}

\bibitem[\protect\citeauthoryear{{Cardone}}{{Cardone}}{2004}]{Cardone_2004}
{Cardone} V.~F.,  2004, \mn@doi [\aap] {10.1051/0004-6361:20031696}, \href
  {https://ui.adsabs.harvard.edu/abs/2004A&A...415..839C} {415, 839}

\bibitem[\protect\citeauthoryear{{Cerny} et~al.,}{{Cerny}
  et~al.}{2021}]{Cerny_2021}
{Cerny} W.,  et~al., 2021, \mn@doi [\apjl] {10.3847/2041-8213/ac2d9a}, \href
  {https://ui.adsabs.harvard.edu/abs/2021ApJ...920L..44C} {920, L44}

\bibitem[\protect\citeauthoryear{{Chatterjee} \& {Koopmans}}{{Chatterjee} \&
  {Koopmans}}{2018}]{Chatterjee:2018ast}
{Chatterjee} S.,  {Koopmans} L.~V.~E.,  2018, \mn@doi [\mnras]
  {10.1093/mnras/stx2674}, \href
  {http://adsabs.harvard.edu/abs/2018MNRAS.474.1762C} {474, 1762}

\bibitem[\protect\citeauthoryear{{Chiba}}{{Chiba}}{2002}]{Chiba:aa}
{Chiba} M.,  2002, \mn@doi [\apj] {10.1086/324493}, \href
  {http://adsabs.harvard.edu/abs/2002ApJ...565...17C} {565, 17}

\bibitem[\protect\citeauthoryear{{Clark}, {Lewis}  \& {Scott}}{{Clark}
  et~al.}{2016}]{Clark_2016}
{Clark} H.~A.,  {Lewis} G.~F.,   {Scott} P.,  2016, \mn@doi [\mnras]
  {10.1093/mnras/stv2743}, \href
  {https://ui.adsabs.harvard.edu/abs/2016MNRAS.456.1394C} {456, 1394}

\bibitem[\protect\citeauthoryear{Collette}{Collette}{2013}]{h5py}
Collette A.,  2013, Python and HDF5.
O'Reilly Media Inc.

\bibitem[\protect\citeauthoryear{Colton \& Kress}{Colton \&
  Kress}{1998}]{Colton98inverseacoustic}
Colton D.,  Kress R.,  1998, Inverse Acoustic and Electromagnetic Scattering
  Theory.
Applied Mathematical Sciences, Springer Berlin Heidelberg

\bibitem[\protect\citeauthoryear{{Cooray} \& {Sheth}}{{Cooray} \&
  {Sheth}}{2002}]{Cooray_halo_models}
{Cooray} A.,  {Sheth} R.,  2002, \mn@doi [\physrep]
  {10.1016/S0370-1573(02)00276-4}, \href
  {https://ui.adsabs.harvard.edu/abs/2002PhR...372....1C} {372, 1}

\bibitem[\protect\citeauthoryear{{Cornachione} et~al.,}{{Cornachione}
  et~al.}{2018}]{Cornachione_2018}
{Cornachione} M.~A.,  et~al., 2018, \mn@doi [\apj] {10.3847/1538-4357/aaa412},
  \href {https://ui.adsabs.harvard.edu/abs/2018ApJ...853..148C} {853, 148}

\bibitem[\protect\citeauthoryear{{Cremonese}, {Ezquiaga}  \&
  {Salzano}}{{Cremonese} et~al.}{2021}]{Cremonese_2021}
{Cremonese} P.,  {Ezquiaga} J.~M.,   {Salzano} V.,  2021, \mn@doi [\prd]
  {10.1103/PhysRevD.104.023503}, \href
  {https://ui.adsabs.harvard.edu/abs/2021PhRvD.104b3503C} {104, 023503}

\bibitem[\protect\citeauthoryear{{Cyr-Racine}, {Keeton}  \&
  {Moustakas}}{{Cyr-Racine} et~al.}{2019}]{Cyr-Racine_2019}
{Cyr-Racine} F.-Y.,  {Keeton} C.~R.,   {Moustakas} L.~A.,  2019, \mn@doi [\prd]
  {10.1103/PhysRevD.100.023013}, \href
  {https://ui.adsabs.harvard.edu/abs/2019PhRvD.100b3013C} {100, 023013}

\bibitem[\protect\citeauthoryear{{Dalal} \& {Kochanek}}{{Dalal} \&
  {Kochanek}}{2002}]{Dalal_2002}
{Dalal} N.,  {Kochanek} C.~S.,  2002, \mn@doi [\apj] {10.1086/340303}, \href
  {https://ui.adsabs.harvard.edu/abs/2002ApJ...572...25D} {572, 25}

\bibitem[\protect\citeauthoryear{{Delos} \& {Schmidt}}{{Delos} \&
  {Schmidt}}{2022}]{Delos_2021}
{Delos} M.~S.,  {Schmidt} F.,  2022, \mn@doi [\mnras] {10.1093/mnras/stac1022},
  \href {https://ui.adsabs.harvard.edu/abs/2022MNRAS.513.3682D} {513, 3682}

\bibitem[\protect\citeauthoryear{Despali, Vegetti, White, Giocoli  \& van~den
  Bosch}{Despali et~al.}{2018}]{Despali:2017ksx}
Despali G.,  Vegetti S.,  White S. D.~M.,  Giocoli C.,   van~den Bosch F.~C.,
  2018, \mn@doi [\mnras] {10.1093/mnras/sty159}, 475, 5424

\bibitem[\protect\citeauthoryear{Diaz~Rivero, Cyr-Racine  \&
  Dvorkin}{Diaz~Rivero et~al.}{2018a}]{DiazRivero:2017xkd}
Diaz~Rivero A.,  Cyr-Racine F.-Y.,   Dvorkin C.,  2018a, \mn@doi [Phys. Rev. D]
  {10.1103/PhysRevD.97.023001}, 97, 023001

\bibitem[\protect\citeauthoryear{D\'\i{}az~Rivero, Dvorkin, Cyr-Racine, Zavala
  \& Vogelsberger}{D\'\i{}az~Rivero et~al.}{2018b}]{DiazRivero:2018oxk}
D\'\i{}az~Rivero A.,  Dvorkin C.,  Cyr-Racine F.-Y.,  Zavala J.,   Vogelsberger
  M.,  2018b, \mn@doi [Phys. Rev. D] {10.1103/PhysRevD.98.103517}, 98, 103517

\bibitem[\protect\citeauthoryear{{D{\'\i}az-S{\'a}nchez}, {Dannerbauer},
  {Sulzenauer}, {Iglesias-Groth}  \& {Rebolo}}{{D{\'\i}az-S{\'a}nchez}
  et~al.}{2021}]{Diaz-Sanchez_2021}
{D{\'\i}az-S{\'a}nchez} A.,  {Dannerbauer} H.,  {Sulzenauer} N.,
  {Iglesias-Groth} S.,   {Rebolo} R.,  2021, \mn@doi [\apj]
  {10.3847/1538-4357/ac0f75}, \href
  {https://ui.adsabs.harvard.edu/abs/2021ApJ...919...48D} {919, 48}

\bibitem[\protect\citeauthoryear{{Drlica-Wagner} et~al.,}{{Drlica-Wagner}
  et~al.}{2015}]{Drlica-Wagner_2015}
{Drlica-Wagner} A.,  et~al., 2015, \mn@doi [\apj]
  {10.1088/0004-637X/813/2/109}, \href
  {https://ui.adsabs.harvard.edu/abs/2015ApJ...813..109D} {813, 109}

\bibitem[\protect\citeauthoryear{{Drlica-Wagner} et~al.,}{{Drlica-Wagner}
  et~al.}{2020}]{Drlica-Wagner_2020}
{Drlica-Wagner} A.,  et~al., 2020, \mn@doi [\apj] {10.3847/1538-4357/ab7eb9},
  \href {https://ui.adsabs.harvard.edu/abs/2020ApJ...893...47D} {893, 47}

\bibitem[\protect\citeauthoryear{Enzi et~al.}{Enzi et~al.}{2021}]{Enzi:2020ieg}
Enzi W.,  et~al., 2021, \mn@doi [\mnras] {10.1093/mnras/stab1960}, 506, 5848

\bibitem[\protect\citeauthoryear{{Falco}, {Gorenstein}  \& {Shapiro}}{{Falco}
  et~al.}{1985}]{Falco_1985}
{Falco} E.~E.,  {Gorenstein} M.~V.,   {Shapiro} I.~I.,  1985, \mn@doi [\apjl]
  {10.1086/184422}, \href
  {https://ui.adsabs.harvard.edu/abs/1985ApJ...289L...1F} {289, L1}

\bibitem[\protect\citeauthoryear{{Gavazzi}, {Treu}, {Koopmans}, {Bolton},
  {Moustakas}, {Burles}  \& {Marshall}}{{Gavazzi} et~al.}{2008}]{Gavazzi_2008}
{Gavazzi} R.,  {Treu} T.,  {Koopmans} L. V.~E.,  {Bolton} A.~S.,  {Moustakas}
  L.~A.,  {Burles} S.,   {Marshall} P.~J.,  2008, \mn@doi [\apj]
  {10.1086/529541}, \href
  {https://ui.adsabs.harvard.edu/abs/2008ApJ...677.1046G} {677, 1046}

\bibitem[\protect\citeauthoryear{{Geha} et~al.,}{{Geha}
  et~al.}{2017}]{Geha_2017}
{Geha} M.,  et~al., 2017, \mn@doi [\apj] {10.3847/1538-4357/aa8626}, \href
  {https://ui.adsabs.harvard.edu/abs/2017ApJ...847....4G} {847, 4}

\bibitem[\protect\citeauthoryear{Gilman, Birrer, Treu, Nierenberg  \&
  Benson}{Gilman et~al.}{2019}]{Gilman:2019vca}
Gilman D.,  Birrer S.,  Treu T.,  Nierenberg A.,   Benson A.,  2019, \mn@doi
  [\mnras] {10.1093/mnras/stz1593}, 487, 5721

\bibitem[\protect\citeauthoryear{Gilman, Birrer, Nierenberg, Treu, Du  \&
  Benson}{Gilman et~al.}{2020a}]{Gilman:2019nap}
Gilman D.,  Birrer S.,  Nierenberg A.,  Treu T.,  Du X.,   Benson A.,  2020a,
  \mn@doi [\mnras] {10.1093/mnras/stz3480}, 491, 6077

\bibitem[\protect\citeauthoryear{Gilman, Du, Benson, Birrer, Nierenberg  \&
  Treu}{Gilman et~al.}{2020b}]{Gilman:2019bdm}
Gilman D.,  Du X.,  Benson A.,  Birrer S.,  Nierenberg A.,   Treu T.,  2020b,
  \mn@doi [\mnras] {10.1093/mnrasl/slz173}, 492, L12

\bibitem[\protect\citeauthoryear{{Gilman}, {Bovy}, {Treu}, {Nierenberg},
  {Birrer}, {Benson}  \& {Sameie}}{{Gilman} et~al.}{2021}]{gilman2021strong}
{Gilman} D.,  {Bovy} J.,  {Treu} T.,  {Nierenberg} A.,  {Birrer} S.,  {Benson}
  A.,   {Sameie} O.,  2021, \mn@doi [\mnras] {10.1093/mnras/stab2335}, \href
  {https://ui.adsabs.harvard.edu/abs/2021MNRAS.507.2432G} {507, 2432}

\bibitem[\protect\citeauthoryear{{Gilman}, {Benson}, {Bovy}, {Birrer}, {Treu}
  \& {Nierenberg}}{{Gilman} et~al.}{2022}]{Gilman_2021_pwr}
{Gilman} D.,  {Benson} A.,  {Bovy} J.,  {Birrer} S.,  {Treu} T.,   {Nierenberg}
  A.,  2022, \mn@doi [\mnras] {10.1093/mnras/stac670}, \href
  {https://ui.adsabs.harvard.edu/abs/2022MNRAS.512.3163G} {512, 3163}

\bibitem[\protect\citeauthoryear{Graus, Bullock, Boylan-Kolchin  \&
  Nierenberg}{Graus et~al.}{2018}]{Graus:2017rrr}
Graus A.~S.,  Bullock J.~S.,  Boylan-Kolchin M.,   Nierenberg A.~M.,  2018,
  \mn@doi [\mnras] {10.1093/mnras/sty1924}, 480, 1322

\bibitem[\protect\citeauthoryear{Harris et~al.,}{Harris et~al.}{2020}]{Numpy}
Harris C.~R.,  et~al., 2020, \mn@doi [Nature] {10.1038/s41586-020-2649-2}, 585,
  357

\bibitem[\protect\citeauthoryear{{Hayashi}, {Ibe}, {Kobayashi}, {Nakayama}  \&
  {Shirai}}{{Hayashi} et~al.}{2021}]{Hayashi_2021}
{Hayashi} K.,  {Ibe} M.,  {Kobayashi} S.,  {Nakayama} Y.,   {Shirai} S.,  2021,
  \mn@doi [\prd] {10.1103/PhysRevD.103.023017}, \href
  {https://ui.adsabs.harvard.edu/abs/2021PhRvD.103b3017H} {103, 023017}

\bibitem[\protect\citeauthoryear{{He} et~al.,}{{He} et~al.}{2022a}]{He_2022}
{He} Q.,  et~al., 2022a, arXiv e-prints, \href
  {https://ui.adsabs.harvard.edu/abs/2022arXiv220210191H} {p. arXiv:2202.10191}

\bibitem[\protect\citeauthoryear{{He} et~al.,}{{He}
  et~al.}{2022b}]{He_et_al_2020}
{He} Q.,  et~al., 2022b, \mn@doi [\mnras] {10.1093/mnras/stac191}, \href
  {https://ui.adsabs.harvard.edu/abs/2022MNRAS.511.3046H} {511, 3046}

\bibitem[\protect\citeauthoryear{Hezaveh et~al.,}{Hezaveh
  et~al.}{2016a}]{Hezaveh2016}
Hezaveh Y.~D.,  et~al., 2016a, \mn@doi [\apj] {10.3847/0004-637X/823/1/37},
  823, 37

\bibitem[\protect\citeauthoryear{Hezaveh, Dalal, Holder, Kisner, Kuhlen  \&
  Perreault~Levasseur}{Hezaveh et~al.}{2016b}]{Hezaveh_2014}
Hezaveh Y.,  Dalal N.,  Holder G.,  Kisner T.,  Kuhlen M.,
  Perreault~Levasseur L.,  2016b, \mn@doi [J. Cosmol. Astropart. Phys.]
  {10.1088/1475-7516/2016/11/048}, 1611, 048

\bibitem[\protect\citeauthoryear{{Hsueh}, {Fassnacht}, {Vegetti}, {McKean},
  {Spingola}, {Auger}, {Koopmans}  \& {Lagattuta}}{{Hsueh}
  et~al.}{2016}]{Hsueh_2016}
{Hsueh} J.-W.,  {Fassnacht} C.~D.,  {Vegetti} S.,  {McKean} J.~P.,  {Spingola}
  C.,  {Auger} M.~W.,  {Koopmans} L.~V.~E.,   {Lagattuta} D.~J.,  2016, \mn@doi
  [\mnras] {10.1093/mnrasl/slw146}, \href
  {https://ui.adsabs.harvard.edu/abs/2016MNRAS.463L..51H} {463, L51}

\bibitem[\protect\citeauthoryear{{Hsueh} et~al.,}{{Hsueh}
  et~al.}{2017}]{Hsueh_2017}
{Hsueh} J.-W.,  et~al., 2017, \mn@doi [\mnras] {10.1093/mnras/stx1082}, \href
  {https://ui.adsabs.harvard.edu/abs/2017MNRAS.469.3713H} {469, 3713}

\bibitem[\protect\citeauthoryear{Hsueh, Enzi, Vegetti, Auger, Fassnacht,
  Despali, Koopmans  \& McKean}{Hsueh et~al.}{2020}]{Hsueh:2019ynk}
Hsueh J.-W.,  Enzi W.,  Vegetti S.,  Auger M.~W.,  Fassnacht C.~D.,  Despali
  G.,  Koopmans L. V.~E.,   McKean J.~P.,  2020, \mn@doi [\mnras]
  {10.1093/mnras/stz3177}, 492, 3047

\bibitem[\protect\citeauthoryear{Hunter}{Hunter}{2007}]{matplotlib}
Hunter J.~D.,  2007, \mn@doi [Comput. Sci. Eng.] {10.1109/MCSE.2007.55}, 9, 90

\bibitem[\protect\citeauthoryear{{Jose}, {Baugh}, {Lacey}  \&
  {Subramanian}}{{Jose} et~al.}{2017}]{Jose_2017}
{Jose} C.,  {Baugh} C.~M.,  {Lacey} C.~G.,   {Subramanian} K.,  2017, \mn@doi
  [\mnras] {10.1093/mnras/stx1014}, \href
  {https://ui.adsabs.harvard.edu/abs/2017MNRAS.469.4428J} {469, 4428}

\bibitem[\protect\citeauthoryear{{Keeton}}{{Keeton}}{2001}]{Keeton_2001}
{Keeton} C.~R.,  2001, arXiv e-prints, \href
  {https://ui.adsabs.harvard.edu/abs/2001astro.ph..2341K} {pp
  astro--ph/0102341}

\bibitem[\protect\citeauthoryear{{Keeton}}{{Keeton}}{2003}]{Keeton:2003aa}
{Keeton} C.~R.,  2003, \mn@doi [\apj] {10.1086/345717}, \href
  {https://ui.adsabs.harvard.edu/abs/2003ApJ...584..664K} {584, 664}

\bibitem[\protect\citeauthoryear{{Kirby}, {Cohen}, {Guhathakurta}, {Cheng},
  {Bullock}  \& {Gallazzi}}{{Kirby} et~al.}{2013}]{Kirby_2013}
{Kirby} E.~N.,  {Cohen} J.~G.,  {Guhathakurta} P.,  {Cheng} L.,  {Bullock}
  J.~S.,   {Gallazzi} A.,  2013, \mn@doi [\apj] {10.1088/0004-637X/779/2/102},
  \href {https://ui.adsabs.harvard.edu/abs/2013ApJ...779..102K} {779, 102}

\bibitem[\protect\citeauthoryear{Kochanek \& Dalal}{Kochanek \&
  Dalal}{2003}]{Kochanek_2003}
Kochanek C.~S.,  Dalal N.,  2003, \mn@doi [AIP Conf. Proc.]
  {10.1063/1.1581778}, 666, 103

\bibitem[\protect\citeauthoryear{{Kochanek} \& {Dalal}}{{Kochanek} \&
  {Dalal}}{2004}]{Kochanek_2004}
{Kochanek} C.~S.,  {Dalal} N.,  2004, \mn@doi [\apj] {10.1086/421436}, \href
  {https://ui.adsabs.harvard.edu/abs/2004ApJ...610...69K} {610, 69}

\bibitem[\protect\citeauthoryear{{Koopmans}}{{Koopmans}}{2005}]{koopmans2005}
{Koopmans} L.~V.~E.,  2005, \mn@doi [\mnras]
  {10.1111/j.1365-2966.2005.09523.x}, \href
  {http://adsabs.harvard.edu/abs/2005MNRAS.363.1136K} {363, 1136}

\bibitem[\protect\citeauthoryear{{Koposov}, {Belokurov}, {Torrealba}  \&
  {Evans}}{{Koposov} et~al.}{2015}]{Koposov_2015}
{Koposov} S.~E.,  {Belokurov} V.,  {Torrealba} G.,   {Evans} N.~W.,  2015,
  \mn@doi [\apj] {10.1088/0004-637X/805/2/130}, \href
  {https://ui.adsabs.harvard.edu/abs/2015ApJ...805..130K} {805, 130}

\bibitem[\protect\citeauthoryear{Lam, Pitrou  \& Seibert}{Lam
  et~al.}{2015}]{numba}
Lam S.~K.,  Pitrou A.,   Seibert S.,  2015, in Proceedings of the Second
  Workshop on the LLVM Compiler Infrastructure in HPC. pp~1--6

\bibitem[\protect\citeauthoryear{{Laporte}, {Walker}  \&
  {Penarrubia}}{{Laporte} et~al.}{2013}]{Laporte_2013}
{Laporte} C.~F.~P.,  {Walker} M.~G.,   {Penarrubia} J.,  2013, \mn@doi [\mnras]
  {10.1093/mnrasl/slt057}, \href
  {https://ui.adsabs.harvard.edu/abs/2013MNRAS.433L..54L} {433, L54}

\bibitem[\protect\citeauthoryear{Lazar, Bullock, Boylan-Kolchin, Feldmann,
  Çatmabacak  \& Moustakas}{Lazar et~al.}{2021}]{Lazar_2021}
Lazar A.,  Bullock J.~S.,  Boylan-Kolchin M.,  Feldmann R.,  Çatmabacak O.,
  Moustakas L.,  2021, \mn@doi [\mnras] {10.1093/mnras/stab448}, 502, 6064

\bibitem[\protect\citeauthoryear{{Lewis}}{{Lewis}}{2019}]{getdist}
{Lewis} A.,  2019, arXiv e-prints, \href
  {https://ui.adsabs.harvard.edu/abs/2019arXiv191013970L} {p. arXiv:1910.13970}

\bibitem[\protect\citeauthoryear{{Li}}{{Li}}{2019}]{mcfit}
{Li} Y.,  2019, Astrophysics Source Code Library, record ascl:1906.017

\bibitem[\protect\citeauthoryear{Li, Frenk, Cole, Wang  \& Gao}{Li
  et~al.}{2017}]{Li:2016afu}
Li R.,  Frenk C.~S.,  Cole S.,  Wang Q.,   Gao L.,  2017, \mn@doi [\mnras]
  {10.1093/mnras/stx554}, 468, 1426

\bibitem[\protect\citeauthoryear{{Lin} \& {Ishak}}{{Lin} \&
  {Ishak}}{2016}]{Lin_2016}
{Lin} W.,  {Ishak} M.,  2016, \mn@doi [J. Cosmol. Astropart. Phys.]
  {10.1088/1475-7516/2016/10/025}, \href
  {https://ui.adsabs.harvard.edu/abs/2016JCAP...10..025L} {2016, 025}

\bibitem[\protect\citeauthoryear{{Lyskova}, {Churazov}  \& {Naab}}{{Lyskova}
  et~al.}{2018}]{Lyskova_2018}
{Lyskova} N.,  {Churazov} E.,   {Naab} T.,  2018, \mn@doi [\mnras]
  {10.1093/mnras/sty018}, \href
  {https://ui.adsabs.harvard.edu/abs/2018MNRAS.475.2403L} {475, 2403}

\bibitem[\protect\citeauthoryear{{Mao} \& {Schneider}}{{Mao} \&
  {Schneider}}{1998}]{Mao:1998aa}
{Mao} S.,  {Schneider} P.,  1998, \mn@doi [\mnras]
  {10.1046/j.1365-8711.1998.01319.x}, \href
  {https://ui.adsabs.harvard.edu/abs/1998MNRAS.295..587M} {295, 587}

\bibitem[\protect\citeauthoryear{{Mao}, {Geha}, {Wechsler}, {Weiner},
  {Tollerud}, {Nadler}  \& {Kallivayalil}}{{Mao} et~al.}{2021}]{Mao_2021}
{Mao} Y.-Y.,  {Geha} M.,  {Wechsler} R.~H.,  {Weiner} B.,  {Tollerud} E.~J.,
  {Nadler} E.~O.,   {Kallivayalil} N.,  2021, \mn@doi [\apj]
  {10.3847/1538-4357/abce58}, \href
  {https://ui.adsabs.harvard.edu/abs/2021ApJ...907...85M} {907, 85}

\bibitem[\protect\citeauthoryear{McCully, Keeton, Wong  \& Zabludoff}{McCully
  et~al.}{2014}]{McCully_2014}
McCully C.,  Keeton C.~R.,  Wong K.~C.,   Zabludoff A.~I.,  2014, \mn@doi
  [\mnras] {10.1093/mnras/stu1316}, 443, 3631

\bibitem[\protect\citeauthoryear{{Metcalf} \& {Zhao}}{{Metcalf} \&
  {Zhao}}{2002}]{Metcalf:ac}
{Metcalf} R.~B.,  {Zhao} H.,  2002, \mn@doi [\apjl] {10.1086/339798}, \href
  {http://adsabs.harvard.edu/abs/2002ApJ...567L...5M} {567, L5}

\bibitem[\protect\citeauthoryear{{Minor}, {Kaplinghat}  \& {Li}}{{Minor}
  et~al.}{2017}]{minor2016}
{Minor} Q.~E.,  {Kaplinghat} M.,   {Li} N.,  2017, \mn@doi [\apj]
  {10.3847/1538-4357/aa7fee}, \href
  {https://ui.adsabs.harvard.edu/abs/2017ApJ...845..118M} {845, 118}

\bibitem[\protect\citeauthoryear{{Moffat}}{{Moffat}}{1969}]{moffat_1969}
{Moffat} A.~F.~J.,  1969, \aap, \href
  {https://ui.adsabs.harvard.edu/abs/1969A&A.....3..455M} {3, 455}

\bibitem[\protect\citeauthoryear{Nadler, Mao, Wechsler, Garrison-Kimmel  \&
  Wetzel}{Nadler et~al.}{2018}]{Nadler:2017dxq}
Nadler E.~O.,  Mao Y.-Y.,  Wechsler R.~H.,  Garrison-Kimmel S.,   Wetzel A.,
  2018, \mn@doi [\apj] {10.3847/1538-4357/aac266}, 859, 129

\bibitem[\protect\citeauthoryear{Nadler, Gluscevic, Boddy  \& Wechsler}{Nadler
  et~al.}{2019}]{Nadler:2019zrb}
Nadler E.~O.,  Gluscevic V.,  Boddy K.~K.,   Wechsler R.~H.,  2019, \mn@doi
  [\apj] {10.3847/2041-8213/ab1eb2}, 878, L32

\bibitem[\protect\citeauthoryear{Nadler et~al.}{Nadler
  et~al.}{2020}]{DES:2019ltu}
Nadler E.~O.,  et~al., 2020, \mn@doi [\apj] {10.3847/1538-4357/ab846a}, 893, 48

\bibitem[\protect\citeauthoryear{{Nadler} et~al.,}{{Nadler}
  et~al.}{2021a}]{Nadler_2021}
{Nadler} E.~O.,  et~al., 2021a, \mn@doi [\prl]
  {10.1103/PhysRevLett.126.091101}, \href
  {https://ui.adsabs.harvard.edu/abs/2021PhRvL.126i1101N} {126, 091101}

\bibitem[\protect\citeauthoryear{{Nadler}, {Birrer}, {Gilman}, {Wechsler},
  {Du}, {Benson}, {Nierenberg}  \& {Treu}}{{Nadler}
  et~al.}{2021b}]{Nadler:2021dft}
{Nadler} E.~O.,  {Birrer} S.,  {Gilman} D.,  {Wechsler} R.~H.,  {Du} X.,
  {Benson} A.,  {Nierenberg} A.~M.,   {Treu} T.,  2021b, \mn@doi [\apj]
  {10.3847/1538-4357/abf9a3}, \href
  {https://ui.adsabs.harvard.edu/abs/2021ApJ...917....7N} {917, 7}

\bibitem[\protect\citeauthoryear{{Newton}, {Cautun}, {Jenkins}, {Frenk}  \&
  {Helly}}{{Newton} et~al.}{2018}]{Newton_2018}
{Newton} O.,  {Cautun} M.,  {Jenkins} A.,  {Frenk} C.~S.,   {Helly} J.~C.,
  2018, \mn@doi [\mnras] {10.1093/mnras/sty1085}, \href
  {https://ui.adsabs.harvard.edu/abs/2018MNRAS.479.2853N} {479, 2853}

\bibitem[\protect\citeauthoryear{{Nierenberg}, {Treu}, {Wright}, {Fassnacht}
  \& {Auger}}{{Nierenberg} et~al.}{2014}]{Nierenberg_2014}
{Nierenberg} A.~M.,  {Treu} T.,  {Wright} S.~A.,  {Fassnacht} C.~D.,   {Auger}
  M.~W.,  2014, \mn@doi [\mnras] {10.1093/mnras/stu862}, \href
  {https://ui.adsabs.harvard.edu/abs/2014MNRAS.442.2434N} {442, 2434}

\bibitem[\protect\citeauthoryear{{Nierenberg} et~al.,}{{Nierenberg}
  et~al.}{2017}]{Nierenberg_2017}
{Nierenberg} A.~M.,  et~al., 2017, \mn@doi [\mnras] {10.1093/mnras/stx1400},
  \href {https://ui.adsabs.harvard.edu/abs/2017MNRAS.471.2224N} {471, 2224}

\bibitem[\protect\citeauthoryear{{O'Riordan}, {Warren}  \&
  {Mortlock}}{{O'Riordan} et~al.}{2021}]{O'Riordan_2021}
{O'Riordan} C.~M.,  {Warren} S.~J.,   {Mortlock} D.~J.,  2021, \mn@doi [\mnras]
  {10.1093/mnras/staa3747}, \href
  {https://ui.adsabs.harvard.edu/abs/2021MNRAS.501.3687O} {501, 3687}

\bibitem[\protect\citeauthoryear{{Okumura}, {Hand}, {Seljak}, {Vlah}  \&
  {Desjacques}}{{Okumura} et~al.}{2015}]{Okumura_2015}
{Okumura} T.,  {Hand} N.,  {Seljak} U.,  {Vlah} Z.,   {Desjacques} V.,  2015,
  \mn@doi [\prd] {10.1103/PhysRevD.92.103516}, \href
  {https://ui.adsabs.harvard.edu/abs/2015PhRvD..92j3516O} {92, 103516}

\bibitem[\protect\citeauthoryear{{Oldham} et~al.,}{{Oldham}
  et~al.}{2017}]{oldham_2017}
{Oldham} L.,  et~al., 2017, \mn@doi [\mnras] {10.1093/mnras/stw2832}, \href
  {https://ui.adsabs.harvard.edu/abs/2017MNRAS.465.3185O} {465, 3185}

\bibitem[\protect\citeauthoryear{Padfield}{Padfield}{2010}]{Padfield2010}
Padfield D.,  2010, in 2010 IEEE Computer Society Conference on Computer Vision
  and Pattern Recognition. pp 2918--2925, \mn@doi{10.1109/CVPR.2010.5540032}

\bibitem[\protect\citeauthoryear{Padfield}{Padfield}{2012}]{Padfield2012}
Padfield D.,  2012, \mn@doi [IEEE Trans. Image Process.]
  {10.1109/TIP.2011.2181402}, 21, 2706

\bibitem[\protect\citeauthoryear{{Pawase}, {Courbin}, {Faure}, {Kokotanekova}
  \& {Meylan}}{{Pawase} et~al.}{2014}]{Pawase_2014}
{Pawase} R.~S.,  {Courbin} F.,  {Faure} C.,  {Kokotanekova} R.,   {Meylan} G.,
  2014, \mn@doi [\mnras] {10.1093/mnras/stu179}, \href
  {https://ui.adsabs.harvard.edu/abs/2014MNRAS.439.3392P} {439, 3392}

\bibitem[\protect\citeauthoryear{{Peng}, {Ho}, {Impey}  \& {Rix}}{{Peng}
  et~al.}{2010}]{Peng_2010}
{Peng} C.~Y.,  {Ho} L.~C.,  {Impey} C.~D.,   {Rix} H.-W.,  2010, \mn@doi [\aj]
  {10.1088/0004-6256/139/6/2097}, \href
  {https://ui.adsabs.harvard.edu/abs/2010AJ....139.2097P} {139, 2097}

\bibitem[\protect\citeauthoryear{{Planck Collaboration} et~al.,}{{Planck
  Collaboration} et~al.}{2020}]{planck_2018}
{Planck Collaboration} et~al., 2020, \mn@doi [\aap]
  {10.1051/0004-6361/201833910}, \href
  {https://ui.adsabs.harvard.edu/abs/2020A&A...641A...6P} {641, A6}

\bibitem[\protect\citeauthoryear{{Ramani}, {Trickle}  \& {Zurek}}{{Ramani}
  et~al.}{2020}]{Ramani_2020}
{Ramani} H.,  {Trickle} T.,   {Zurek} K.~M.,  2020, \mn@doi [J. Cosmol.
  Astropart. Phys.] {10.1088/1475-7516/2020/12/033}, \href
  {https://ui.adsabs.harvard.edu/abs/2020JCAP...12..033R} {2020, 033}

\bibitem[\protect\citeauthoryear{Refregier}{Refregier}{2003}]{Refregier:2001fd}
Refregier A.,  2003, \mn@doi [\mnras] {10.1046/j.1365-8711.2003.05901.x}, 338,
  35

\bibitem[\protect\citeauthoryear{{Refregier} \& {Bacon}}{{Refregier} \&
  {Bacon}}{2003}]{Refregier_shp_ii}
{Refregier} A.,  {Bacon} D.,  2003, \mn@doi [\mnras]
  {10.1046/j.1365-8711.2003.05902.x}, \href
  {https://ui.adsabs.harvard.edu/abs/2003MNRAS.338...48R} {338, 48}

\bibitem[\protect\citeauthoryear{{Rexroth}, {Natarajan}  \& {Kneib}}{{Rexroth}
  et~al.}{2016}]{Rexroth_2016}
{Rexroth} M.,  {Natarajan} P.,   {Kneib} J.-P.,  2016, \mn@doi [\mnras]
  {10.1093/mnras/stw1017}, \href
  {https://ui.adsabs.harvard.edu/abs/2016MNRAS.460.2505R} {460, 2505}

\bibitem[\protect\citeauthoryear{Ritondale, Vegetti, Despali, Auger, Koopmans
  \& McKean}{Ritondale et~al.}{2019}]{Ritondale:2018cvp}
Ritondale E.,  Vegetti S.,  Despali G.,  Auger M.~W.,  Koopmans L. V.~E.,
  McKean J.~P.,  2019, \mn@doi [\mnras] {10.1093/mnras/stz464}, 485, 2179

\bibitem[\protect\citeauthoryear{{Salucci}, {Wilkinson}, {Walker}, {Gilmore},
  {Grebel}, {Koch}, {Frigerio Martins}  \& {Wyse}}{{Salucci}
  et~al.}{2012}]{Salucci_2012}
{Salucci} P.,  {Wilkinson} M.~I.,  {Walker} M.~G.,  {Gilmore} G.~F.,  {Grebel}
  E.~K.,  {Koch} A.,  {Frigerio Martins} C.,   {Wyse} R. F.~G.,  2012, \mn@doi
  [\mnras] {10.1111/j.1365-2966.2011.20144.x}, \href
  {https://ui.adsabs.harvard.edu/abs/2012MNRAS.420.2034S} {420, 2034}

\bibitem[\protect\citeauthoryear{{Samushia} et~al.,}{{Samushia}
  et~al.}{2014}]{Samushia_2014}
{Samushia} L.,  et~al., 2014, \mn@doi [\mnras] {10.1093/mnras/stu197}, \href
  {https://ui.adsabs.harvard.edu/abs/2014MNRAS.439.3504S} {439, 3504}

\bibitem[\protect\citeauthoryear{Schneider}{Schneider}{2019}]{Schneider_2014}
Schneider P.,  2019, \mn@doi [\aap] {10.1051/0004-6361/201424881}, 624, A54

\bibitem[\protect\citeauthoryear{{Schneider} \& {Seitz}}{{Schneider} \&
  {Seitz}}{1995}]{Schneider_1995}
{Schneider} P.,  {Seitz} C.,  1995, \aap, \href
  {https://ui.adsabs.harvard.edu/abs/1995A&A...294..411S} {294, 411}

\bibitem[\protect\citeauthoryear{{Schneider} \& {Sluse}}{{Schneider} \&
  {Sluse}}{2013}]{Schneider_2013}
{Schneider} P.,  {Sluse} D.,  2013, \mn@doi [\aap]
  {10.1051/0004-6361/201321882}, \href
  {https://ui.adsabs.harvard.edu/abs/2013A&A...559A..37S} {559, A37}

\bibitem[\protect\citeauthoryear{{Schneider} \& {Sluse}}{{Schneider} \&
  {Sluse}}{2014}]{Schneider:2014hsi}
{Schneider} P.,  {Sluse} D.,  2014, \mn@doi [\aap]
  {10.1051/0004-6361/201322106}, \href
  {https://ui.adsabs.harvard.edu/abs/2014A&A...564A.103S} {564, A103}

\bibitem[\protect\citeauthoryear{{Sheth} \& {Tormen}}{{Sheth} \&
  {Tormen}}{1999}]{Sheth_torman_1999}
{Sheth} R.~K.,  {Tormen} G.,  1999, \mn@doi [\mnras]
  {10.1046/j.1365-8711.1999.02692.x}, \href
  {https://ui.adsabs.harvard.edu/abs/1999MNRAS.308..119S} {308, 119}

\bibitem[\protect\citeauthoryear{Sheth, Mo  \& Tormen}{Sheth
  et~al.}{2001}]{Sheth_2001}
Sheth R.~K.,  Mo H.~J.,   Tormen G.,  2001, \mn@doi [\mnras]
  {10.1046/j.1365-8711.2001.04006.x}, 323, 1

\bibitem[\protect\citeauthoryear{{Shu} et~al.,}{{Shu} et~al.}{2016}]{Shu_2016}
{Shu} Y.,  et~al., 2016, \mn@doi [\apj] {10.3847/1538-4357/833/2/264}, \href
  {https://ui.adsabs.harvard.edu/abs/2016ApJ...833..264S} {833, 264}

\bibitem[\protect\citeauthoryear{{Siegel}, {Hertzberg}  \& {Fry}}{{Siegel}
  et~al.}{2007}]{Siegel_2007}
{Siegel} E.~R.,  {Hertzberg} M.~P.,   {Fry} J.~N.,  2007, \mn@doi [\mnras]
  {10.1111/j.1365-2966.2007.12435.x}, \href
  {https://ui.adsabs.harvard.edu/abs/2007MNRAS.382..879S} {382, 879}

\bibitem[\protect\citeauthoryear{{Springel} et~al.,}{{Springel}
  et~al.}{2008}]{Springel_2008}
{Springel} V.,  et~al., 2008, \mn@doi [\mnras]
  {10.1111/j.1365-2966.2008.14066.x}, \href
  {https://ui.adsabs.harvard.edu/abs/2008MNRAS.391.1685S} {391, 1685}

\bibitem[\protect\citeauthoryear{{Tessore} \& {Metcalf}}{{Tessore} \&
  {Metcalf}}{2015}]{Tessore_2015}
{Tessore} N.,  {Metcalf} R.~B.,  2015, \mn@doi [\aap]
  {10.1051/0004-6361/201526773}, \href
  {https://ui.adsabs.harvard.edu/abs/2015A&A...580A..79T} {580, A79}

\bibitem[\protect\citeauthoryear{{Teyssier}, {Pontzen}, {Dubois}  \&
  {Read}}{{Teyssier} et~al.}{2013}]{Teyssier_2013}
{Teyssier} R.,  {Pontzen} A.,  {Dubois} Y.,   {Read} J.~I.,  2013, \mn@doi
  [\mnras] {10.1093/mnras/sts563}, \href
  {https://ui.adsabs.harvard.edu/abs/2013MNRAS.429.3068T} {429, 3068}

\bibitem[\protect\citeauthoryear{{Unruh}, {Schneider}  \& {Sluse}}{{Unruh}
  et~al.}{2017}]{Unruh:2017yhi}
{Unruh} S.,  {Schneider} P.,   {Sluse} D.,  2017, \mn@doi [\aap]
  {10.1051/0004-6361/201629048}, \href
  {https://ui.adsabs.harvard.edu/abs/2017A&A...601A..77U} {601, A77}

\bibitem[\protect\citeauthoryear{{Vale} \& {Ostriker}}{{Vale} \&
  {Ostriker}}{2006}]{Vale_2006}
{Vale} A.,  {Ostriker} J.~P.,  2006, \mn@doi [\mnras]
  {10.1111/j.1365-2966.2006.10605.x}, \href
  {https://ui.adsabs.harvard.edu/abs/2006MNRAS.371.1173V} {371, 1173}

\bibitem[\protect\citeauthoryear{{Vegetti} \& {Koopmans}}{{Vegetti} \&
  {Koopmans}}{2009}]{vegetti2009a}
{Vegetti} S.,  {Koopmans} L.~V.~E.,  2009, \mn@doi [\mnras]
  {10.1111/j.1365-2966.2008.14005.x}, \href
  {http://adsabs.harvard.edu/abs/2009MNRAS.392..945V} {392, 945}

\bibitem[\protect\citeauthoryear{{Vegetti} \& {Vogelsberger}}{{Vegetti} \&
  {Vogelsberger}}{2014}]{vegetti_2014}
{Vegetti} S.,  {Vogelsberger} M.,  2014, \mn@doi [\mnras]
  {10.1093/mnras/stu1284}, \href
  {https://ui.adsabs.harvard.edu/abs/2014MNRAS.442.3598V} {442, 3598}

\bibitem[\protect\citeauthoryear{{Vegetti}, {Czoske}  \& {Koopmans}}{{Vegetti}
  et~al.}{2010a}]{vegetti2010a}
{Vegetti} S.,  {Czoske} O.,   {Koopmans} L.~V.~E.,  2010a, \mn@doi [\mnras]
  {10.1111/j.1365-2966.2010.16952.x}, \href
  {http://adsabs.harvard.edu/abs/2010MNRAS.407..225V} {407, 225}

\bibitem[\protect\citeauthoryear{{Vegetti}, {Koopmans}, {Bolton}, {Treu}  \&
  {Gavazzi}}{{Vegetti} et~al.}{2010b}]{vegetti2010b}
{Vegetti} S.,  {Koopmans} L.~V.~E.,  {Bolton} A.,  {Treu} T.,   {Gavazzi} R.,
  2010b, \mn@doi [\mnras] {10.1111/j.1365-2966.2010.16865.x}, \href
  {http://adsabs.harvard.edu/abs/2010MNRAS.408.1969V} {408, 1969}

\bibitem[\protect\citeauthoryear{{Vegetti}, {Lagattuta}, {McKean}, {Auger},
  {Fassnacht}  \& {Koopmans}}{{Vegetti} et~al.}{2012}]{vegetti2012}
{Vegetti} S.,  {Lagattuta} D.~J.,  {McKean} J.~P.,  {Auger} M.~W.,  {Fassnacht}
  C.~D.,   {Koopmans} L.~V.~E.,  2012, \mn@doi [\nat] {10.1038/nature10669},
  \href {http://adsabs.harvard.edu/abs/2012Natur.481..341V} {481, 341}

\bibitem[\protect\citeauthoryear{{Vegetti}, {Koopmans}, {Auger}, {Treu}  \&
  {Bolton}}{{Vegetti} et~al.}{2014}]{vegetti2014}
{Vegetti} S.,  {Koopmans} L.~V.~E.,  {Auger} M.~W.,  {Treu} T.,   {Bolton}
  A.~S.,  2014, \mn@doi [\mnras] {10.1093/mnras/stu943}, \href
  {http://adsabs.harvard.edu/abs/2014MNRAS.442.2017V} {442, 2017}

\bibitem[\protect\citeauthoryear{{Vegetti}, {Despali}, {Lovell}  \&
  {Enzi}}{{Vegetti} et~al.}{2018}]{Vegetti_2018}
{Vegetti} S.,  {Despali} G.,  {Lovell} M.~R.,   {Enzi} W.,  2018, \mn@doi
  [\mnras] {10.1093/mnras/sty2393}, \href
  {https://ui.adsabs.harvard.edu/abs/2018MNRAS.481.3661V} {481, 3661}

\bibitem[\protect\citeauthoryear{{Vernardos} \& {Koopmans}}{{Vernardos} \&
  {Koopmans}}{2022}]{Vernardos_2022}
{Vernardos} G.,  {Koopmans} L.~V.~E.,  2022, \mn@doi [\mnras]
  {10.1093/mnras/stac1924}, \href
  {https://ui.adsabs.harvard.edu/abs/2022MNRAS.516.1347V} {516, 1347}

\bibitem[\protect\citeauthoryear{{Vernardos}, {Tsagkatakis}  \&
  {Pantazis}}{{Vernardos} et~al.}{2020}]{Vernardos_2020}
{Vernardos} G.,  {Tsagkatakis} G.,   {Pantazis} Y.,  2020, \mn@doi [\mnras]
  {10.1093/mnras/staa3201}, \href
  {https://ui.adsabs.harvard.edu/abs/2020MNRAS.499.5641V} {499, 5641}

\bibitem[\protect\citeauthoryear{Virtanen et~al.,}{Virtanen
  et~al.}{2020}]{SciPy}
Virtanen P.,  et~al., 2020, \mn@doi [Nature Methods]
  {10.1038/s41592-019-0686-2}, \href {https://rdcu.be/b08Wh} {17, 261}

\bibitem[\protect\citeauthoryear{{Walker} \& {Pe{\~n}arrubia}}{{Walker} \&
  {Pe{\~n}arrubia}}{2011}]{Walker_2011}
{Walker} M.~G.,  {Pe{\~n}arrubia} J.,  2011, \mn@doi [\apj]
  {10.1088/0004-637X/742/1/20}, \href
  {https://ui.adsabs.harvard.edu/abs/2011ApJ...742...20W} {742, 20}

\bibitem[\protect\citeauthoryear{{Wertz}, {Orthen, Bastian}  \& {Schneider,
  Peter}}{{Wertz} et~al.}{2018}]{Wertz_2018}
{Wertz} O.,  {Orthen, Bastian}  {Schneider, Peter} 2018, \mn@doi [A\&A]
  {10.1051/0004-6361/201732240}, 617, A140

\bibitem[\protect\citeauthoryear{Xu, Mao, Cooper, Gao, Frenk, Angulo  \&
  Helly}{Xu et~al.}{2012}]{Xu:2011ru}
Xu D.~D.,  Mao S.,  Cooper A.~P.,  Gao L.,  Frenk C.~S.,  Angulo R.~E.,   Helly
  J.,  2012, \mn@doi [\mnras] {10.1111/j.1365-2966.2012.20484.x}, 421, 2553

\bibitem[\protect\citeauthoryear{{Zheng}}{{Zheng}}{2004}]{Zheng_2004}
{Zheng} Z.,  2004, \mn@doi [\apj] {10.1086/421542}, \href
  {https://ui.adsabs.harvard.edu/abs/2004ApJ...610...61Z} {610, 61}

\bibitem[\protect\citeauthoryear{\c{S}eng\"ul, Tsang, Diaz~Rivero, Dvorkin, Zhu
   \& Seljak}{\c{S}eng\"ul et~al.}{2020}]{CaganSengul:2020nat}
\c{S}eng\"ul A.~{\c{C}}.,  Tsang A.,  Diaz~Rivero A.,  Dvorkin C.,  Zhu H.-M.,
   Seljak U.,  2020, \mn@doi [Phys. Rev. D] {10.1103/PhysRevD.102.063502}, 102,
  063502

\bibitem[\protect\citeauthoryear{{{\c{S}}eng{\"u}l}, {Dvorkin}, {Ostdiek}  \&
  {Tsang}}{{{\c{S}}eng{\"u}l} et~al.}{2022}]{Sengul_2021}
{{\c{S}}eng{\"u}l} A.~{\c{C}}.,  {Dvorkin} C.,  {Ostdiek} B.,   {Tsang} A.,
  2022, \mn@doi [\mnras] {10.1093/mnras/stac1967}, \href
  {https://ui.adsabs.harvard.edu/abs/2022MNRAS.515.4391S} {515, 4391}

\makeatother
\end{thebibliography}




\appendix 

\section{The Power Spectrum Multipoles}\label{App.A}

The goal of this appendix is to derive an expression that relates the power spectrum
multipoles to the correlation function multipoles.

 In general, the power spectrum can be written as the Fourier transform of the correlation function
\be \label{Eq.29}
P(k, \mu_\kk)=\int {\rm d}^2\rr e^{-i\kk \cdot \rr} \xi(r, \mu_\rr),
\ee where $\mu_\rr = \hat{\rr}\cdot\hat{\rr}_{\rm h}$ and $\mu_\kk = \hat{\kk}\cdot\hat{\rr}_{\rm h}$. Now we substitute for $e^{-i\kk \cdot \rr}$ using the Jacobi--Anger expansion \citep{abramowitz+stegun, Colton98inverseacoustic}

\begin{align} \label{Eq.30}
 e^{-i \kk \cdot \rr} & = e^{-ikr\cos(\theta_\rr -\theta_\kk)} \en
 & =  \sum_{n=0}^{\infty} (2-\delta_{n0}) (-i)^n J_n(kr)T_n(\cos(\theta_\rr -\theta_\kk)) \en
 & =  \sum_{n=0}^{\infty} (2-\delta_{n0})(-i)^n J_n(kr)T_n(\mu_{\kk\rr}) ,
\end{align} where $T_n(x)$ denotes the Chebyshev polynomial of order $n$, $J_n(x)$ the Bessel functions of the first kind and $\mu_{\kk\rr}=\hat{\kk}\cdot\hat{\rr}$. We decompose the two-point correlation function as

\be \label{Eq.31}
\xi(r, \mu_\rr)=\sum_{\ell=0}^\infty \xi_\ell(r)T_\ell(\mu_\rr),
\ee where the correlation multipoles are given by

\be \label{Eq.32}
\xi_\ell(r) = \frac{2-\de_{\ell 0}}{\pi}\int_{-1}^1 {\rm d} \mu_\rr \,\, \frac{\xi(r, \mu_\rr)T_\ell(\mu_\rr)}{\sqrt{1-\mu_\rr^2}}\,\,.
\ee Combining equations~\eqref{Eq.29},~\eqref{Eq.30}, and \eqref{Eq.31} gives

\begin{align}
    P(k, \mu_\kk) 
    & = \sum_{\ell=0}^\infty \sum_{n=0}^{\infty} (2-\delta_{n0})  (-i)^n\int {\rm d}r \,\,r  J_n(kr)\xi_\ell(r) \en
    & \hspace{3.5cm} \times \int_0 ^{2\pi} {\rm d} \theta_\rr T_n(\mu_{\kk\rr}) T_\ell(\mu_\rr)\en
    & = \sum_{\ell=0}^\infty P_\ell(k)T_\ell(\mu_\kk),
\end{align} where

\begin{align} \label{Eq.34}
    P_\ell(\kk)T_\ell(\mu_\kk) & = \sum_{n=0}^{\infty}(2-\delta_{n0})  (-i)^n\int {\rm d}r \,\,r  J_n(kr)\xi_\ell(r)\en
    & \hspace{2cm} \times \int_0 ^{2\pi} {\rm d} \theta_\rr T_n(\hat{\kk}\cdot\hat{\rr}) T_\ell(\hat{\rr}\cdot\hat{\rr}_{\rm h}) \,\,.
\end{align}  Now consider the following integration:

\begin{align} \label{Eq.35}
    & \int_0 ^{2\pi} {\rm d} \theta_\rr T_n(\hat{\kk}\cdot\hat{\rr}) T_\ell(\hat{\rr}\cdot\hat{\rr}_{\rm h}) \en
    & \hspace{15mm} = \int_0 ^{2\pi} {\rm d} \theta_\rr \,\,  T_n[\cos(\theta_\rr -\theta_\kk)] \,\, T_\ell[\cos(\theta_\rr -\theta_{\rr_{\rm h}})] \en
    & \hspace{15mm} = \int_0 ^{2\pi} {\rm d} \theta_\rr \,\, \cos[n(\theta_\rr -\theta_\kk)] \,\,  \cos[\ell(\theta_\rr -\theta_{\rr_{\rm h}})]\en
    & \hspace{15mm} = \begin{cases} \vspace{5pt}
    2\pi T_0(\hat{\kk}\cdot\hat{\rr}_{\rm h}) = 2\pi T_0(\mu_\kk)     & \quad \text{if } n=\ell=0\\ \vspace{5pt}
    \pi T_\ell(\hat{\kk}\cdot\hat{\rr}_{\rm h}) = \pi T_\ell(\mu_\kk)       & \quad \text{if } n=\ell \neq 0\\ 
    0     & \quad \text{if } n \neq \ell\\ 
\end{cases} \en
    & \hspace{15mm} =  \frac{2\pi}{2-\delta_{\ell0} } T_\ell(\mu_\kk) \de_{n\ell} 
\end{align}  Then equation~\eqref{Eq.35} reduces equation~\eqref{Eq.34} to

\be \label{Eq.36}
 P_\ell(k)T_\ell(\mu_\kk)  = \left[2\pi (-i)^\ell \int {\rm d}r \,\,r  J_\ell(kr)\xi_\ell(r)\right]  T_\ell(\mu_\kk)  \,\,.
\ee Therefore, the power spectrum multipoles are given by 

\be
P_\ell(k) = 2\pi(-i)^\ell \int {\rm d}r \,\, r \, J_\ell(kr) \xi_\ell(r) \,\,.
\ee

\section{Masked Normalized Correlation Function}\label{App.B}

The unmasked normalized autocorrelation function can be simply written as

\be \label{Eq.mnc}
\tilde{\Xi}(\rr)=\frac{\int {\rm d}^2\rr_1  \big[\kappa(\rr_1) - \left<\kappa(\rr_1) \right> \big] \,\, \big[\kappa(\rr_2) - \left<\kappa(\rr_2) \right>\big]}{\sqrt{\int {\rm d}^2\rr_1  \big[\kappa(\rr_1) - \left<\kappa(\rr_1) \right> \big]^2\int {\rm d}^2\rr_1  \big[\kappa(\rr_2) - \left<\kappa(\rr_2) \right> \big]^2}},
\ee where $\rr_2 = \rr + \rr_1$ and $\left<\kappa(\rr) \right>$ is the mean of the convergence map. The two-point correlation function is normalized with the zero-shifted autocorrelation (similar to the zero-lag autocorrelation in signal processing) of the mean subtracted convergence fields.

We consider an annular region on the convergence map to make sure that there is no influence from the central part of the convergence map and that the correlation computation is carried out inside the region of interest. In the binary annular mask we used, the domain of interest is given a value of 1, and the regions to be masked out (regions outside the annulus) are 0. A detailed study of the computation of masked normalized cross-correlation was presented in \cite{Padfield2010, Padfield2012}. 

Following the method described in \cite{Padfield2010, Padfield2012}, we introduce the domain $D_\rr=D_{\rr_2}(\rr)\cap D_{\rr_1}$ to express the overlap region of the two convergence maps where neither of the masked convergence maps is 0, during the shifting of the moving convergence map, $\kappa(\rr_2)$, with respect to the fixed map, $\kappa(\rr_1)$, by a distance $\rr$, to consider only the masked regions. In the pre-described domain ($D_\rr$), the term $D_{\rr_1}$ represents the domain of $\kappa(\rr_1)$ in the regions that $W(\rr_1)\neq 0$, and $D_{\rr_2}(\rr)=\{\rr_1|(\rr_2 = \rr_1 + \rr) \in D_{\rr_2}\}$ is the domain $D_{\rr_2}$ of $\kappa(\rr_2=\rr_1+\rr)$ displaced by $\rr$ in the regions where $W(\rr_1+\rr) \neq 0$. This approach makes certain that the regions outside the annulus have no impact on the overlap regions. Contrary to this method, if the convergence maps are masked prior to correlate them using equation ~\eqref{Eq.mnc}, all pixels in the overlap areas, including zero regions in the mask, will be used in the computation and thus influence the computation of the correlation. Since we need to carry out the computation only using the valid regions on the convergence maps, the method described in \cite{Padfield2010, Padfield2012} is an optimal approach to remove the effect of the mask. 

Using the pointwise multiplications of masks, $W(\rr)$, and convergence maps such that $\tilde{\kappa}(\rr_1)= W(\rr_1) \cdot \kappa(\rr_1)$ and $\tilde{\kappa}(\rr+\rr_1)= W(\rr+\rr_1) \cdot \kappa(\rr+\rr_1)$, the masked normalized correlation function is presented in \cite{Padfield2010, Padfield2012} as

\be \label{Eq.B2}
\tilde{\xi}(\rr) = \frac{\rm Num}{\sqrt{{\rm Den}_1 \cdot {\rm Den}_2}},
\ee where

\begin{align}\label{Eq.Num}
   {\rm Num} & = \int {\rm d}^2 \rr_1 \,\,  \ktil(\rr_1) \, \ktil(\rr+\rr_1) \en
    & \hspace{1cm} - \frac{\int {\rm d}^2 \rr_1 \,\,  \ktil(\rr_1) \, W(\rr+\rr_1) \cdot \int {\rm d}^2 \rr_1 \,\,  W(\rr_1) \, \ktil(\rr+\rr_1)}{\int {\rm d}^2 \rr_1 \,\,  W(\rr_1) \, W(\rr+\rr_1)},
\end{align}

\begin{align}\label{Eq.Den_1}
   {\rm Den}_1 & = \int {\rm d}^2 \rr_1 \,\,  \ktil^2(\rr_1) \, W(\rr+\rr_1) \en
   & \hspace{3cm} -\frac{\left[\int {\rm d}^2 \rr_1 \,\,  \ktil(\rr_1) \, W(\rr+\rr_1) \right]^2}{\int {\rm d}^2 \rr_1 \,\,  W(\rr_1) \, W(\rr+\rr_1)},
\end{align} and

\begin{align}\label{Eq.Den_2}
   {\rm Den}_2 & = \int {\rm d}^2 \rr_1 \,\,  W(\rr_1) \, \ktil^2(\rr+\rr_1) \en
   & \hspace{3cm} -\frac{\left[\int {\rm d}^2 \rr_1 \,\,  W(\rr_1) \, \ktil(\rr+\rr_1) \right]^2}{\int {\rm d}^2 \rr_1 \,\,  W(\rr_1) \, W(\rr+\rr_1)}.
\end{align}

The first term of equation~\eqref{Eq.Num} is the autocorrelation of the two masked convergence maps, $\ktil^2(\rr_1)$ and $\ktil^2(\rr+\rr_1)$. The local summation presented by the integration, $\int {\rm d}^2 \rr_1 \,\,  W(\rr_1) \, W(\rr+\rr_1)$, counts the number of pixels in the overlap region for a particular choice of $\rr$. The integration, $\int {\rm d}^2 \rr_1 \,\,  \ktil(\rr_1) \, W(\rr+\rr_1)$, represents the sum of the convergence map $\kappa(\rr_1)$ values within the overlap region given as the pointwise multiplication of the fixed and moving masks $W(\rr_1)$ and $W(\rr_1+\rr)$. The term $\int {\rm d}^2 \rr_1 \,\,  W(\rr_1) \, \ktil(\rr+\rr_1)$ given in equations~\eqref{Eq.Num}--\eqref{Eq.Den_2} can be explained analogously. Similarly, $\int {\rm d}^2 \rr_1 \,\,  W(\rr_1) \, \ktil^2(\rr+\rr_1)$ and $\int {\rm d}^2 \rr_1 \,\,  \ktil^2(\rr_1) \, W(\rr+\rr_1)$ given in equations~\eqref{Eq.Den_1} and \eqref{Eq.Den_2}, calculate the local sum of the elementwise square of the masked convergence maps $ \ktil(\rr+\rr_1)$ and $ \ktil(\rr_1)$ inside the overlapping region, respectively. Finally, we refer the reader to \cite{Padfield2010} and \cite{Padfield2012} for an in-depth description of the procedure of computing the masked normalized cross-correlation.

\section{The effect of the main-lens model} \label{App.C}

To study the effects of external shear due to structure near the main-lens in a gravitational lens system, the logarithmic slope of the EPL main-lens mass profile, and the eccentricity components of the main-lens on the multipole moments of $\kappa_{\rm div}$ fields, we use 100 full CDM populations with the rescaling factor for the line of sight halo mass function of $\de_{\rm LOS}=1.0$, and the subhalo mass function normalization of $\Sigma_{\rm sub}=0.05 \,\,{\rm kpc}^{-2}$. The values of all other physical parameters, including the main deflector and source redshifts, are also similar to the parameter values given in Section~\ref{sec:sec2.1}, except for the quantity that we vary in a particular study.

\begin{figure}
\begin{center}
	\includegraphics[clip, trim=0.25cm 0cm 0cm 0.2cm, width=0.48\textwidth]{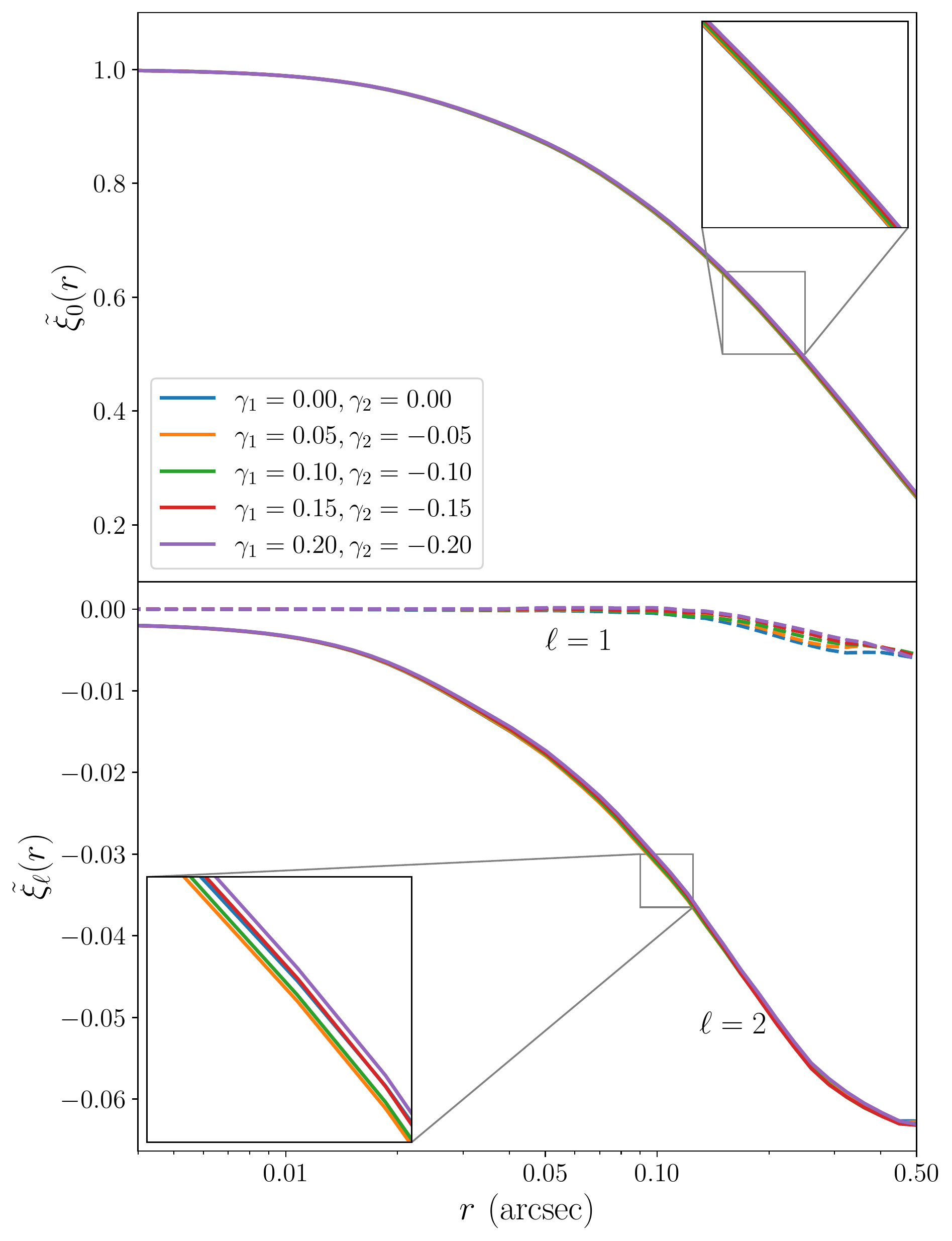}
	\includegraphics[clip, trim=0.22cm 0.2cm 0cm 0cm, width=0.48\textwidth]{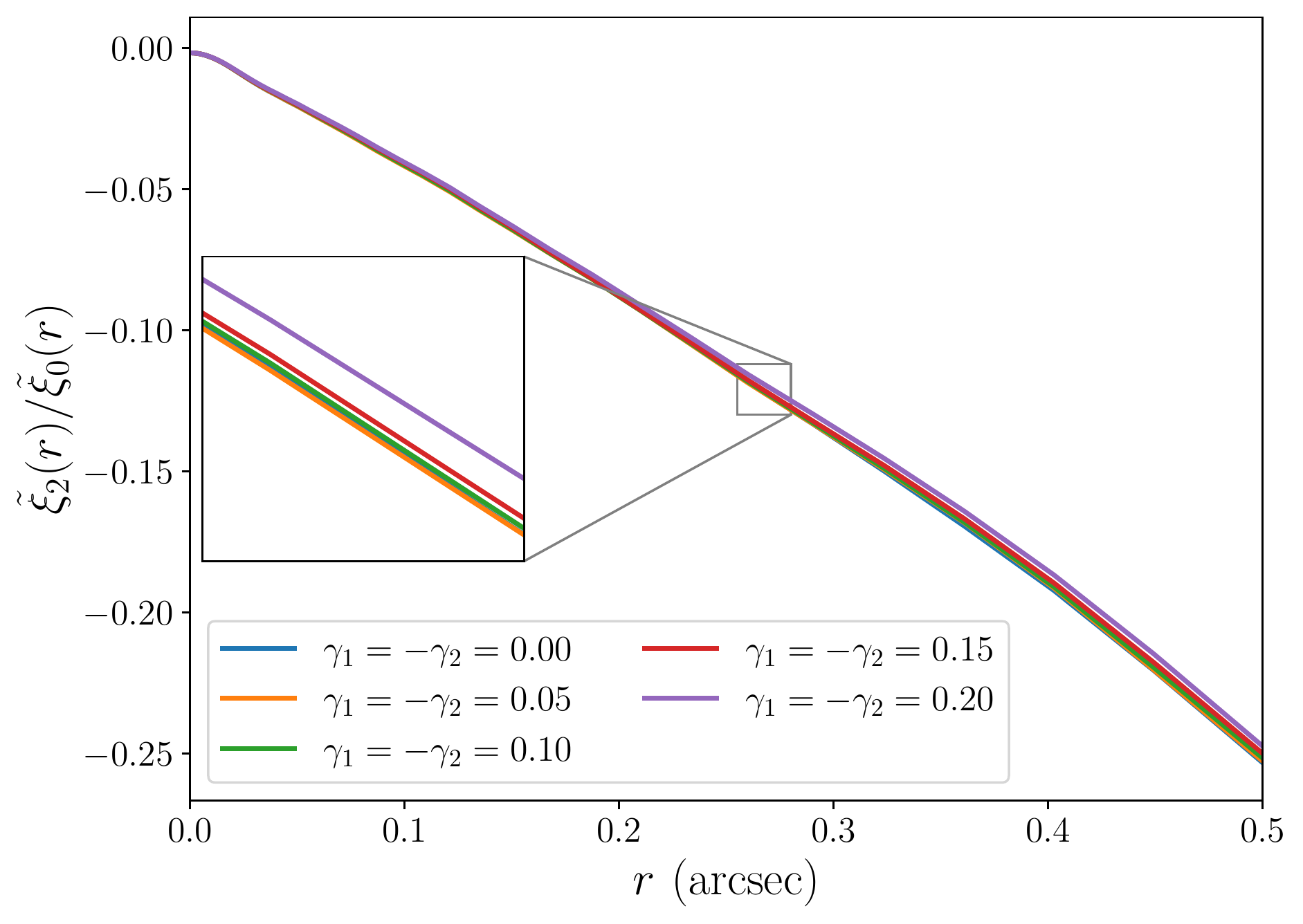}
\end{center}
\caption{\label{fig:corr_multi_external_shear_effect}The effect of external shear on multipole moments of the normalized two-point correlation function of the masked $\kappa_{\rm div}$ field of full CDM halo realizations with $\de_{\rm LOS}=1.0$ and $\Sigma_{\rm sub}= 0.05 \,\,{\rm kpc}^{-2}$ as a function of shear tensor considering a less eccentric macrolens with the eccentricity of $e=0.01$. The changes generated by the external shear field in the amplitudes of multipoles of the effective convergence field are too small to be noticeable. The bottom panel shows how the external shear affects the quadrupole to monopole ratio for the effective projected mass density.}
\end{figure}

\begin{figure}
\begin{center}
	\includegraphics[clip, trim=0.25cm 0cm 0cm 0.2cm, width=0.48\textwidth]{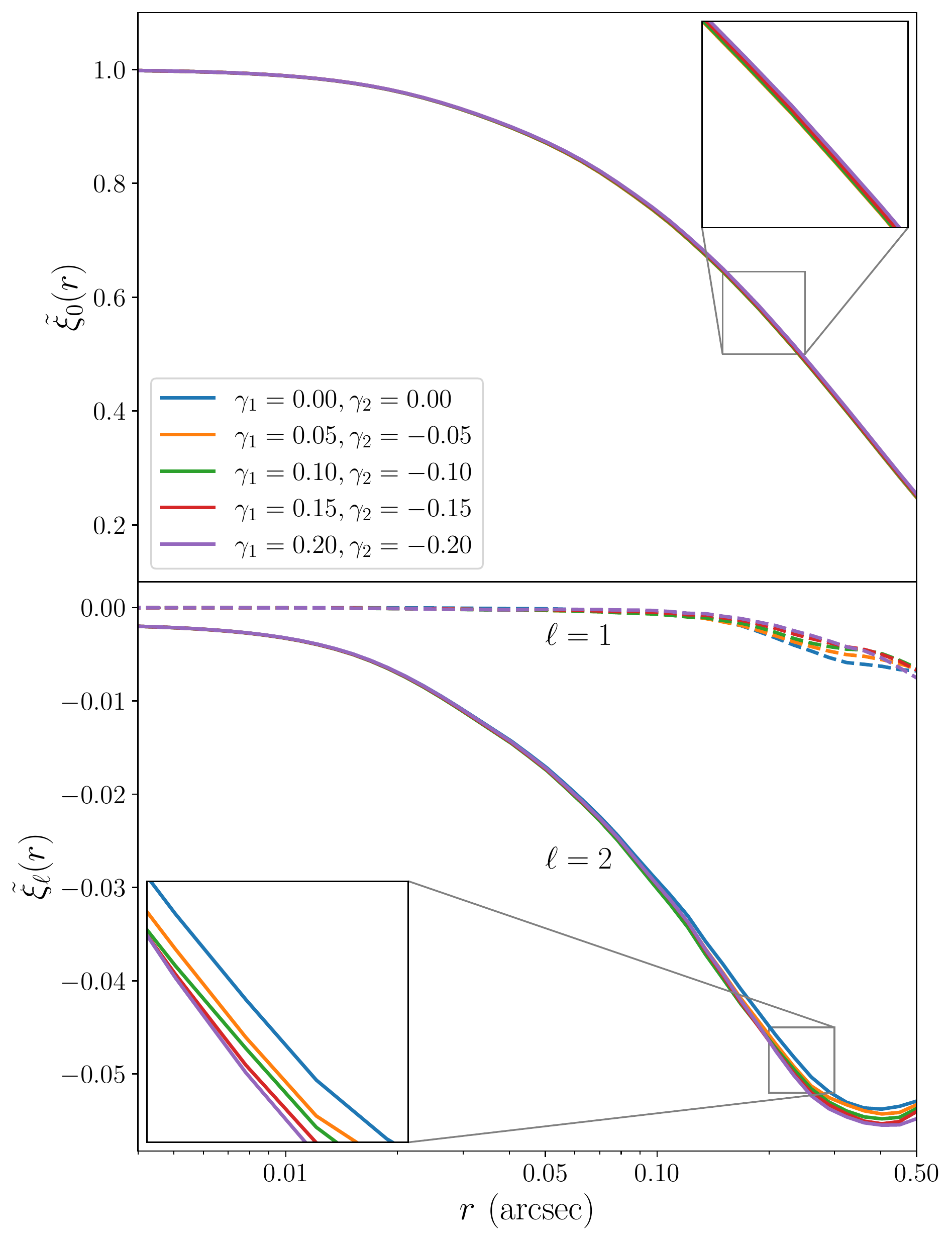}
	\includegraphics[clip, trim=0.22cm 0.2cm 0cm 0cm, width=0.48\textwidth]{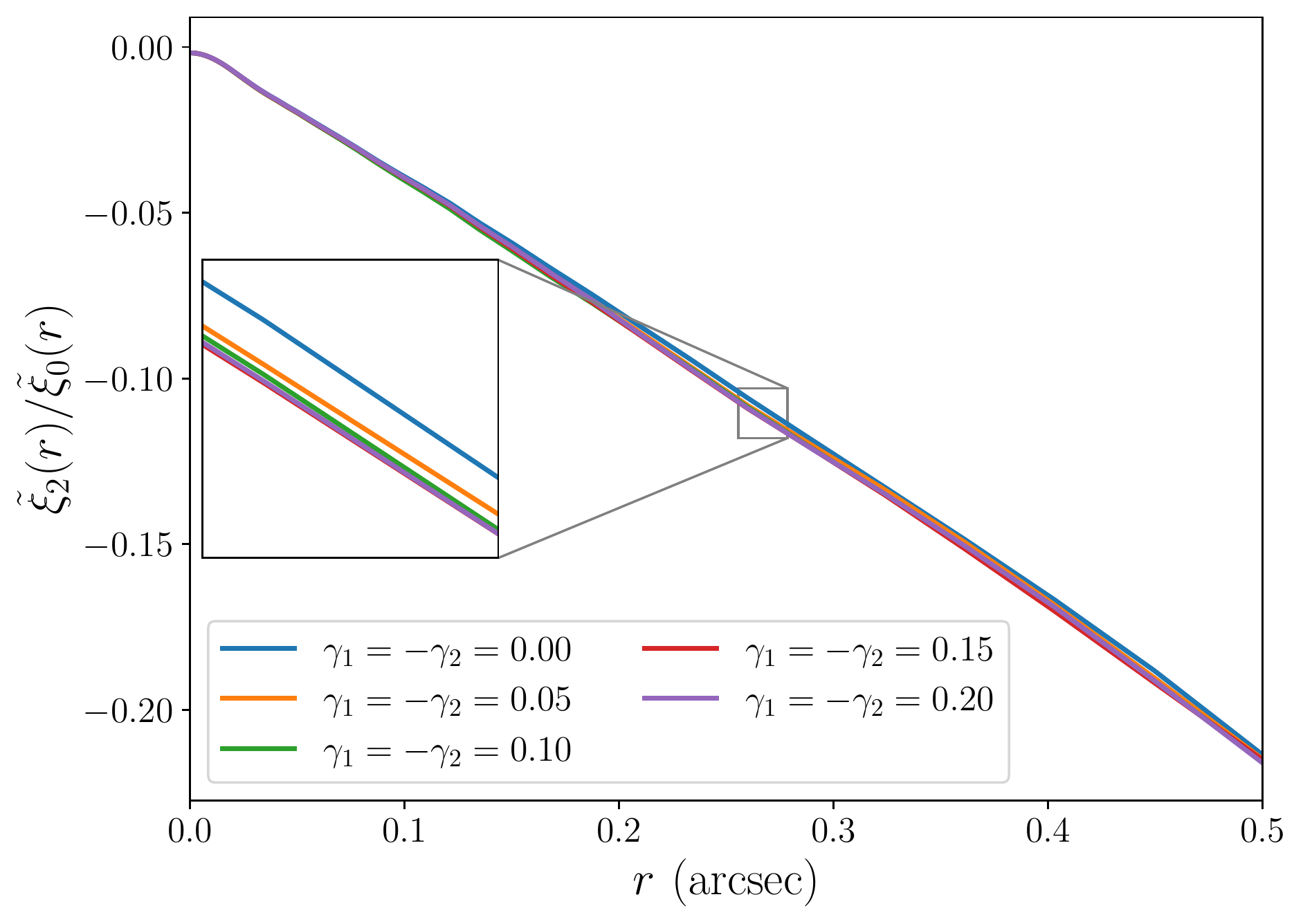}
\end{center}
\caption{\label{fig:corr_multi_external_shear_effect_high_ecc}Same as Fig. \ref{fig:corr_multi_external_shear_effect}, for strong lens systems with a very elliptical main-lens with the eccentricity of $e=0.3$.}
\end{figure}

To assess the effect of external shear, we keep the orientation of the shear constant and only change its magnitude, $\gamma = \sqrt{\gamma_1^2+\gamma_2^2}$. The top panel of Fig.~\ref{fig:corr_multi_external_shear_effect} shows how the amplitudes and shapes of multipole moments change as the shear field changes for strong lens systems with a less eccentric EPL macro lens with the eccentricity components of $(e_1, e_2)=(0.01, 0.0)$. Since there is a degeneracy between the eccentricity of the main-lens and its shear field, Fig.~\ref{fig:corr_multi_external_shear_effect_high_ecc} depicts the effect of external shear for a strong lens system with a very elliptical macro lens with eccentricity components of $(e_1, e_2) = (0.3, 0.0) $ and dark matter haloes. In both cases, the effect of external shear on the quadrupole is not significantly apparent. It hence does not directly affect the studies based on the shape of the quadrupole moment. 
The bottom panels of Fig.~\ref{fig:corr_multi_external_shear_effect} and Fig.~\ref{fig:corr_multi_external_shear_effect_high_ecc} show the shear field effect on the quadrupole to monopole ratio for the effective convergence field, $\kappa_{\rm div}$, where the mean of 100 full CDM halo realizations is taken into account. We discuss how this ratio can be used to quantify subhalo and line-of-sight halo abundances in Section~\ref{sec:sec4.2}, and it is clear that the shear effect lies within $1-\sigma$ of the mean and is not strong enough to cause a noticeable change in a ratio plot similar to Fig.~\ref{fig:abunadances} and make a change in our measurements within the feasible limits of the magnitude of the external shear.

\begin{figure}
\begin{center}
	\includegraphics[clip, trim=0.22cm 0cm 0cm 0.2cm, width=0.48\textwidth]{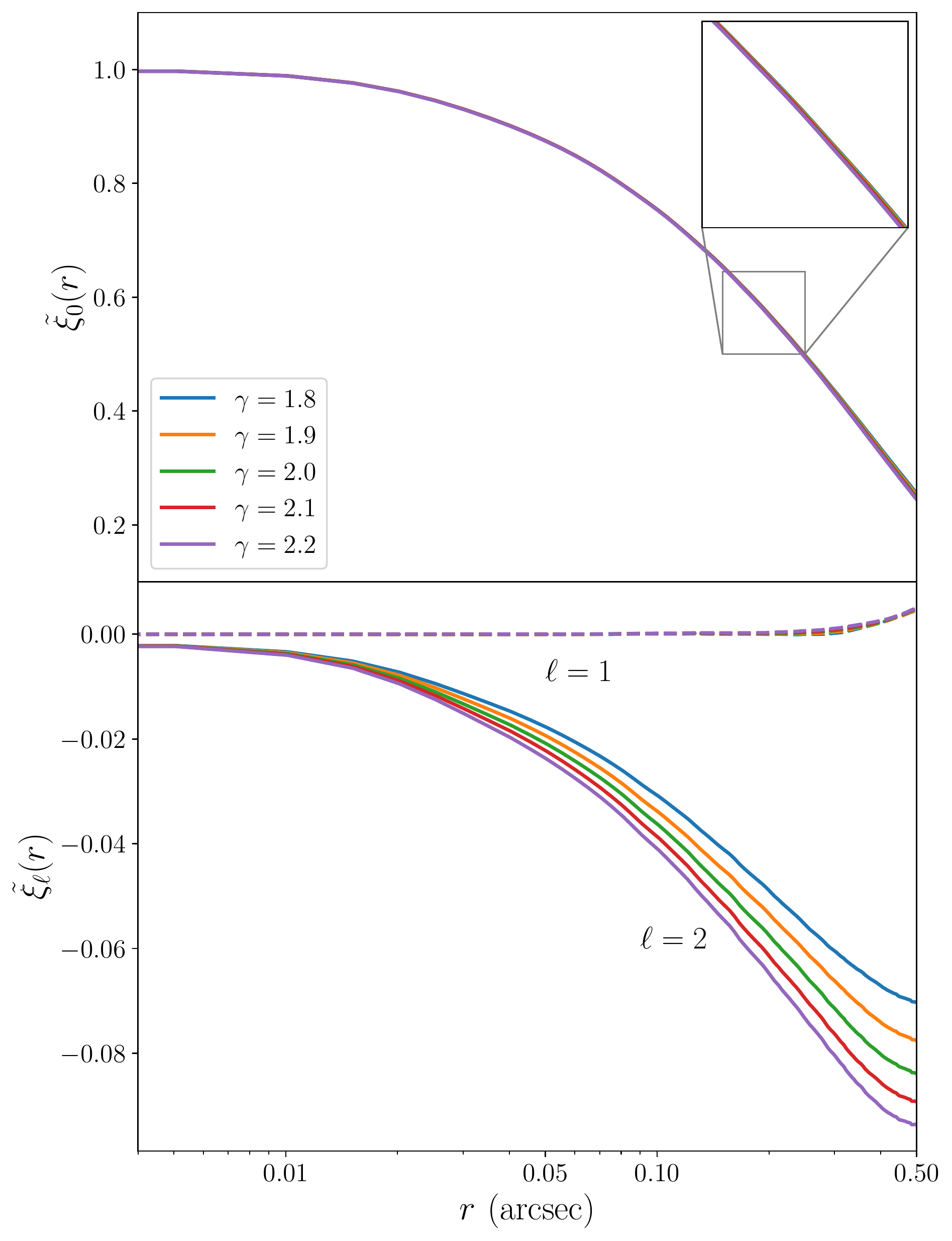}
	\includegraphics[clip, trim=0.22cm 0.2cm 0cm 0cm,
	width=0.48\textwidth]{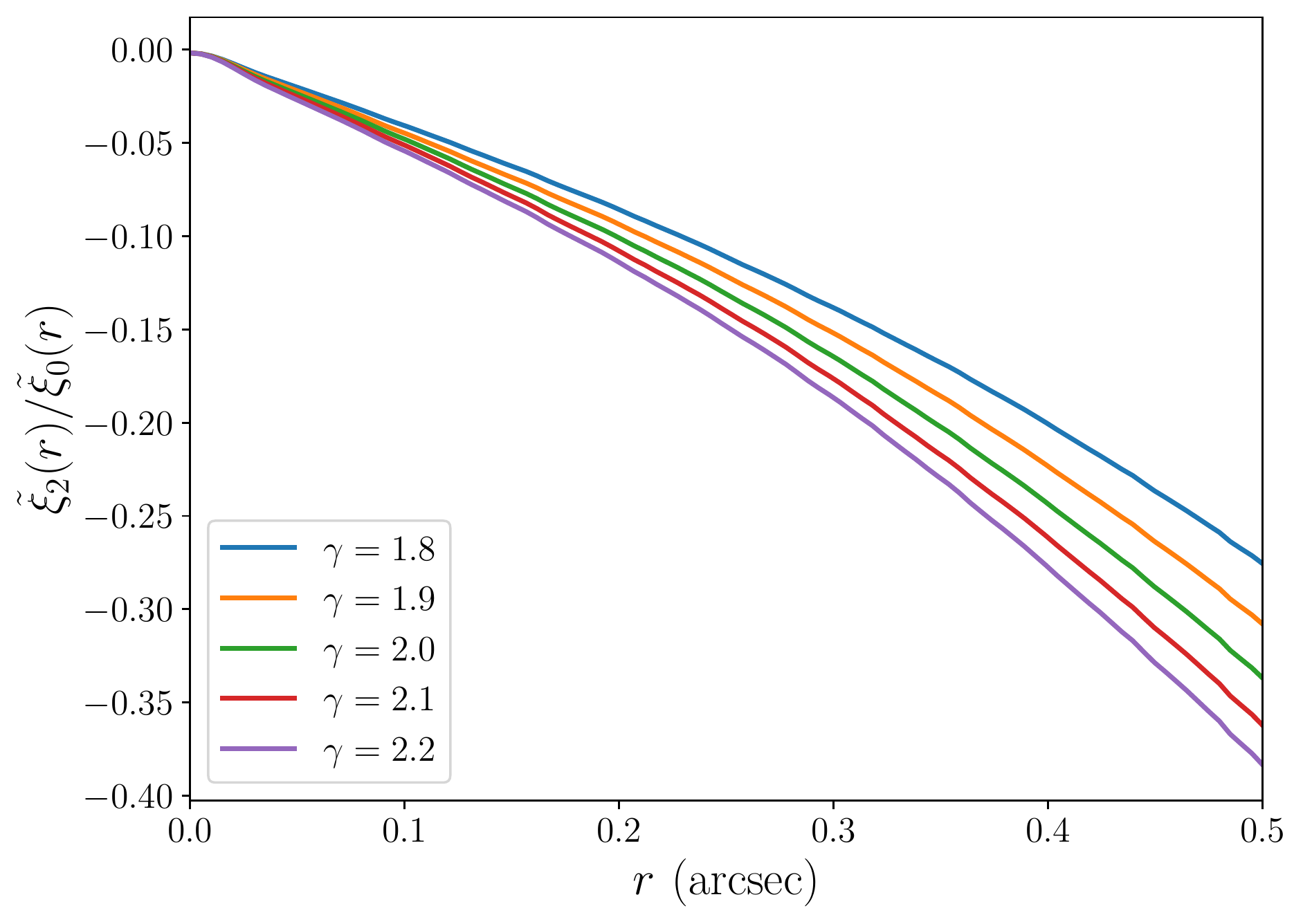}
\end{center}
\caption{\label{fig:corr_multi_gamma_effect}The effect of the logarithmic slope of the EPL main-lens mass profile on multipole moments of the normalized two-point correlation function of the masked $\kappa_{\rm div}$ field of full CDM halo realizations (top panel). Because the logarithmic slope of the mass profile directly affects the central density of the main-lens and thus changes the magnitude of the anisotropies of projected mass densities of line-of-sight haloes, changes in the amplitudes of quadrupoles in the effective convergence field are noticeable. The effect on the quadrupole to monopole ratios is shown in the bottom panel.}
\end{figure}

\begin{figure}
\begin{center}
	\includegraphics[clip, trim=0.22cm 0cm 0cm 0.2cm, width=0.48\textwidth]{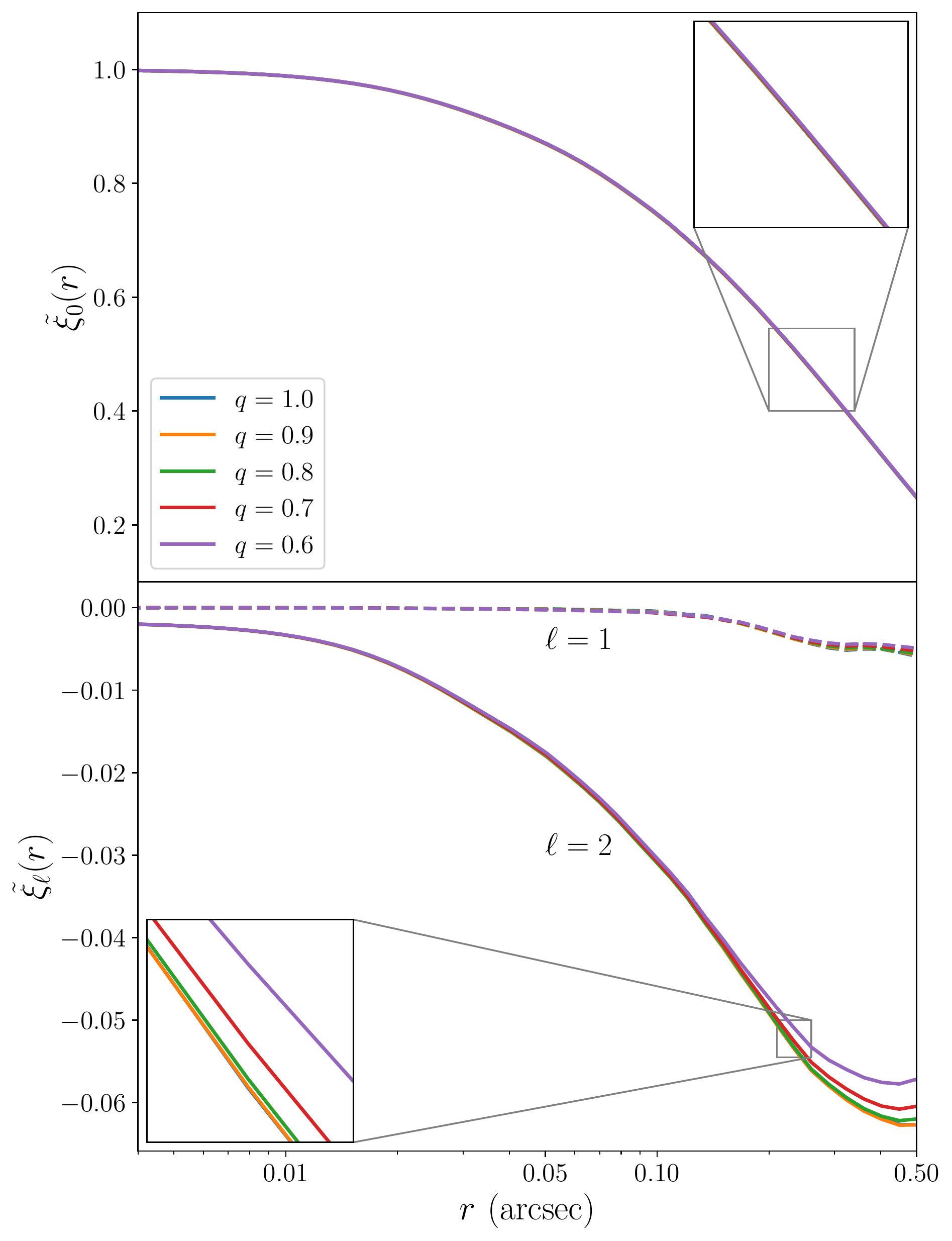}
	\includegraphics[clip, trim=0.22cm 0.2cm 0cm 0cm, width=0.48\textwidth]{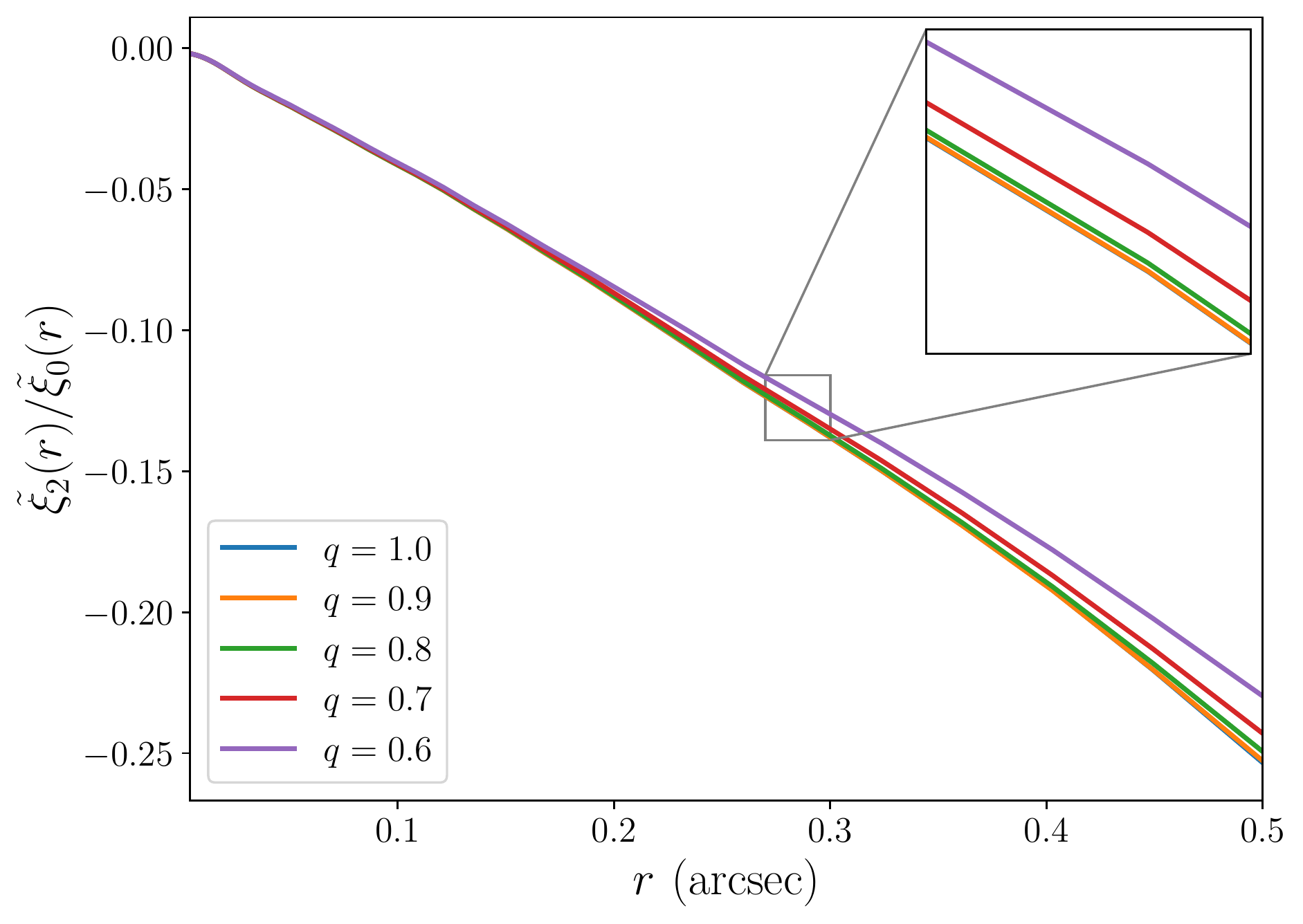}
\end{center}
\caption{\label{fig:corr_multi_ellipticity_effect}The top panel shows the effect of main-lens ellipticity on multipole moments of the normalized two-point correlation function of the masked $\kappa_{\rm div}$ field of full CDM halo population of subhaloes and line-of-sight haloes as a function of the minor to major axis ratio ($q$) of the macrolens mass profile. As shown in these plots, the eccentricities of the main-lens mass profile (here the EPL mass profile) have a very small effect on the quadrupole and quadrupole to monopole ratios.}
\end{figure}

Before we study how the logarithmic slope of the EPL mass profile and eccentricity components impact the multipole measurements, we first review the mathematical structure of the main-lens mass profile. The convergence of a main-lens with the forementioned EPL mass profile is given by \citep{Tessore_2015} 

\be 
\kappa_0(x,y)=\frac{3-\gamma}{2}\left(\frac{\theta_{\rm E}}{\sqrt{qx^2+\frac{y^2}{q}}} \right)^{\gamma-1},
\ee where $1 < \gamma < 3$ is the power-law slope of the mass distribution, $\theta_{\rm E}$ is the Einstein radius, $0 < q \leq 1$ is the ratio of minor and major axis. Here the Cartesian coordinates $x$ and $y$ are defined along the major and minor axis of the main-lens. For a given main deflector with the eccentricity components $e_1$ and $e_2$, the axis ratio can be written as

\be
q=\frac{1-e}{1+e},
\ee where $e=\sqrt{e_1^2+e_2^2}$ is the eccentricity of the mass profile.

The top panel of Fig.~\ref{fig:corr_multi_gamma_effect} shows how the
multipole moments of the normalized two-point correlation function of the masked $\kappa_{\rm div}$ field behave as a function of the power-law slope ($\gamma$) of EPL mass profile. As the logrithmic slope of the profile becomes steeper, the enclosed mass within a given radius and central density of the macrolens increase, whereas the dispersed mass spread and low central density values for smaller power-law slope values decrease \citep{Lyskova_2018}. Therefore, the lensing effect increases as the logarithmic slope of the main-lens increases.  Since the quadrupole moment reflects the coupling between different lens planes, the stretch of the line-of-sight halo effective convergence field along the tangential direction increases with the $\gamma$. Therefore, these increasing anisotropic signatures can directly be noticed as a gain in the amplitude of the quadrupole moment as anticipated. Since the multiplane coupling does not involve the monopole term, the monopole remains unchanged. The bottom panel of Fig.~\ref{fig:corr_multi_gamma_effect} illustrates the effect of $\gamma$ on the subhalo and line-of-sight halo abundance measurements. Considering the top and bottom panels we may conclude that the logarithmic slope of the mass profile is a crucial measurement to more precise dark halo physics predictions using our approach.

Fig.~\ref{fig:corr_multi_ellipticity_effect} illustrates the results of our study based on the effect of the main-lens mass profile ellipticity on the multipole moments. The increasing eccentricity of the main-lens decreases the axis ratio, $q$. However, using a constant logarithmic slope ($\gamma$) in this study keeps the central density of the macrolens and total enclosed mass within radius $r$ constant, allowing an ellipticity invariant lensing effect to be observed. As a result, the effect on the non-linear coupling term associated with the quadrupole moment remains constant, resulting in very small or negligible changes in the quadrupole amplitudes. Furthermore, the eccentricities of the main-lens mass profile have a very small effect (lies within $1-\sigma$) on the quadrupole to monopole ratios, as shown in the bottom panel.

\begin{figure}
\begin{center}
	\includegraphics[clip, trim=0.22cm 0cm 0cm 0.2cm, width=0.48\textwidth]{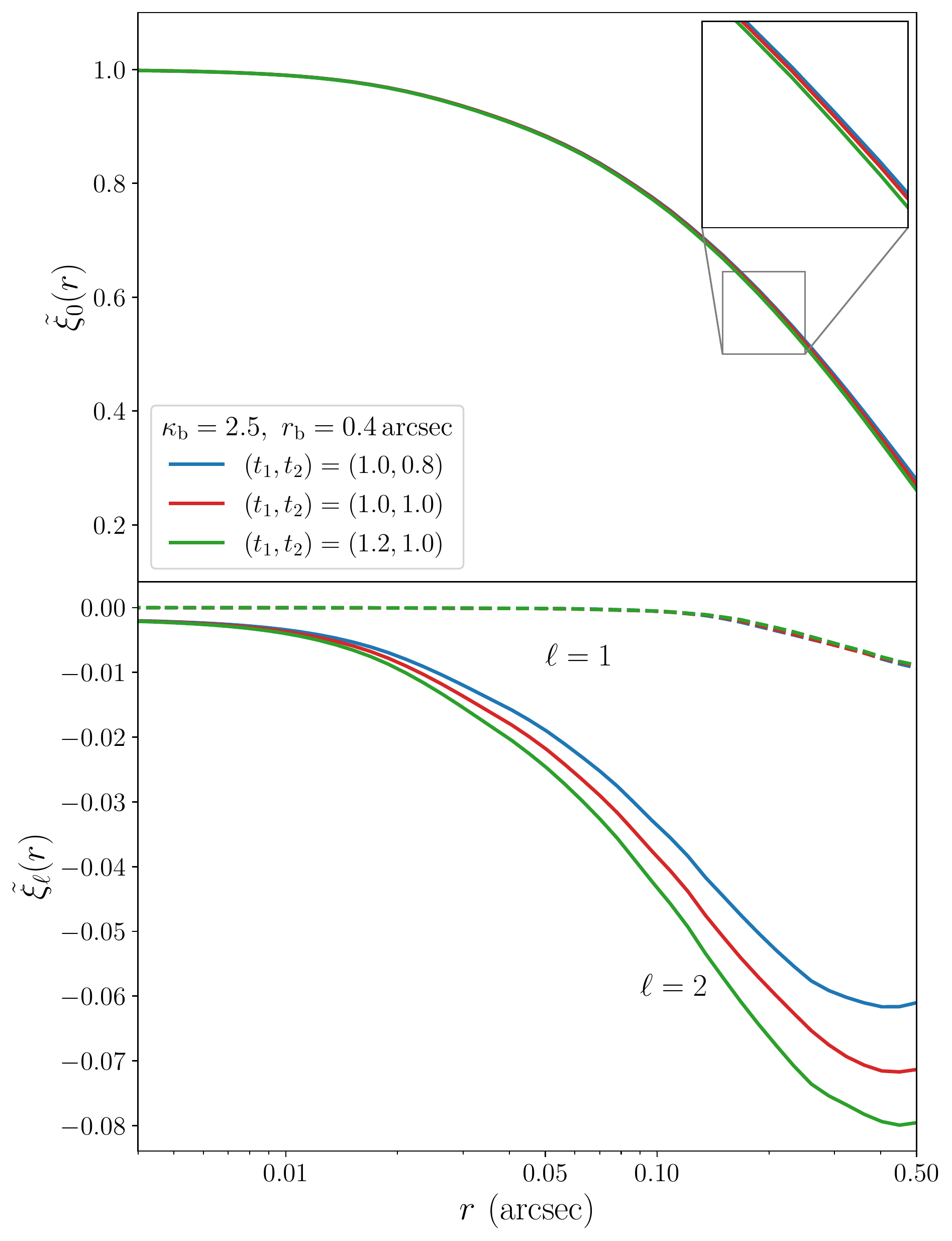}
	\includegraphics[clip, trim=0.22cm 0.2cm 0cm 0cm, width=0.48\textwidth]{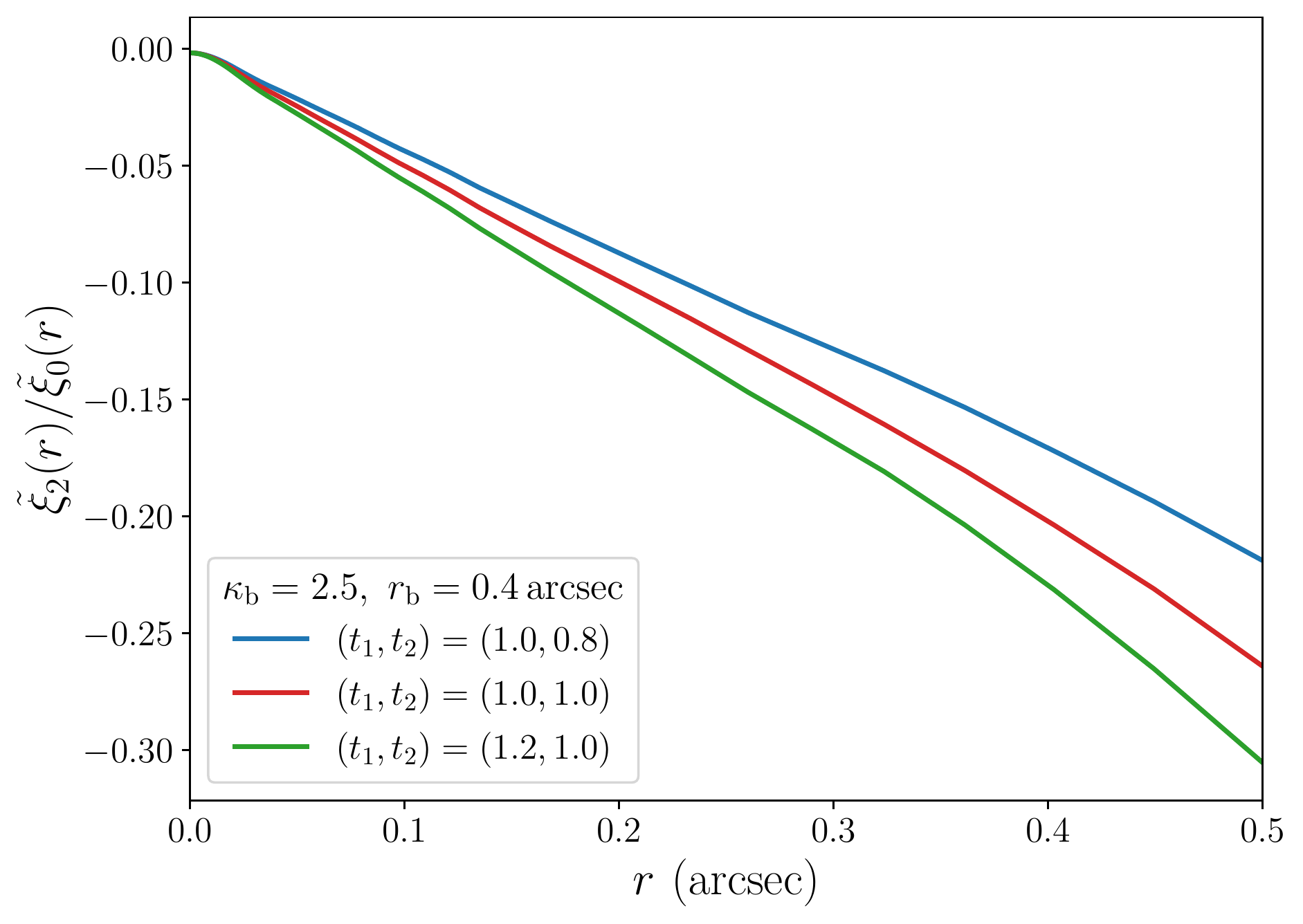}
\end{center}
\caption{\label{fig:corr_multi_BPL_effect}The effects of using an elliptical BPL density profile as the main-lens on the multipole moments of the normalized two-point correlation function and the quadrupole to monopole ratios of the masked $\kappa_{\rm div}$ field of the full CDM halo population as a function of the inner ($t_1$) and outer ($t_2$) slopes are shown in the top and bottom panels, respectively. According to these plots, there is a clear degeneracy between the choice of $t_1$ and $t_2$ and the correlation function multipoles due to the change in the central density of the main-lens.}
\end{figure}

To add a complex radial structure to the main-lens galaxy and see how this implementation affects the multipole moments in the two-point correlation function, we use the broken power-law (BPL) mass profile as our main-lens \citep{O'Riordan_2021, He_2022}. The convergence of the elliptical BPL profile is given by 
\be 
\kappa_0(r)=    \begin{cases}
      \kappa_{\rm b}\left({r_{\rm b}}/{r} \right)^{t_1}, & \text{if}\ r \leq r_{\rm b} \\
      \kappa_{\rm b}\left({r_{\rm b}}/{r} \right)^{t_2}, & \text{if}\ r > r_{\rm b}
    \end{cases},
\ee 
where $r_{\rm b}$ represents the break radius and $\kappa_{\rm b}$ represents the convergence at $r_{\rm b}$. The inner and outer slopes are denoted by the parameters $t_1$ and $t_2$, and when $t_1 = t_2=t$, the elliptical BPL mass profile reduces to the EPL density profile with $\gamma = t+1$. In our study, we change the slopes $t_1$ and $t_2$ while keeping $\kappa_{\rm b}$ and $r_{\rm b}$ constant at 2.5 and 0.4 arcsec, respectively. The effect of using a macrolens model with a complex radial structure on correlation function multipoles is shown in Fig.~\ref{fig:corr_multi_BPL_effect}. Since the choice of $t_1$ and $t_2$ directly affects the enclosed mass and thus the central density of the macro lens, we see significant degeneracies between the shape or amplitudes of multipoles and the inner ($t_1$) and outer ($t_2$) slopes with respect to the EPL mass profile (red line). This confirms that the selection of the main-lens mass profile during the best-fitting image reconstruction process is critical for probing dark matter microphysics.

\begin{figure}
\begin{center}
	\includegraphics[clip, trim=0.22cm 0cm 0cm 0.2cm, width=0.48\textwidth]{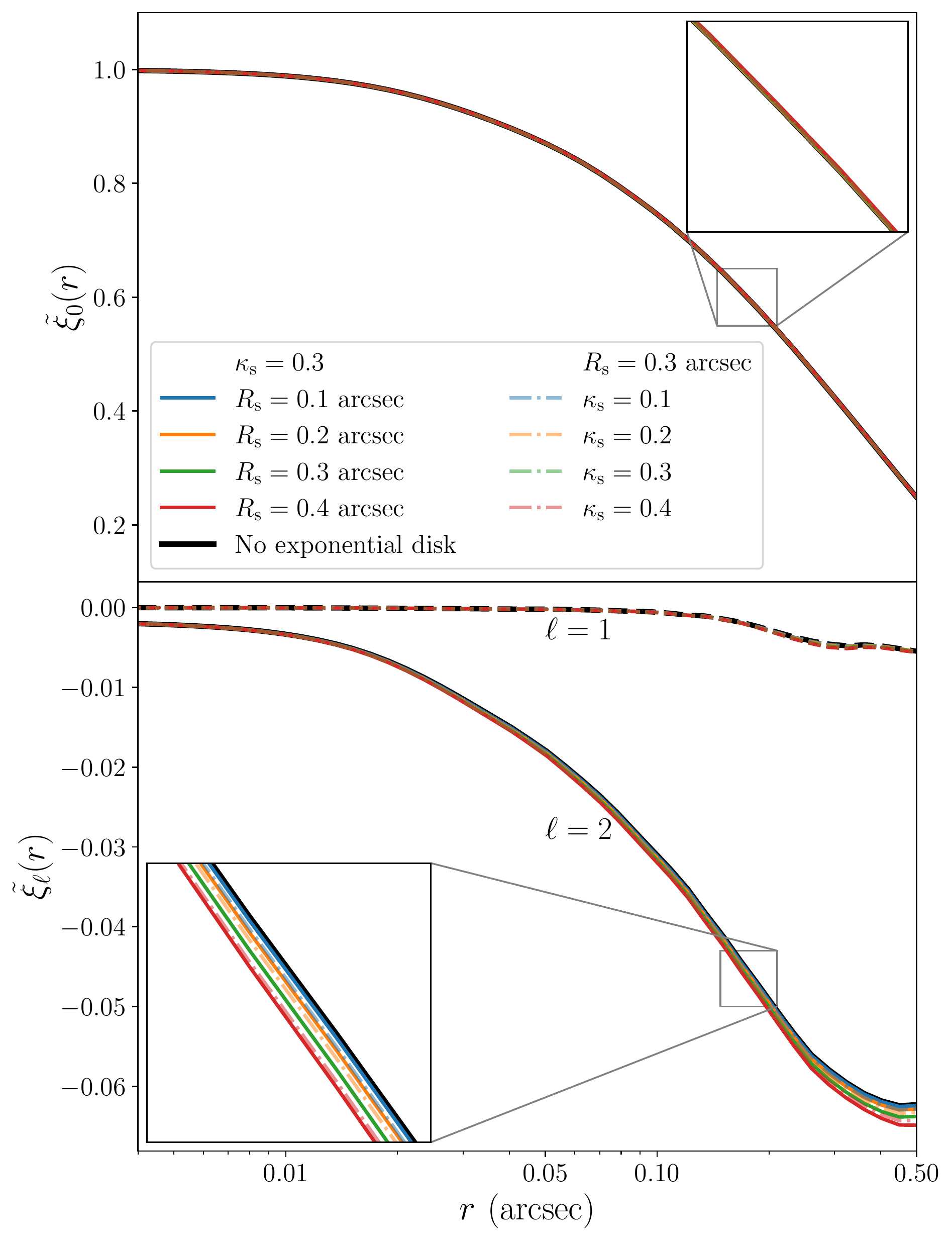}
	\includegraphics[clip, trim=0.22cm 0.2cm 0cm 0cm, width=0.48\textwidth]{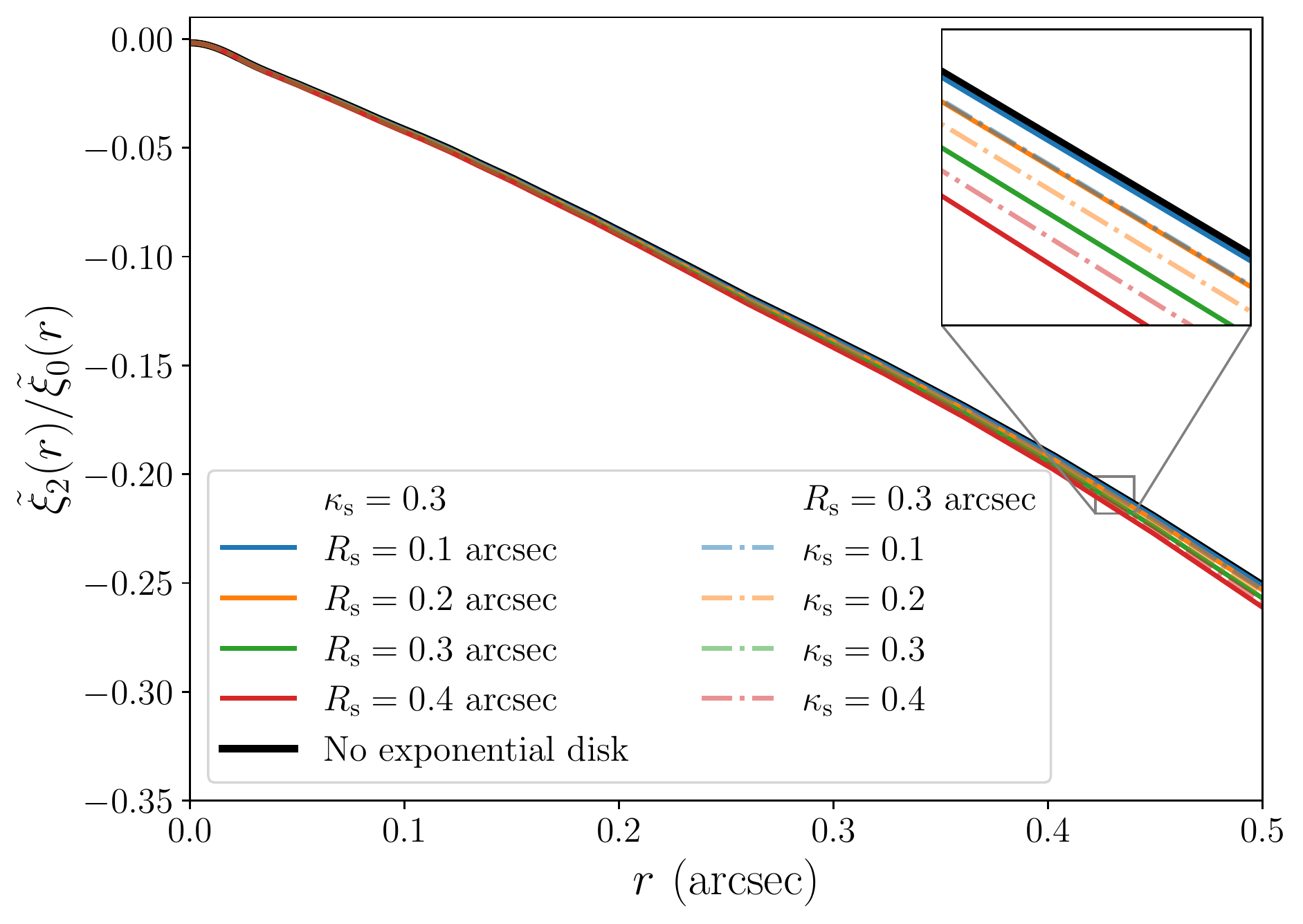}
\end{center}
\caption{\label{fig:corr_exp_disk_effect}The top and bottom panels show the effects of adding an exponential disc to the smooth lens potential on the multipole moments of the normalized two-point correlation function and the quadrupole to monopole ratios of the masked $\kappa_{\rm div}$ field as a function of the parameters of the exponential disc, respectively. The effects of a complex structure in the main-lens model on multipole amplitudes are negligible. }
\end{figure}

Recent quasar lens analyses have shown that adding an exponential disc to a smooth macrolens model, which accounts for the mass of the main-lens galaxy's luminous bulge and its dark matter halo, successfully reconstructs the observed flux ratio anomaly in the lens system \citep{Hsueh_2016, Hsueh_2017}. To test the effects of this complex main-lens structure on multipole measurements, we add an additional mass component in the form of an exponential disc to our previous simple EPL mass model with an external shear. The exponential disc lens model is a subset of the $\rm S\acute{e}rsic$ profile \citep{Cardone_2004} with the $\rm S\acute{e}rsic$ index of $n = 1$. The functional form of the convergence of the face-on thin circular exponential disc with intrinsic central density $\kappa_{\rm s}$ and scale length $R_{\rm s}$ \citep{Keeton_2001, Peng_2010} is given by
\be 
\kappa (r) = \kappa_{\rm s} \exp\left[-\frac{r}{R_{\rm s}} \right].
\ee
First, we keep the value of $\kappa_{\rm s}$ constant at 0.30 and vary the value of $R_{\rm s}$; second, we keep the value of $R_{\rm s}$ constant at $0.3$ arcsec and investigate the changes in multipoles as a function of $\kappa_{\rm s}$ while ensuring that these parameter values are consistent with the best-fitting model parameters reported in \cite{Hsueh_2016, Hsueh_2017}. We also consider a main-lens system without an exponential disc as a benchmark. The effect of having this complex lens structure for the main-lens model on the multipole amplitudes and quadrupole to monopole ratio, which reflects the abundance of line-of-sight haloes, is depicted in Fig.~\ref{fig:corr_exp_disk_effect}. The effect of using an exponential disc in addition to the smooth lens potential is clearly within $1 - \sigma$ of the mean and is not strong enough to cause a change in our measurements within the feasible limits of the exponential disc's intrinsic central convergence and scale length.

\section{The Covariance Matrix of mode amplitudes} \label{App.D}

In this appendix, we derive an expression for an element in the covariance matrix of mode amplitudes given in equation~\eqref{Eq.31_}. 

Using equations~\eqref{Eq.24} and \eqref{Eq.31_}, the variance of the mode amplitudes can be written as

\begin{align} \label{Eq. 36}
   {\bf C}_{{\rm div,}mm'}^{(\ell)}  &= \frac{4}{ A^2}\int_{ A} {\rm d}^2  \rr_2  \,\,  \int_{A}  \,\, {\rm d}^2  \rr_1\,\,  \varphi^*_{m} (\rr_2)\,\, \varphi_{m'} (\rr_1) \en
   & \hspace{4.5cm}\times\left< \kappa_{\rm div}(\rr_2) \kappa_{\rm div}(\rr_1) \right> \en
   &= \frac{4}{ A^2}\int_{ A} {\rm d}^2  \rr  \,\,  \int_{A}  \,\, {\rm d}^2  \rr_1\,\,  \varphi^*_{m} (\rr_2)\,\, \varphi_{m'} (\rr_1) \,\, \xi(\rr) \en
   & = \frac{4}{ A^2} \sum_{\ell=0}^\infty \int_{ A} {\rm d}^2  \rr  \,\, \xi_\ell(r) I \,\,,
\end{align} where

\be \label{Eq.C_2}
I =  \int_{A}  \,\, {\rm d}^2  \rr_{\rm h}\,\,  \varphi_m^*\left(\rr_2=\rr_{\rm h} + \frac{\rr}{2}\right)\varphi_{m'}\left(\rr_1 = \rr_{\rm h} - \frac{\rr}{2}\right) T_\ell(\mu_\rr) \,\,.
\ee

\begin{figure}
\begin{center}
\includegraphics[clip, trim=3.1cm 3.5cm 9.5cm 5.1cm, width=0.48\textwidth]{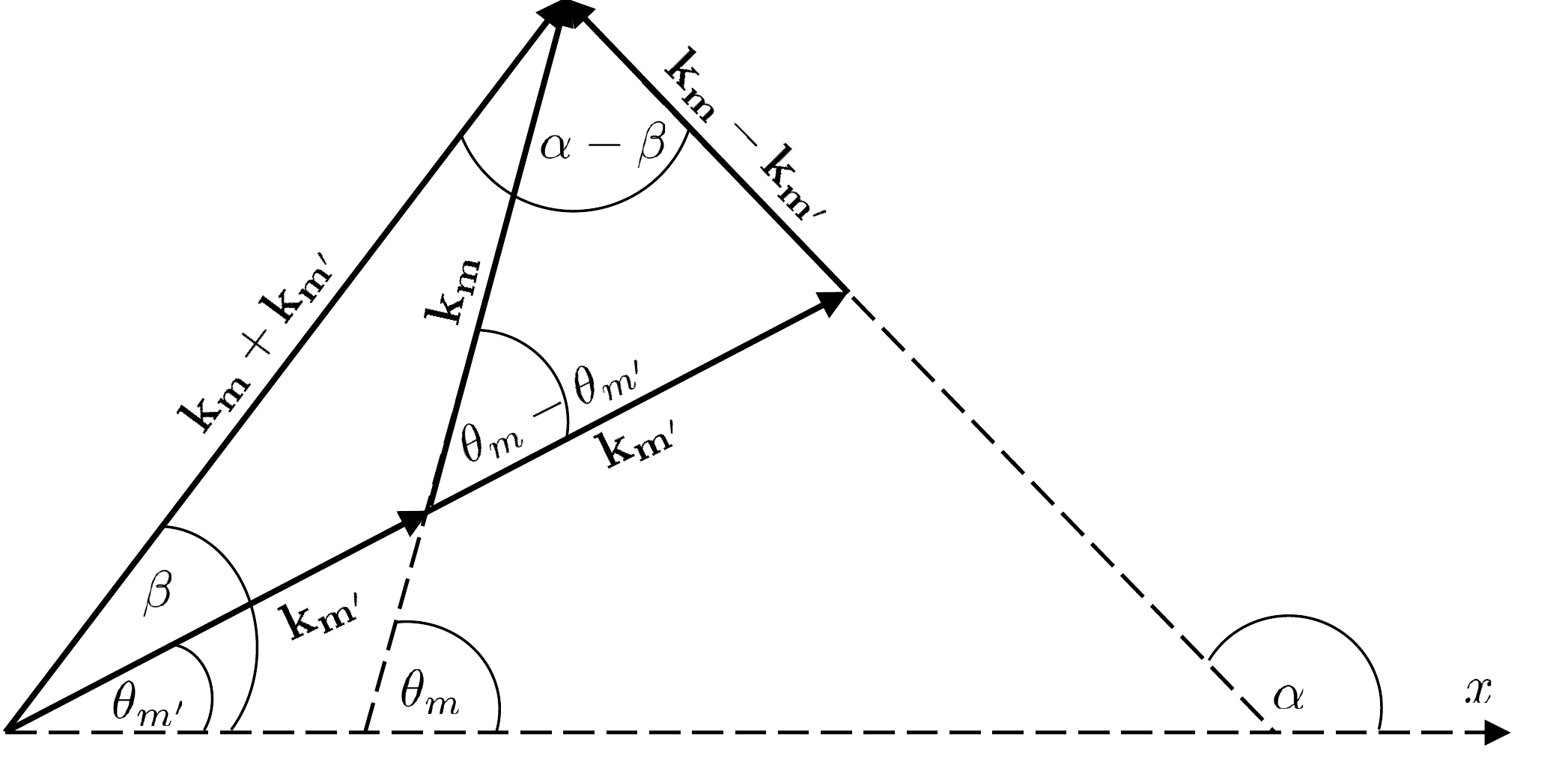}
\end{center}
\caption{\label{fig:kappa_vec} A diagram illustrating the wave vectors $\kk_m$ and $\kk_{m'}$, as well as the angles $\theta_m$, $\theta_{m'}$, $\alpha$, and $\beta$.
}
\end{figure}

Using the discrete orthonormal Fourier basis functions defined in equation~\eqref{Eq.23_} for $\varphi_m(\xx)$ yields

\begin{align}
\varphi_m^*\left(\rr_{\rm h} + \frac{\rr}{2}\right)\varphi_{m'}\left(\rr_{\rm h} - \frac{\rr}{2}\right) & = \frac{1}{k_mk_{m'}} e^{-i(\kk_m-\kk_{m'})\cdot \rr_{\rm h}}\,\, e^{-i(\kk_m+\kk_{m'})\cdot \frac{\rr}{2}}  \en
& = \frac{1}{k_mk_{m'}} e^{-ik'r_{\rm h}\cos(\theta_{\rm \rr_h}-\alpha)}\en 
&\hspace{2.5cm} \times e^{-ik''\frac{r}{2}\cos(\theta_\rr-\beta)},
\end{align} where $k' = |\kk_m-\kk_{m'}|$ and $k'' = |\kk_m+\kk_{m'}|$. Also, $\theta_{\rr_{\rm h}}$, $\theta_\rr$, $\alpha$, and $\beta$ are the angles formed by the vectors $\rr_{\rm h}$, $\rr$, $\kk_m-\kk_{m'}$, and $\kk_m+\kk_{m'}$ with the x -axis, respectively (see Fig.~\ref{fig:2} and Fig.~\ref{fig:kappa_vec}). 

\noindent Now consider the integration given in equation~\eqref{Eq.C_2}.

\begin{align} \label{Eq. 38}
 I & = \frac{e^{-ik''\frac{r}{2}\cos(\theta_\rr-\beta)}}{k_mk_{m'}} \int_{A}  \,\, {\rm d}^2  \rr_{\rm h}\,\, e^{-ik'r_{\rm h}\cos(\theta_{\rm \rr_h}-\alpha)}T_\ell(\cos(\theta_{\rm \rr_h}-\theta_\rr)).
\end{align}
Using the Jacobi--Anger expansion given in equation~\eqref{Eq.30}, we can expand equation~\eqref{Eq. 38} as

\begin{align}\label{Eq. 40}
    I & = \frac{e^{-ik''\frac{r}{2}\cos(\theta_\rr-\beta)}}{k_mk_{m'}}\sum_{n=0}^{\infty} (2-\delta_{n0}) (-i)^n  \int {\rm d}r_{\rm h} \, r_{\rm h} \, J_n(k'r_{\rm h}) \en
    & \hspace{1.5cm}\times \int_0^{2\pi}{\rm d}\theta_{\rm \rr_h}T_\ell(\cos(\theta_{\rm \rr_h}-\theta_\rr)) T_n(\cos(\theta_{\rm \rr_h}-\alpha)) \en
    & = \frac{e^{-ik''\frac{r}{2}\cos(\theta_\rr-\beta)}}{k_mk_{m'}}\sum_{n=0}^{\infty} (2-\delta_{n0}) (-i)^n  \int {\rm d}r_{\rm h} \, r_{\rm h} \, J_n(k'r_{\rm h}) \en
    & \hspace{1.5cm} \times \frac{2\pi}{2-\delta_{\ell0} } T_\ell(\cos(\theta_\rr-\alpha)) \de_{n\ell}\en
    & = 2\pi (-i)^\ell \frac{e^{-ik''\frac{r}{2}\cos(\theta_\rr-\beta)}}{k_mk_{m'}} T_\ell(\cos(\theta_\rr-\alpha)) \int {\rm d}r_{\rm h} \, r_{\rm h} \, J_\ell(k'r_{\rm h})\en
    & =\frac{2\pi (-i)^\ell}{k_mk_{m'}} \sum_{n=0}^{\infty} (2-\delta_{n0}) (-i)^n J_n\left(\frac{k''r}{2}\right) \en
    & \hspace{1.5cm} \times T_n(\cos(\theta_\rr -\beta)) T_\ell(\cos(\theta_\rr-\alpha)) \int {\rm d}r_{\rm h} \, r_{\rm h} \, J_\ell(k'r_{\rm h})
\end{align}

\noindent Now use the result given in equation~\eqref{Eq.35} to integrate equation~\eqref{Eq. 40} over $\theta_\rr$. 

\begin{align}
    \int_0^{2\pi}{\rm d}\theta_\rr\, I & = \frac{2\pi (-i)^\ell}{k_mk_{m'}} \sum_{n=0}^{\infty} (2-\delta_{n0}) (-i)^n J_n\left(\frac{k''r}{2}\right ) \en
    & \hspace{1.4cm} \times \int_0^{2\pi}{\rm d}\theta_\rr\, T_n(\cos(\theta_\rr -\beta)) T_\ell(\cos(\theta_\rr-\alpha)) \en 
    & \hspace{4.3cm} \times  \int {\rm d}r_{\rm h} \, r_{\rm h} \, J_\ell(k'r_{\rm h}) \en
    & = \frac{(2\pi)^2 (-i)^{2\ell}}{k_mk_{m'}} J_\ell\left(\frac{k''r}{2}\right )\, T_\ell(\cos(\alpha-\beta))  \en
     & \hspace{4.3cm} \times \int {\rm d}r_{\rm h} \, r_{\rm h} \, J_\ell(k'r_{\rm h})\,\,. \en
\end{align}

\noindent Then equation~\eqref{Eq. 36} can be finalized as

\begin{align} \label{Eq. 42}
   {\bf C}_{{\rm div,}mm'}^{(\ell)}  & = \frac{16\pi^2 (-1)^\ell}{A^2k_mk_{m'}} \int {\rm d}r \, r\,  \xi_\ell(r) \, J_\ell\left(\frac{k''r}{2}\right )\, T_\ell(\cos(\alpha-\beta)) \en
   & \hspace{4.6cm} \times \int {\rm d}r_{\rm h} \, r_{\rm h} \, J_\ell(k'r_{\rm h})\en
   & = \frac{8\pi(-i)^\ell}{A^2k_mk_{m'}} P_\ell\left(\frac{k''}{2}\right )\, T_\ell(\cos(\alpha-\beta)) \int {\rm d}r_{\rm h} \, r_{\rm h} \, J_\ell(k'r_{\rm h})
\end{align} where 

\begin{align}
   k' & = |\kk_m-\kk_{m'}| = \sqrt{k_m^2 + k_{m'}^2 - 2k_mk_{m'}\cos(\theta_m-\theta_{m'})},\en
   k'' & = |\kk_m+\kk_{m'}| = \sqrt{k_m^2 + k_{m'}^2 + 2k_mk_{m'}\cos(\theta_m-\theta_{m'})},\en
{\rm and} \en
   \cos(\alpha-\beta) & = \frac{k_m^2-k_{m'}^2}{\sqrt{k_m^4 + k_{m'}^4 - 2k_m^2k_{m'}^2\cos[2(\theta_m-\theta_{m'})]}}.
\end{align}

Consider the diagonal elements of the covariance matrix. Since $m=m'$, $k'=0$ and $k''=2k_m$. We know that $J_\ell(k'r_{\rm h})=J_\ell(0)=\de_{0\ell}$. Then equation~\eqref{Eq. 42} can be simplified as

\begin{align} \label{Eq. 44}
   {\bf C}_{{\rm div,}mm}^{(\ell)} & = \frac{4P_0\left(k_m\right )}{Ak_m^2} \, \de_{0 \ell} \,\,.
\end{align}

It is worth noting that diagonal elements are non-zero only in the monopole ($\ell = 0$) case and zero in all other cases.


\bsp	
\label{lastpage}
\end{document}